%% file: main.tex
\documentclass[pdflatex,sn-mathphys-num]{sn-jnl}

\usepackage{graphicx}%
\usepackage{multirow}%
\usepackage{amsmath,amssymb,amsfonts,bm}%
\usepackage{amsthm}%
\usepackage{mathrsfs}%
\usepackage[title]{appendix}%
\usepackage{xcolor}%
\usepackage{textcomp}%
\usepackage{manyfoot}%
\usepackage{algorithm}%
\usepackage{algorithmicx}%
\usepackage{algpseudocode}%
\usepackage{listings}%
\usepackage{array}
\usepackage{booktabs}
\usepackage{arydshln}

\usepackage{caption}%

\usepackage{tabularx}
\usepackage[export]{adjustbox}

\theoremstyle{thmstyleone}%
\theoremstyle{thmstyletwo}%

\theoremstyle{thmstylethree}%
\renewcommand{\state}[2]{\boldsymbol{x}_{#1}^{#2}}
\newcommand{\pd}[2]{\cfrac{\partial{#1}} {\partial{#2}}}
\newcommand{\obsvec}[2]{\boldsymbol{y}_{#1}^{#2}}
\newcommand{\obsop}[1]{\mathcal{H}\left(#1\right)}
\newcommand{\enkfmean}[2]{\bm{\mu}_{#1}^{#2}}

\newcommand{\enkfanomaly}[2]{\bm{\Lambda}_{#1}^{#2}}
\newcommand{\enkfcovariance}[3]{\bm{\Xi}_{#1}^{#2,#3}}
\newcommand{\model}[3]{\mathcal{M}_{#1}^{#2}\left(#3\right)}

\newcommand{\princ}{\bm{\chi}}
\newcommand{\cont}{\bm{\gamma}}
\newcommand{\anc}{\bm{\eta}}
\newcommand{\tot}{\bm{\varkappa}}

\newcommand{\expectprinc}{\bm{\mu}_{\princ}}
\newcommand{\expectcont}{\bm{\mu}_{\cont}}
\newcommand{\expectanc}{\bm{\mu}_{\anc}}
\newcommand{\expecttot}{\bm{\mu}_{\tot}}

\newcommand{\principalstate}[2]{\boldsymbol{x}_{#1}^{\princ^#2}}
\newcommand{\controlstate}[2]{\boldsymbol{x}_{#1}^{\cont^#2}}
\newcommand{\ancillarystate}[2]{\boldsymbol{x}_{#1}^{\anc^#2}}
\newcommand{\totalstate}[2]{\boldsymbol{x}_{#1}^{\tot^#2}}

\newcommand{\coarsemodel}{\mathcal{M}^{\textbf{C}}}
\newcommand{\finemodel}{\mathcal{M}^{\textbf{F}}}

\newcommand{\princmodel}{\mathcal{M}^{\princ}}
\newcommand{\contmodel}{\mathcal{M}^{\cont}}
\newcommand{\ancmodel}{\mathcal{M}^{\anc}}

\newcommand{\totalobsop}{\widetilde{\mathcal{H}\left(\totalstate{i,k}{f} \right)}}

\newcommand{\projPlus}{\bm{\mathcal{P}^+}}
\newcommand{\projMinus}{\bm{\mathcal{P}^-}}

\newcommand{\mfenkfcov}[2]{\bm{\Upsilon}^{#1,#2}}

\newcommand{\kalmangain}{\bm{K}}

\begin{document}


\title[Assisted Turbulence Modeling via Data Assimilation and Machine Learning for Turbomachinery Flows]{Assisted Turbulence Modeling via Data Assimilation and Machine Learning for Turbomachinery Flows}


\author*[1]{\fnm{Tom} \sur{Moussie}}\email{tom.moussie@ensam.eu}
\equalcont{These authors contributed equally to this work.}

\author[1]{\fnm{Pierric} \sur{Joseph}}
\equalcont{These authors contributed equally to this work.}

\author[1]{\fnm{Marcello} \sur{Meldi}}
\equalcont{These authors contributed equally to this work.}

\affil*[1]{Univ. Lille, CNRS, ONERA, Arts et Métiers ParisTech, Centrale Lille, UMR 9014-LMFL- Laboratoire de Mécanique des fluides de Lille, Kampé de Feriet, F-59000, Lille, France}


\abstract{This work presents the development and assessment of a hybrid framework coupling numerical simulation, experimental samples, data assimilation (DA) and machine learning (ML) to investigate the flow within an axial compressor stage. The methodology first implements a Multi-Fidelity Ensemble Kalman Filter (MFEnKF) which integrates experimental PIV data with a two-fidelity numerical simulation setup, combining steady-state RANS on a fixed grid with unsteady hybrid RANS/LES simulations on a moving grid. This data assimilation step successfully reduces discrepancies between experiments and CFD realizations. Besides, it provides additional high fidelity samples when compared with the available PIV data. This enriched database is used to train machine learning (ML) tools, with the objective to obtain black-box forcing terms to be included in the discretized form of the Navier-Stokes equations. These models are expected to preserve the accuracy of the DA results while not longer requiring the input of experimental data. The neural network architecture used here maps local physical features such as the velocity, the velocity gradients and the turbulent viscosity, to obtain the volume forcing terms. The analysis of the performance of such ML models shows that global accuracy is preserved but localized increased errors in specific regions of the compressor, such as close to the blade tips, can exceed those observed from the baseline RANS simulations. To resolve this issue, a Mixture of Experts (MoE) method is employed, which classifies and combines the trained ML tools for different flow regions. The discrepancy with experimental data of this method is systematically lower than the one observed for the baseline RANS solution and very similar to the one obtained with the MFEnKF.}




\maketitle


\input{sections/intro}

\input{sections/experiments}
\input{sections/numerical_ingredients}
\input{sections/data_assimilation_work}
\input{sections/ann_enhancement}

\input{sections/conclusion}
\input{sections/appendix}

\section*{Acknowledgments}

This research work was carried out using TGCC storage and computing resources on the HPC Joliot Curie's Skylake partition, in the framework of the GENCI grant A020A01741. The present work is funded by the French Ministry of the Armed Forces - Defense Innovation Agency (AID) and by the Hauts-de-France Region.

\bibliography{ref}

\end{document}

%% file: sections/intro.tex
\section{Introduction}

The improvement of the performance of energy devices in many industrial sectors is a key objective in current research activities in fluid mechanics. Among the objectives targeted, one can list increased efficiency as well as strict reduction of the carbon footprint associated with the industrial processes. Among the sectors that are strongly affected by the achievement of such objectives, aircraft engine manufacturing needs continuous advancement due to environmental regulation and fluctuating energy prices. Among all the components of aircraft engines, the axial compressors which equip the vast majority of the current engine generation are under severe scrutiny. In fact, they need to balance requirements for high energy efficiency (with respect to the Brayton-Joule thermodynamic cycle) with aircraft safety regulations due to the potential stability issues encountered at high pressure ratio and low mass flow rates. The importance of these components for the aeronautic industry resulted in extended investigations of several research teams since the 1940s, as comprehensively discussed in the review by Day \citep{Day2015}. Scientific advancement in recent decades highlights the global gains in the use of smaller and globally more compact engine cores to increase thermal efficiency \citep{Nasa_Short_Core}. This produces higher mechanical loads for the compressor stages, which has led to the development and use of unconventional design of the stages such as high turning stator blades \citep{Liu2025} or rotors with tandem blades \citep{Tandem2023}, possibly coupled with active flow control \citep{LiReview}.
The development of these exotic rotor designs raise questions about the ability of the current industrial tools to correctly predict the global performance for such complex designs, in particular in terms of the prediction of the flow field around the blades. These flows are usually at high Reynolds regimes and they exhibit compressibility effects, therefore requiring advanced strategies for accurate prediction. 
These new research directions add to various existing or upcoming problems inherent to the implementation of such machines in aero engines. First, both military applications (with stealth requirement \citep{Rao2002}) and civil ones (such as new architectures with boundary layer ingestion \citep{Yildirim2022}) can require engines to be deeply buried inside the fuselage. This can lead to severe levels of inlet distortion \citep{Cousins2004} with, again, an urgent need to correctly assess the performance of the machines in such degraded conditions. Additionally, axial compressors are known to exhibit diverse and complicated patterns when transitioning to stall \citep{Camp1997} and early detection of such unstable regimes is still an open topic to date \citep{Wilson2025}. This calls, here again, for fast and computationally efficient methods to predict flow across compressor stages. 
Experiments and high fidelity numerical simulations, such as Direct Numerical Simulation (DNS) \cite{Pope2000_cambridge} and Large Eddy Simulation (LES) \cite{Sagaut2006_springer}, are the most reliable tools which can be used to assess and enhance the design strategies previously described. However, these investigative methods are expensive for turbomachinery flows and therefore comprehensive analyses considering several geometric features for multiple working configurations are virtually impossible. To this end, lower-fidelity strategies such as Reduced Order Models (ROM) \cite{Lumley1970_AP} or Reynolds-Averaged Navier-Stokes (RANS) simulation \cite{Wilcox2006_DCW} can be used to investigate a large range of configurations. Computational costs are here reduced permitting a larger degree of freedom on the investigations thanks to the production of larger databases, at the expense of lower accuracy in the prediction of the results. Recent research has aimed to improve the accuracy of such models using tools from Machine Learning and data-driven architectures \cite{Duraisamy2019_arfm}. Several promising techniques have been developed in particular for RANS turbulence modeling \cite{Schmelzer2020_ftc,Volpiani2021_prf}, which have shown increased accuracy for specific test cases. However, difficulties have been observed when applying models trained with specific datasets to different test cases, highlighting a specific challenge in extrapolation and generalizability of such models.   

The difficulties arising from the different features and resources required for investigative tools can be reconciled using methods from Data Assimilation (DA) \cite{Daley1991_cambridge,Asch2016_SIAM,Evensen2022_springer}. This branch of statistics \cite{Simon2006_wiley} is devoted to obtaining accurate state estimation via data fusion from different sources of information. One notable feature of DA tools is the possibility to combine predictions from low-fidelity models, which can usually span very large parametric spaces producing quasi-continuous solutions in space and time, with limited measurements obtained from high-fidelity tools. For the latter, scarcity of data can be associated with limited availability of configurations for selected parameters of the setup of the problem investigated (due to computational cost requirement for DNS and LES) as well as to physical or temporal constraints in data acquisition, due for example to accessibility problems in experiments.      

In the present work, an approach based on the Ensemble Kalman Filter (EnKF) \cite{evensen_sequential_1994,Evensen2009_ieee} is used to investigate the performance of the CME2 compressor test bench. The geometry of this test case is simpler than the ones observed in operative machines, but still permits to fully investigate the problematic aspects observed in realistic industrial and aerospace applications. Data from different sources are combined to obtain a robust state estimation of the flow. The methodology used to this purpose relies on a multifidelity approach \cite{popov_multifidelity_2021} where several RANS realizations and fewer hybrid RANS/LES runs are combined with experimental measurements. The latter are available only on three two-dimensional planes within the compressor blades.  Therefore they are localized on a small part of the domain of investigation. The DA state estimate thus obtained is then used to train corrective ML black box tools, which are generated using Neural Networks (NN), following the approach proposed by \citet{Villiers2025_ftc}. Similar combinations of ensemble DA and ML have recently shown promising results for the state estimation of unsteady flows \citep{Valero2025_cf,Moussie026_cf}. These ML models are trained with the objective to reduce the discrepancy between the DA state estimation and the prediction of the low-fidelity RANS simulations, therefore targeting an increase of the predictive accuracy of the latter. The NN generative tools are then coupled using turbulent models and validated via a posteriori investigation and comparison with experimental data.

The manuscript is structured as follows. In \autoref{sec:expe}, the experimental setup and the data acquisition methods are presented. In \autoref{sec:num_ingredients}, the numerical ingredients, including the dynamic equations and the data assimilation procedure, are introduced. Then, \autoref{sec:num_cme2} provides an introductory numerical overview of the CME2 axial compressor. It is followed by \autoref{sec:data_assimilation_work} in which the data assimilation algorithm is applied, thanks to the experimental and numerical data respectively obtained in \autoref{sec:expe} and \autoref{sec:num_cme2}. Then, \autoref{sec:ann_enhancement} provides an overview of the machine learning tools used, presenting their architecture, training data and post-training metrics, as well as the results of integrating pre-trained tools into CFD solvers. Finally, concluding remarks are drawn in \autoref{sec::conclusions}.

%% file: sections/experiments.tex
\section{Test case of investigation: the CME2} \label{sec:expe}

\begin{figure}[h!]
    \centering
    \includegraphics[width=0.65\linewidth]{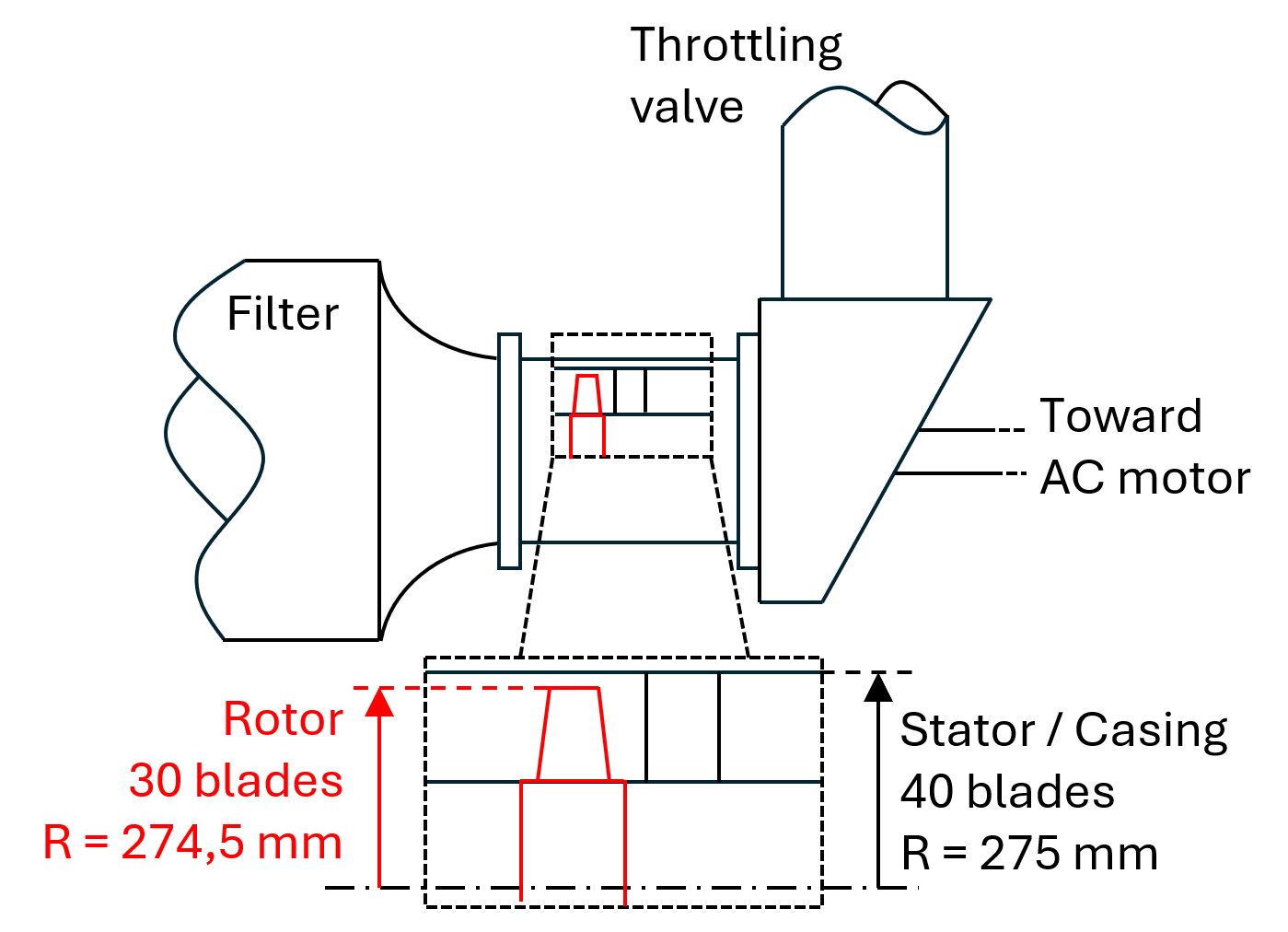}
    \caption{Schematic description and main dimensions of the CME2 axial compressor.}
    \label{fig:cme2_compressor}
\end{figure}

\noindent The test case investigated in the present study is the \textit{Compresseur Mono-Étage, version 2} (CME2) which is a compressor test bench located at Arts et Métiers, Institute of Technology, in Lille, France. This rig, which is shown in \autoref{fig:cme2_compressor}, consists of a single stage, low-speed, axial compressor operated using an electrical AC motor of 200 kW. The whole setup also includes an instrumented inlet section to measure the mass flow rate and to filter the incoming air, along with a throttling valve at the exhaust to control the operating point of the machine.

\noindent The stage is composed of a $30$-blade rotor and a $40$-blade stator, positioned in a casing with a radius of $275$\,mm, as schematized in \autoref{fig:cme2_compressor}. The rotor tip gap is $0.5$\,mm. The mass flow rate for $3\,200\,$rpm is $5.3\,$kg $\cdot$ s$^{-1}$ at the design point, which corresponds to a pressure ratio around $1.03$. The Mach numbers at the inlet and at the rotor tip are $Ma_\infty = 0.12$ and $Ma_T = 0.28$, respectively. 

\begin{figure}[h!]
    \centering
    \begin{tabular}{cc}
       \includegraphics[width=0.45\linewidth]{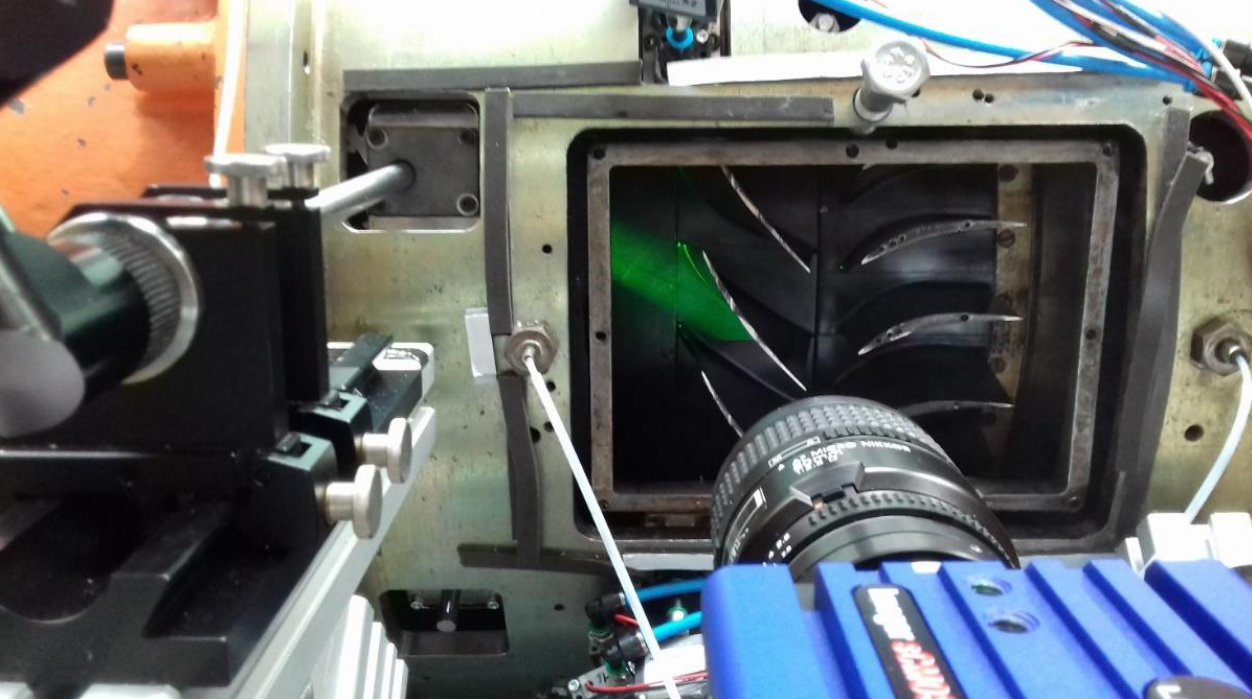} & \includegraphics[width=0.45\linewidth]{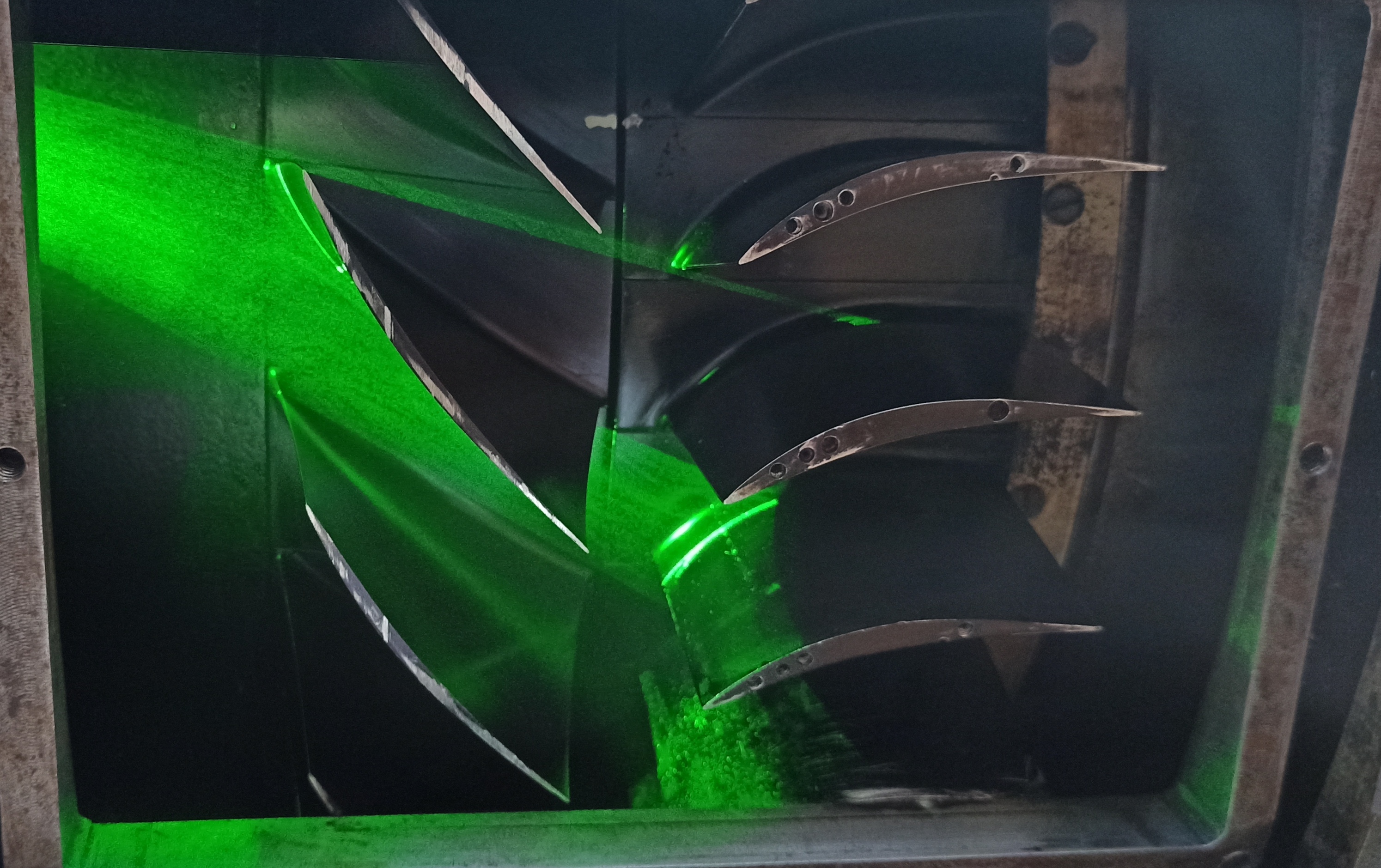} \\
        (a) & (b)
    \end{tabular}
    \caption{(a) Experimental setup showing the laser beam and the camera; (b) Close-up of the laser beam in the compressor.}
    \label{fig:PIV_setup_photo}
\end{figure}

The test bench in Lille has been used for several experimental campaigns to study problematic aspects observed in realistic configurations with different techniques and acquisition systems. Among the different tools used, one of interest for the present analysis is the Particle Image Velocimetry (PIV) which has been used to measure the velocity field inside the stage of the rotating machine. This system has been extensively described in previous articles from the research group (see \citep{MouboghaASME,BaretterJPP,Alekseik2025}), but the main characteristics are here described for the sake of completeness. This is a standard two-dimensional two-component (2D2C) equipment able to sample the velocity field over multiple planes located at a constant radius of the blades in the rotor section of the stage. The system itself is composed of a multi-pulse laser (Innolas Compact 400 PIV) coupled to an sCMOS camera from LaVision (Imager, $2\,560$ $\times$ $2\,160$ pixels). Optical access is performed via a transparent window on the side of the compressor.
The projection of the laser beam into the compressor at the right location requires a sophisticated setup involving an articulated arm fitted with various mirrors and an endoscope at the end. Seeding is usually realized with a smoke generator installed at the compressor inlet. Additional pictures of the setup are provided in \autoref{fig:PIV_setup_photo} for illustration purposes.

\begin{figure}[h!]
    \centering
    \includegraphics[width=0.65\linewidth]{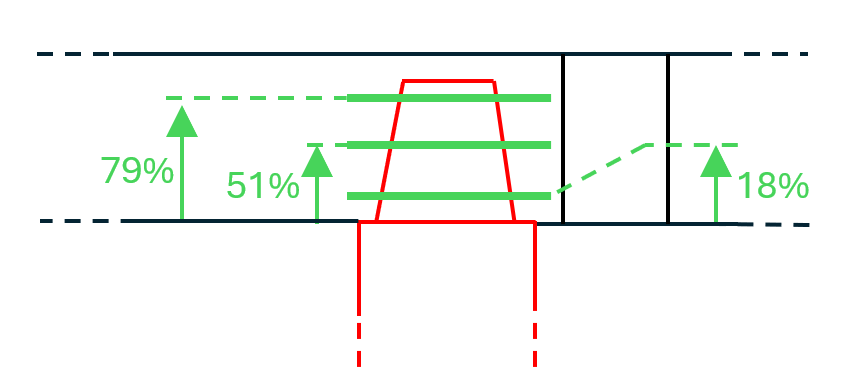}
     \caption{Positioning of PIV planes in CME2, with respect to the blade height, close to the hub (\textbf{P1} at 18\% of the blade's height), mid span (\textbf{P2} at 51\% of the blade's height) and close to the tip (\textbf{P3} at 79\% of the blade's height).}
    \label{fig:PIV_plan_position}
\end{figure}

\begin{figure}[h!]
    \centering
    \begin{tabular}{ccc}
         \includegraphics[width=0.3\linewidth]{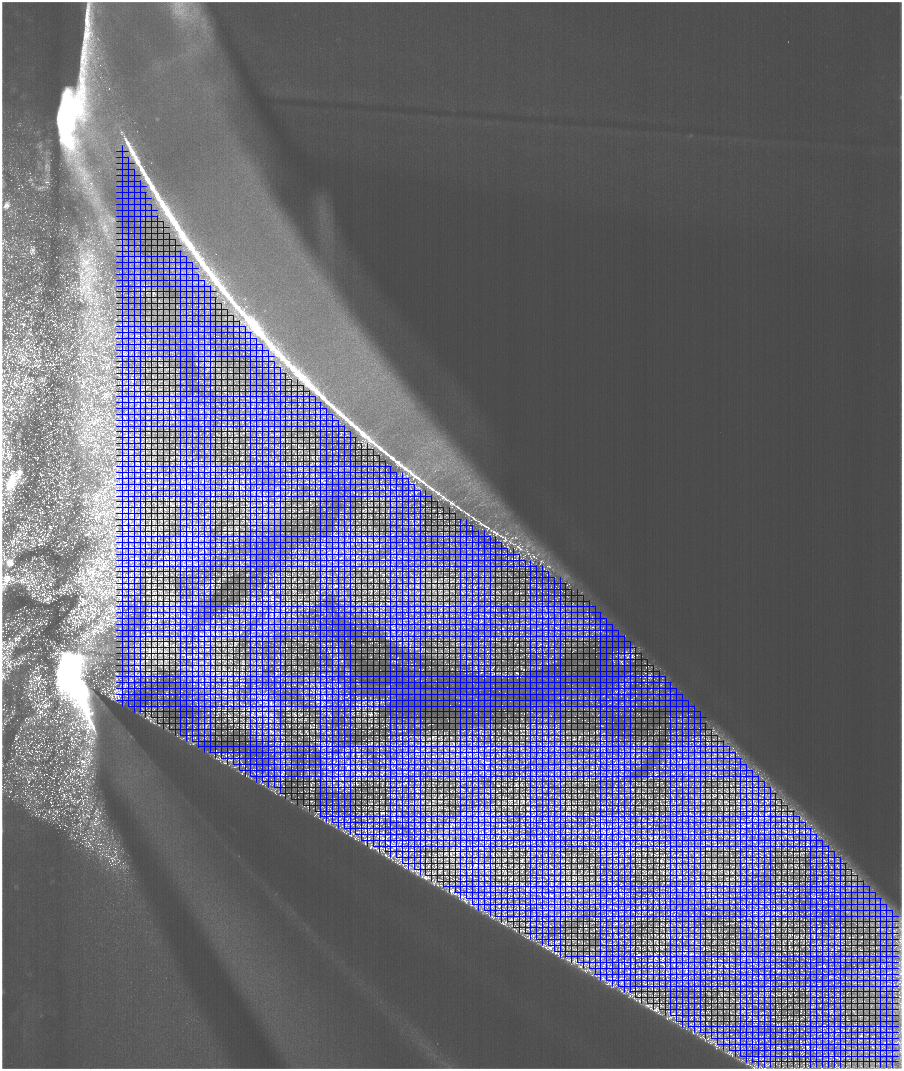}  &  \includegraphics[width=0.3\linewidth]{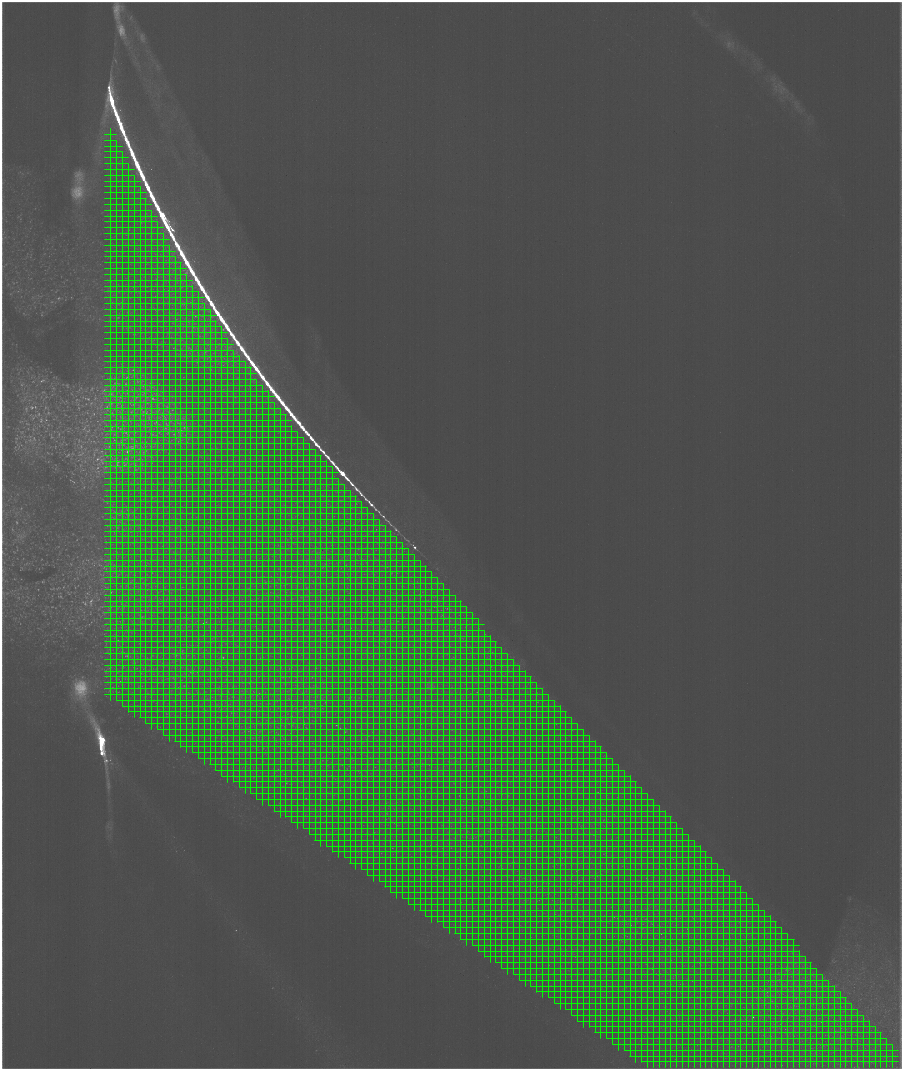} &  \includegraphics[width=0.3\linewidth]{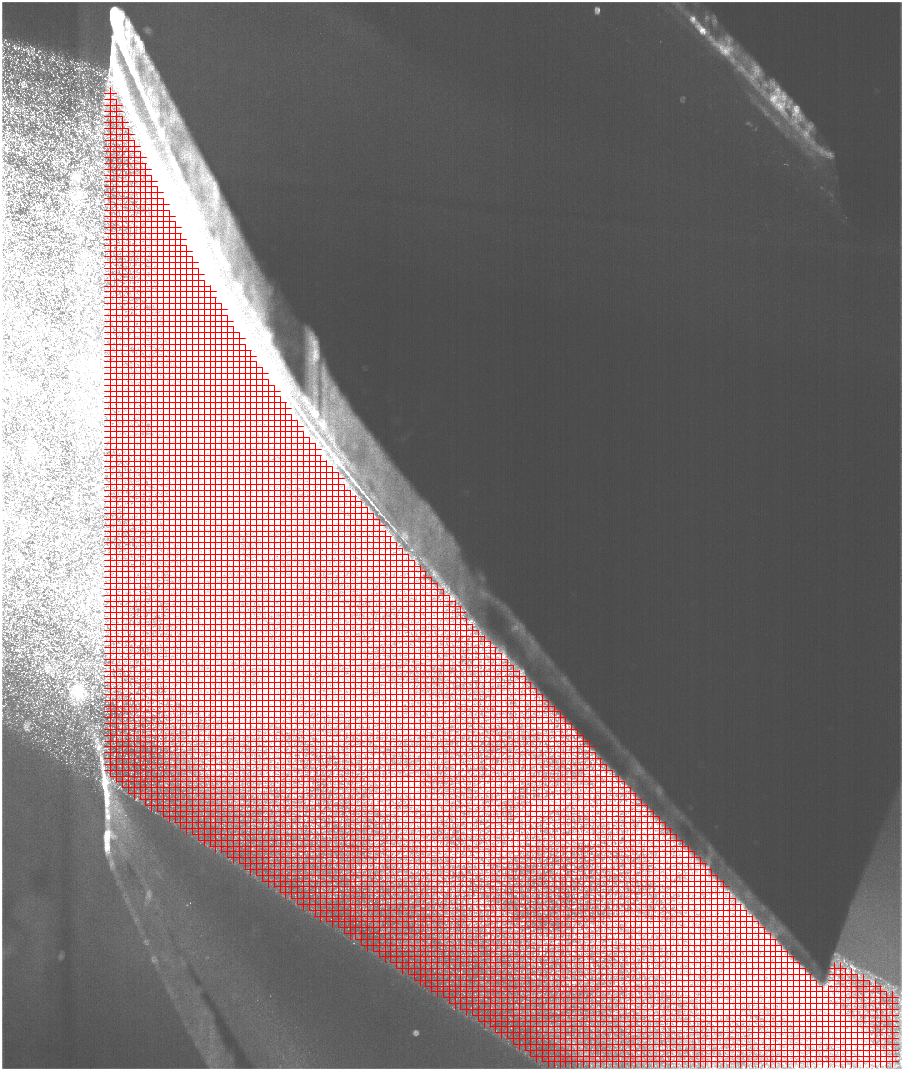} \\
         (a) & (b) & (c) \\[0.2cm]
    \end{tabular}
    \caption{PIV planes shown for (a) \textbf{P1}; (b) \textbf{P2}; (c) \textbf{P3}.}
    \label{fig:plans_piv_bas_mid_haut}
\end{figure}

Measurements used here have been obtained on planes at $18\,\%$, $51\,\%$ and $79\,\%$ of the blade height (see \autoref{fig:PIV_plan_position}), respectively referred to as \textbf{P1}, \textbf{P2} and \textbf{P3}, giving a field of view of approximately $64 \times 76\,\textrm    {mm}^2$. A small variation of the size of this area is reported between the three locations due to the change in blade shape with the radius, and the corresponding planes are shown in \autoref{fig:plans_piv_bas_mid_haut}.

\begin{figure}[h!]
    \centering
    \includegraphics[width=0.99\textwidth]{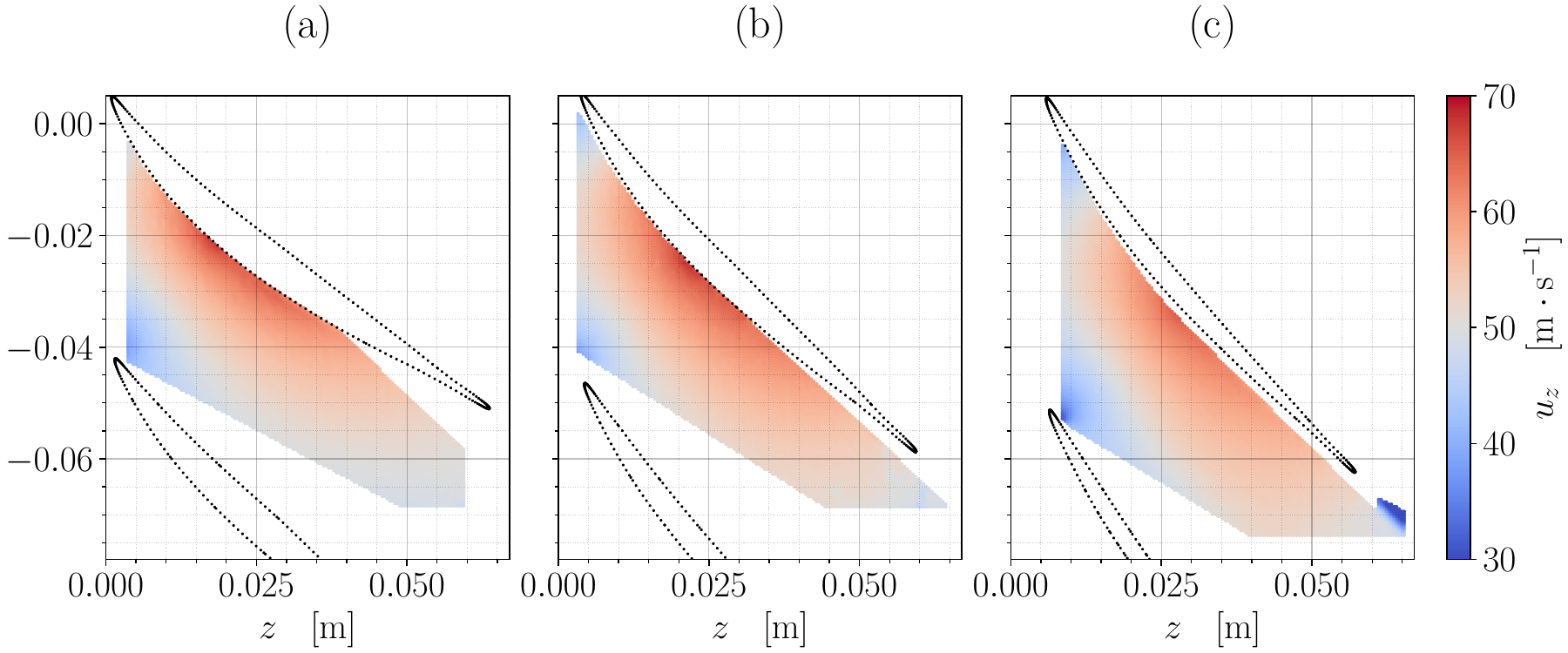}
    \caption{Axial velocity $u_z^{\textrm{PIV}}$ measurements obtained via PIV for (a) \textbf{P1}; (b) \textbf{P2}; (c) \textbf{P3}.}
    \label{fig:maps_axial_velocity_PIV}
\end{figure}

The resulting pairs of images are post-processed using a code developed by the team based on a multi-pass correlation algorithm. In the present case, the final velocity fields have a spacing of $0.5$\,mm between adjacent vectors after using three interrogation windows (subsequently $64 \times 64$ pixels, $32 \times 32$ pixels and $64 \times 64$ pixels with $50\,\%$ overlap). Measurements were repeated 100 times at fixed intervals and averaged to produce the data used in this study. \textbf{P1}, \textbf{P2} and \textbf{P3} maps give access to $8\,180$, $6\,944$, and $8\,090$ data points where the velocity field is measured, respectively. For each point, two components of the velocity field are available. The first corresponds to the flow axial velocity (aligned with the compressor axis) and the second to a projection of the tangential velocity for the vertical component. The projection error (due to the casing curvature) is estimated to be less than $1\,\%$, considering that the PIV observation region is small when compared with the global size of the compressor. The relative velocity can be computed using the local rotational velocity of the blade at each radius. The axial velocity mapped on the PIV point locations is displayed in \autoref{fig:maps_axial_velocity_PIV}. Globally, the measurement uncertainty of the PIV samples is estimated to be around $3\,\%$ (gathering together multiple error sources such as calibration, peak locking and other noise sources). 

An additional uncertainty is related to the operating point of the compressor itself, as the working condition is set using a pair of pressure sensors (First Sensors BTE5000). First, the mass flow rate is measured using the static pressure drop of the converging section, assuming no velocity in the inlet settling chamber. Taking into account the various sources of error (temperature, etc.),  the uncertainty on the mass flow is estimated to $\pm0.06 \,$kg$\cdot$s$^{-1}$ ($\pm1\,\%$ of the design mass flow rate). Second, the pressure rise is directly logged by measuring the differential pressure between both sides of the stage. After calibration, the associated uncertainty is $\pm12.5\,$Pa, corresponding to $\pm0.4\,\%$ of the design pressure rise.   

%% file: sections/numerical_ingredients.tex
\section{Numerical ingredients} \label{sec:num_ingredients}

This section includes the main numerical tools and methods used for this investigation. These include descriptions of the CFD solvers, of the DA strategies and of the ML models. Considering the large amount of information needed to comprehensively describe these tools, the main elements are presented in the core text of the article while detailed explanations are provided in  \autoref{sec:appendix_spalart_allmaras_config}, \autoref{sec:appendix_EnKF} and \autoref{sec:appendix_mfenkf}.

\subsection{Dynamic equations}
\label{sec::CFD}

The current study uses finite-volume-based solvers \citep{Ferziger1996_springer} with discretization of the compressible Navier-Stokes equations for Newtonian fluids. The dynamic equations are written using index notation:
\begin{align} 
    \quad \pd{\rho}{t} + \pd{(\rho u_j)}{x_j} &= 0 \label{eq:NavierStokes_masse} \\
    \pd{(\rho u_i)}{t} + \pd{}{x_j} \left( \rho u_i u_j + p \delta_{ij} - \tau_{ij} \right) &= f_i \label{eq:NavierStokes_moment} \\
    \pd{(\rho E)}{t} + \pd{}{x_j} \left( \rho E u_j + p u_j + q_j - u_i \tau_{ij}  \right) &= 0 \label{eq:NavierStokes_energie}
\end{align}
where \autoref{eq:NavierStokes_masse} represents the mass equation, \autoref{eq:NavierStokes_moment} is the momentum equation and finally \autoref{eq:NavierStokes_energie} is the energy equation. $\rho$ represents the density, $p$ is the pressure, $\mathbf{u}$ is the velocity, $\mathbf{f}$ is a volume forcing and $E$ is the total energy. $\delta_{ij}$ is the Kronecker symbol. For a Newtonian fluid, the components of the viscous stress tensor $\bar{\bar{\bm{\tau}}}$ are expressed as:
\begin{equation}
    \bar{\bar{\bm{\tau}}}_{ij} = 2 \rho \nu_{\textrm{eff}} \bar{\bar{\mathbf{s}}}_{ij}
\end{equation}
where $\nu_{\textrm{eff}}$ is the effective viscosity, which will be described in detail in the next paragraph. The components of the strain rate tensor $\bar{\bar{\mathbf{s}}}$ are defined as:
\begin{equation}
    \bar{\bar{\mathbf{s}}}_{ij} = \frac{1}{2} \left( \pd{u_i}{x_j} + \pd{u_j}{x_i} \right) - \frac{1}{3} \pd{u_k}{x_k}\cdot \delta_{ij}
\end{equation}

Finally, the components of the thermal flux $\mathbf{q}$ are expressed according to the Fourier law as $q_j = -\lambda \left( \partial T / \partial x_j \right)$ with $\lambda$ the thermal conductivity coefficient and $T$ the temperature.

The simulations in the present analysis are performed for Reynolds numbers corresponding to a fully turbulent behavior of the flow. For this reason, turbulence closures \cite{Pope2000_cambridge,Wilcox2006_DCW} are employed to reduce computational costs. Using the Boussinesq approximation one has $\nu_{\textrm{eff}}=\nu +\nu_t$ where $\nu$ is the molecular kinematic viscosity and $\nu_t$ is the turbulent viscosity. For the simulations performed in this work, $\nu_t$ is obtained via the Spalart-Allmaras turbulence model \citep{spalart1992one}. This turbulence closure is a one-equation model formulated by the following transport equation:
{\footnotesize
\begin{equation} 
    \pd{\tilde{\nu}}{t} + \left( \pd{\tilde{\nu}}{x_j} \right) u_j = \underbrace{c_{b1} \left( 1-f_{t2}\right) \tilde{S} \tilde{\nu}}_{\text{Production}} + \underbrace{\frac{1}{\sigma} \left[ \frac{\partial}{\partial x_j} \left( (\nu + \tilde{\nu}) \frac{\partial \tilde{\nu}}{\partial x_j} \right) + c_{b2} \left( \frac{\partial \tilde{\nu}}{\partial x_i} \right)^2 \right]}_{\text{Diffusion}} - \underbrace{c_{w1} f_w \left( \frac{\tilde{\nu}}{\tilde{d}} \right)^2}_{\text{Destruction}} \label{eq:transport_equation_nuTilda}
\end{equation}}

where $\tilde{\nu}$ is a viscosity quantity used to compute the actual eddy kinematic viscosity $\nu_t$:
\begin{equation}
    \nu_t = \tilde{\nu}\,f_{v1}
\end{equation}
The $f_{v1}$ function is computed as:
\begin{equation}
    f_{v1} = \frac{\chi^3}{\chi^3+C_{v1}^3} \, \quad \, \chi = \frac{\tilde{\nu}}{\nu}
\end{equation}

More details on the model constants prescribed for this analysis are provided in \autoref{sec:appendix_spalart_allmaras_config}. The previous numerical model is used to perform two distinct classes of simulations:
\begin{enumerate}
\item The first one includes classical steady RANS runs, in which the rotation of the rotor is accounted for using a Multiple Reference Frame (MRF). Details about this technique are reported in \citet{wilhelm2015rotating}. 
\item The simulations in the second class rely on a hybrid RANS-LES strategy referred to as Delayed Detached Eddy Simulations (DDES) \cite{spalart_new_2006}. These simulations use the same set of equations resolved for the RANS runs, but in this case the variable $\tilde{d}$ in \autoref{eq:transport_equation_nuTilda} is not constant and is calculated as:
\begin{equation}
    \tilde{d} = d - \tilde{f} \left( \max \left( 0, d - C_\textrm{DES} \Delta \right) \right)\label{eq:new_distance_DDES}
\end{equation}

where $d$ is the distance from the closest wall, $C_\textrm{DES} = 0.65$ and $\Delta \equiv \max\left(\Delta x, \Delta y, \Delta z \right)$ is a function of the local grid resolution. \autoref{eq:new_distance_DDES} requires the calculation via empirical equations of two new variables $\tilde{f}$ and $\tilde{r}$:
\begin{equation}
    \tilde{f} \equiv 1 - \tanh \left( \left( 8\tilde{r} \right)^3 \right) \, ; \,     \tilde{r} = \frac{\nu_t + \nu}{\max \left(\sqrt{\left(\pd{u_j}{x_j}\right)^2}, 10^{-10}\right) \cdot \kappa^2 \tilde{d}^2} 
\end{equation}
The expression of the hybrid distance $\tilde{d}$ blends the model behavior between RANS and LES modeling. If the wall distance $d$ increases, $\tilde{r} \rightarrow0$ and $\tilde{f}\rightarrow1$, thus the DDES model switches to LES mode. Conversely, if $d\rightarrow0$, $\tilde{r} \rightarrow1$ and $\tilde{f}\rightarrow0$, which makes the DDES model fully RANS. DDES simulations are scale resolved in LES regions and are unsteady in time. For these cases, an Arbitrary Mesh Interface (AMI) method is used to control the interactions between the moving part of the computational grid (rotor region) and the static part of the grid (the rest of the physical domain). More details are provided in \citet{carneiro2019AMI}. 
\end{enumerate}

The different numerical simulations performed in the current study are realized using the C++ code OpenFOAM \citep{greenshieldsweller2022}. This open-source library is well known in academic studies thanks to its capacities in preprocessing, solving and postprocessing tasks. The current analysis requires the usage of two different solvers. A solver based on the SIMPLE algorithm \citep{caretto_simple_1973} is used to solve steady RANS simulations, while a solver based on the PIMPLE algorithm \citep{caretto_simple_1973, ISSA198666} is used to perform unsteady DDES runs. For the latter, time derivatives are discretized using a second-order backward scheme. 
Advection terms are discretized using a first-order upwind scheme for RANS simulations and a native LUST scheme for DDES calculations. The LUST method is a hybrid scheme combining second-order centered schemes (75\%) with second-order upwind schemes (25\%). For the viscous terms, second-order centered schemes are used for both families of simulations.

\subsection{Data Assimilation and Ensemble Kalman Filter}
\label{sec::EnKF}

Data Assimilation (DA) tools provide a mathematically rigorous framework to combine data from different sources of information, such as measurements from experiments and results of numerical simulations. One of the main advantages of these approaches is that, unlike several data-driven techniques, they are efficient in handling sparse data in space and time. Classical applications usually rely on \textit{observation}, which is sampled using high-fidelity tools on limited parts of the domain for a restricted number of variables, and a \textit{model}, which provides a quasi-continuous representation of the physics investigated but it is usually less accurate than the tools used to sample observation. The DA methods combine observation and model prediction to obtain a \textit{state estimation}. Successful application provides physical states which are as accurate as observation while describing the whole domain of investigation. These applications have been extensively used in meteorological studies \cite{Daley1991_cambridge,Evensen2022_springer} but their usage in fluid mechanics has seen a significant increase in the last two decades \cite{Asch2016_SIAM, jeanney2026ensemble,donato2024integrating,he2020time}. Among notable applications, one can find works using variational techniques \cite{Artana2012_jcp,Foures2014_jfm,Onder2016_cf,Mons2016_jcp,Chandramouli2020_jcp,Padmanaban2026_prf} as well as sequential methods \cite{Rochoux2014_nhess,Leroux2014_ep,Kato2015_jcp,Meldi2017_jcp,Labahn2019_pci,Zhang2020_jcp,Villanueva2024_cof,Moussie2024_ftc,Valero2025_jcp,Villanueva2025_jfm,Ozalp2026_cmame, Villanueva2024_ijhff,Moldovan2024_cf}. The latter have shown very favorable features for the investigation of unsteady, nonlinear problems, such as the high Reynolds flows here investigated.

Among the different sequential DA methods proposed in the literature, the stochastic Ensemble Kalman Filter (EnKF) \cite{Evensen2022_springer} builds upon the fundamental principle of the classic Kalman Filter (KF) \citep{kalman_new_1960}, but is tailored for high-dimensional and non-linear systems. Initially proposed by \citet{evensen_sequential_1994} for state estimation and later generalized for parameter optimization of the model \cite{Evensen2009_ieee}, this methodology utilizes a population of $n_e$ distinct realizations commonly referred to as \textit{ensemble members}. The methodology, which is fully described in \autoref{sec:appendix_EnKF}, is developed in two steps:
\begin{enumerate}
\item The \textit{forecast} step consists of the time advance of the numerical model between two available observations.
\item The \textit{analysis} step includes numerical manipulations of the ensemble predicted state and the available observation to derive the state estimation.
\end{enumerate}
While the generation of the state estimation for the EnKF requires computational resources that are significantly lower when compared with classical KF strategies, the generation of large databases using models describing millions of degrees of freedom (such as the solutions obtained via CFD) is still prohibitive for most realistic applications. A number of research works in the last years have offered solutions to reduce the computational burdens while keeping high accuracy, such as the Multigrid EnKF (MGEnKF) \cite{Moldovan2021_jcp, Moldovan2022_jcp, Moldovan2024_cf}, the Multifidelity EnKF (MFEnKF) \cite{popov_multifidelity_2021}, or multi-level strategies enhanced by neural networks \cite{Moussie026_cf}. The MFEnKF relies on a higher-fidelity model, whose quantities are indicated via a suffix $\princ$, and a computationally efficient lower-fidelity surrogate, identified via a suffix $\cont$. A variable $\cont$ is a control variate for a primary variable $\princ$ if $\cont$ is strongly correlated with $\princ$ and its expected value $\expectcont$ is known \textit{a priori}. By combining a few model realizations  $\princ$ and a larger number of runs $\cont$, the MFEnKF can obtain a state estimation whose accuracy is similar to the classical EnKF application. However, only a limited number of higher-fidelity model runs is required, therefore dramatically reducing the computational costs. The MFEnKF, which will be used in the present analysis, is comprehensively described in \autoref{sec:appendix_mfenkf}.   





\section{Numerical investigation of the CME2 test case} \label{sec:num_cme2}

\begin{figure}[h!]
    \centering
    \includegraphics[width=0.5\textwidth]{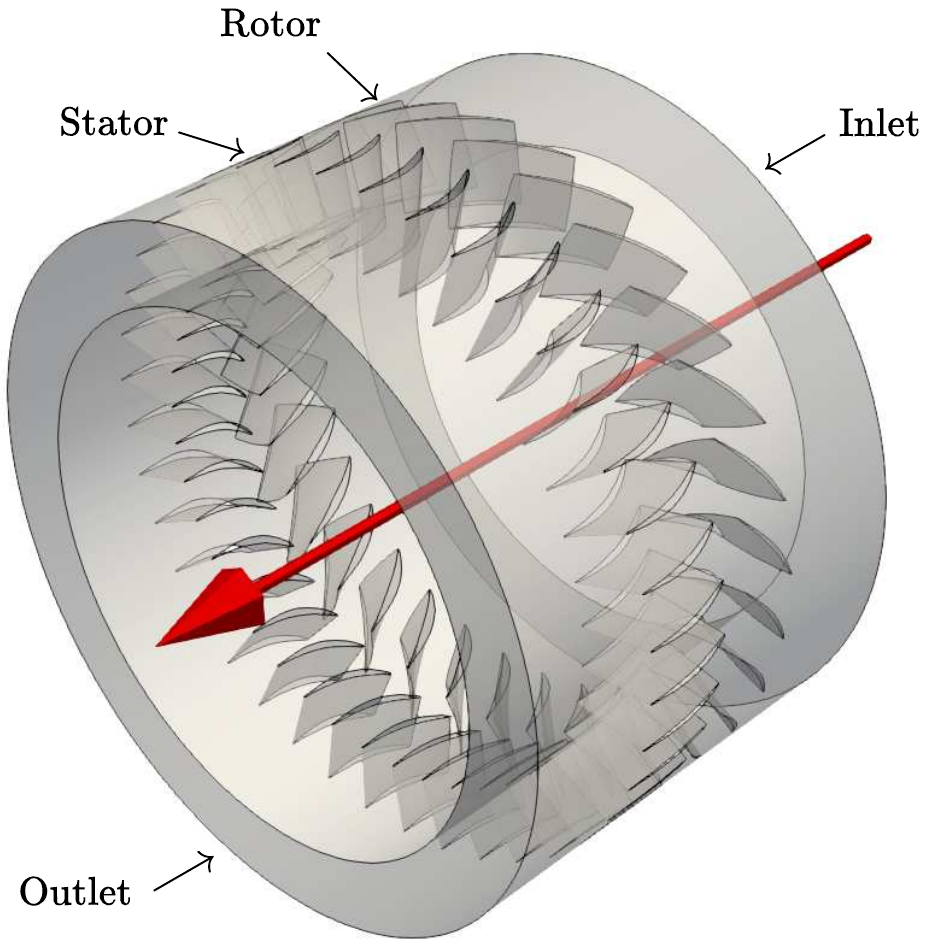}
    \caption{Scheme of the computational domain used to perform numerical simulations of the flow inside the {CME2} compressor. The red arrow represents the asymptotic axial direction of the flow.}
    \label{fig:cme2_comp_scheme}
\end{figure}

\begin{figure}
    \centering
    \begin{tabular}{c}
        \includegraphics[width=0.65\linewidth]{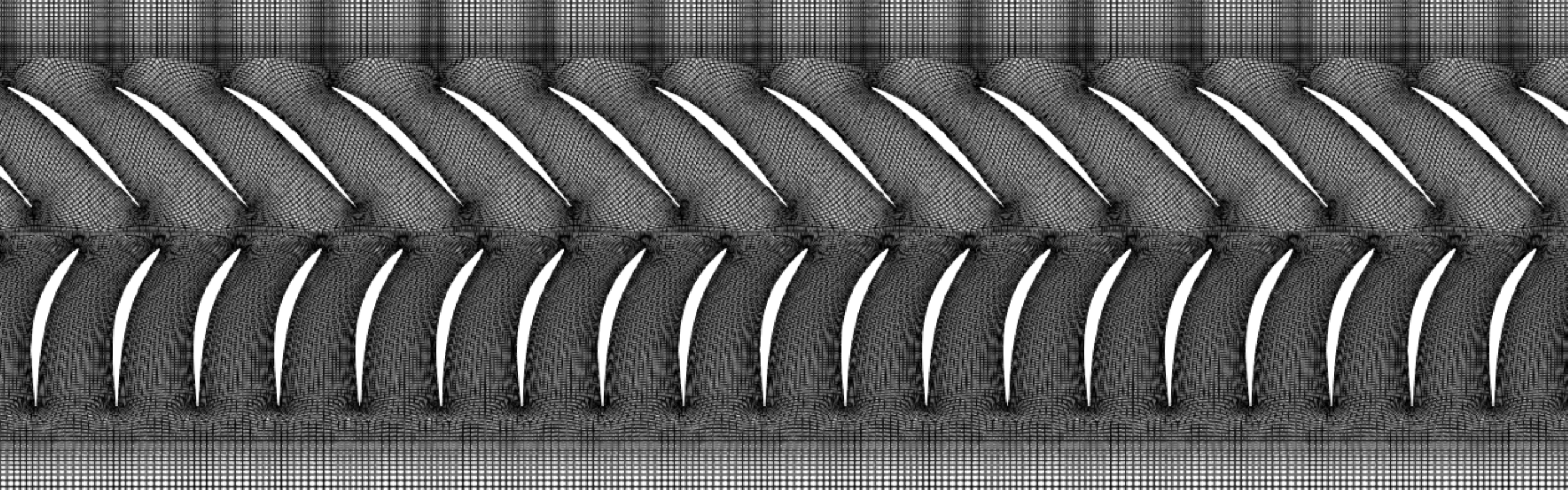} \\
        (a) \\[0.25cm]
        \includegraphics[width=0.65\linewidth]{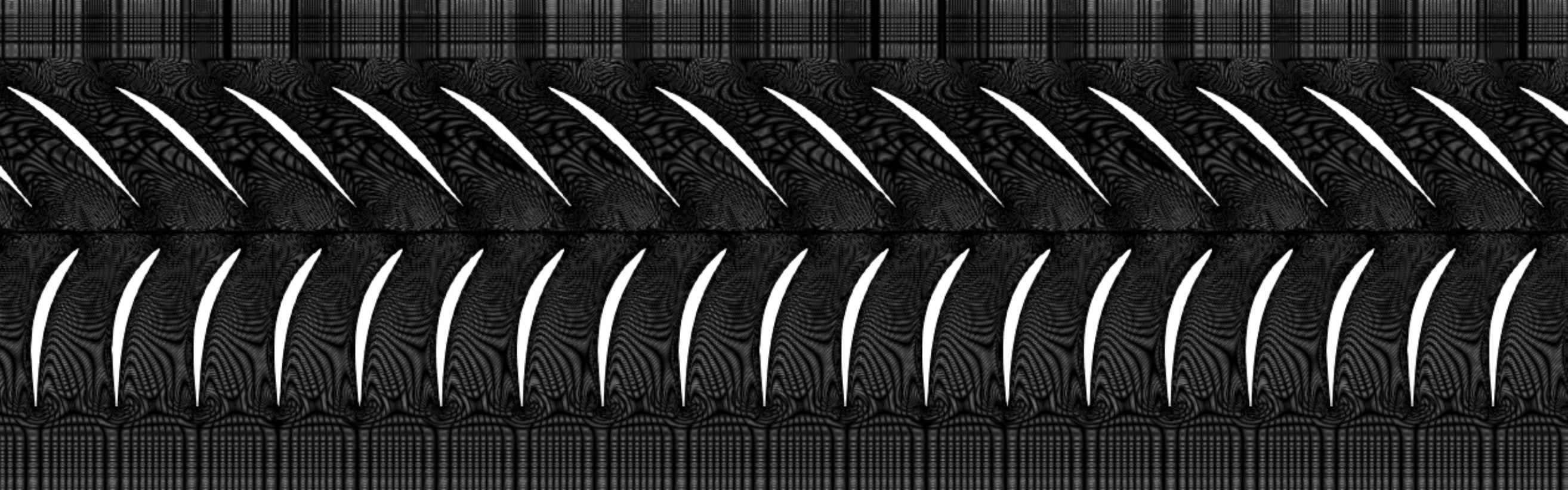} \\
        (b) \\[0.25cm]
    \end{tabular}
    \caption{Snapshot of the computational grids used for calculation. An azimuthal section for $r=0.2475\,$m is shown for (a) the grid \textbf{M1} used for RANS calculations and (b) the grid \textbf{M2} used for DDES calculations.}
    \label{fig:placeholder}
\end{figure}

Numerical simulations of the CME2 compressor are performed for the computational domain shown in \autoref{fig:cme2_comp_scheme}. One can see that $z$ is the axial direction of the flow. The geometry is here simplified when compared with the case used in experiments, but previous analyses in the literature \cite{rannou2022effect, marty2023numerical}  highlighted that these simplifications do not preclude the acquisition of physically meaningful results. Present analyses are performed for the operating condition presented in \autoref{sec:expe}, for which very small separation regions are observed. Therefore, the Spalart-Allmaras closure presented in \autoref{sec::CFD} has been selected for both RANS and DDES simulations.
Two numerical grids have been prepared using the \textit{Cadence AutoGrid} software, which is specifically designed to obtain exportable grids for the OpenFOAM format. The two grids, which encompass the physical domain shown in \autoref{fig:cme2_comp_scheme} and are qualitatively shown in \autoref{fig:placeholder}, have the following features:
\begin{enumerate}
    \item The first grid \textbf{M1} is made of $ \approx 5 \cdot 10^6$ cells. The resolution in the proximity of the wall is chosen to provide an adequate trade-off between computational costs and accuracy. An initial estimation of the size of the elements was performed via dimensional arguments and results were checked a posteriori after a few runs where modifications to boundary conditions were also performed. Calculations of the quantity $y^+ = y_{\textrm{cell}} / \delta_\nu$ showed that in the rotor region we have $y^+_{\textrm{rotor}} \in [1.67, \, 38]$. A part of the grid elements has its center in the viscous sublayer ($y^+ \leq 5$), providing an adequate resolution of the near-wall gradients. However, a maximum value of $y^+=38$ indicates that a part of the grid elements is within the buffer layer ($5<y^+<30$). The mesh resolution is now briefly described, indicating the size of the grid elements in the axial direction $\Delta_z$, in the radial direction $\Delta_r$ and in the azimuthal direction $\Delta_{\theta}$. The resolution of the grid elements in the rotor area is $\Delta_z \in [1.46 \cdot 10^{-2}, 3.7\cdot10^{-2}]\,c$, $\Delta_{\theta} \in [1.85 \cdot 10^{-2}, 3.35\cdot10^{-2}]\,c$ and $\Delta_r \in [1.55\cdot10^{-3}, 5\cdot10^{-2}]\,c$, where $c$ is the average chord of the blade profiles.
    In the stator region, the wall resolution is $y^+_{\textrm{stator}} \in [7.6, \, 54]$. Here, since the minimum $y^+$ is $7.6$, the grid does not fully resolve the viscous sublayer and the center of the near-wall cells is mainly located in the buffer layer, extending into the logarithmic zone for $y^+=54$. For the stator region, $\Delta_z \in [1.07 \cdot 10^{-2}, 6.7\cdot10^{-2}]\,c$, $\Delta_\theta \in [3.63 \cdot 10^{-2}, 4.87\cdot10^{-2}]\,c$ and $\Delta_r \in [3.43\cdot10^{-3}, 9.84\cdot10^{-2}]\,c$.
    The grid \textbf{M1} is specifically used for the RANS simulations and the MRF technique is here used to account for the movement of the rotor. Native OpenFOAM Spalding wall functions were used to improve near-wall accuracy of the calculations. 
    
    \item The second grid \textbf{M2} consists of $\approx 50 \cdot 10^6$ elements. For this second grid, we observe $y^+_{\textrm{rotor}} \in [0.5, \, 32]$, which indicates that the viscous sublayer is fully resolved over a part of the surface of the rotor blades. The upper bound at $y^+=32$ shows that the center of a part of the near-wall grid elements is located around the beginning of the logarithmic zone. The resolution of the grid elements in the rotor area is $\Delta_z \in [6.04 \cdot 10^{-3}, 2.14\cdot10^{-2}]\,c$, $\Delta_\theta \in [8 \cdot 10^{-3}, 1.28\cdot10^{-2}]\,c$ and $\Delta_r \in [1.49\cdot10^{-3}, 2.37\cdot10^{-2}]\,c$. In the stator region, $y^+_{\textrm{stator}} \in [0.7, \, 21]$, therefore exhibiting similar to that of the rotor region. 
    The size of the grid elements in the stator area is $\Delta_z \in [9.01 \cdot 10^{-3}, 1.92\cdot10^{-2}]\,c$, $\Delta_\theta = 9.1 \cdot 10^{-3}\,c$ and $\Delta_r \in [3.12\cdot10^{-3}, 3.32\cdot10^{-2}]\,c$. The grid \textbf{M2} is used to perform DDES simulations and the AMI tool permits for motion of the rotor region.
\end{enumerate}

The boundary conditions, which are identical for both RANS and DDES simulations except for the rotor region, are now described. At the inlet, Dirichlet conditions for the total pressure $P_t = 101\,325\,$Pa, the static temperature $T=293.15\,$K, and $\tilde{\nu}=10^{-4}\,\textrm{m}^2/\textrm{s}$ are set using the native OpenFOAM boundary condition \textit{pressureInletVelocity}. The shaft, casing, and stator blades are walls and they are represented using a \textit{noSlip} condition for the velocity, a \textit{zeroGradient} condition for the pressure and a \textit{Spalding wall function} for the turbulent viscosity. 
For the steady RANS simulations using the MRF method, the rotor boundary condition for velocity is also set to \textit{noSlip}. On the other hand, the unsteady DDES simulation using the AMI technique imposes a \textit{movingWallVelocity} boundary condition to ensure mesh displacement in the rotor region. Finally, the outlet utilizes a custom static-pressure condition, referred to as \textit{valve law}, which aims to replicate the mechanical behavior of the valve located at the outlet of the experimental setup (see \autoref{sec:expe}). This boundary condition has been previously used in the analyses by \citet{rannou2023methodologie}. The static outlet pressure is defined as:
\begin{equation}
    P_{s_{\textrm{outlet}}} = P_{\textrm{ref}} + P_{i,0} \cdot \alpha \left( \frac{\dot{q}}{\dot{q}_{\textrm{ref}}} \right)^2 \label{eq:loi2vanne}
\end{equation}

where the stagnation pressure $P_{i,0} = 101\,325\,\textrm{Pa}$ and the reference pressure $P_{\textrm{ref}} = 0.8 \cdot P_{i,0}$. The reference mass flow rate is $\dot{q}_{\textrm{ref}} = 5.3\,\textrm{kg}\cdot\textrm{s}^{-1}$, and $\dot{q}$ is the current mass flow rate calculated at the outlet. Finally, the \textit{valve law} is governed by the free coefficient $\alpha$ which defines operating conditions at a given mass flow rate. The nominal operating conditions are obtained when $(\dot{q}/\dot{q}_{\textrm{ref}})=1$ and $\alpha=0.23$, which leads to $P_{s_{\textrm{outlet}}} = 104\,364.75\,\textrm{Pa}$. \\

\begin{figure}[h!]
    \centering
    \includegraphics[width=0.45\textwidth]{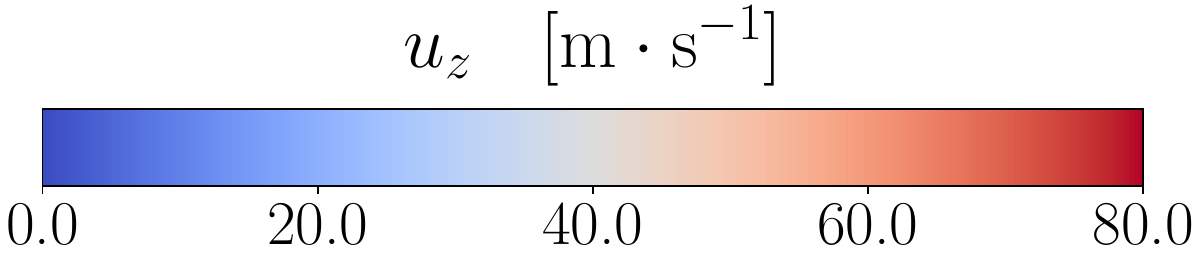}
    \begin{tabular}{cc}
    \includegraphics[width=0.45\textwidth]{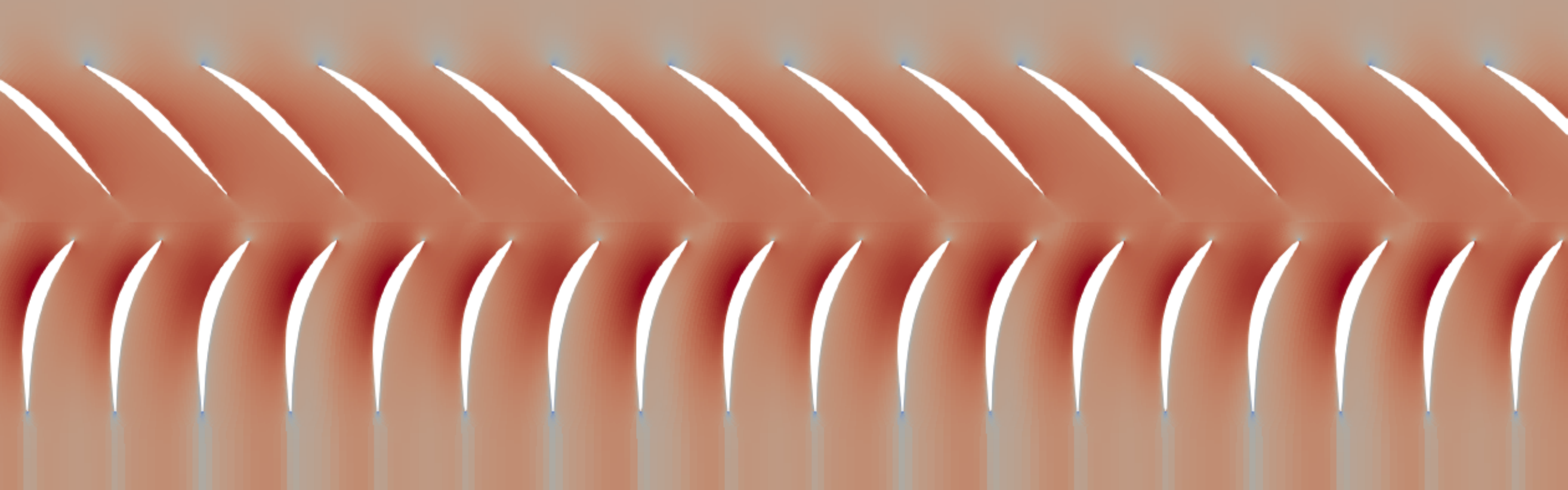} & \includegraphics[width=0.45\textwidth]{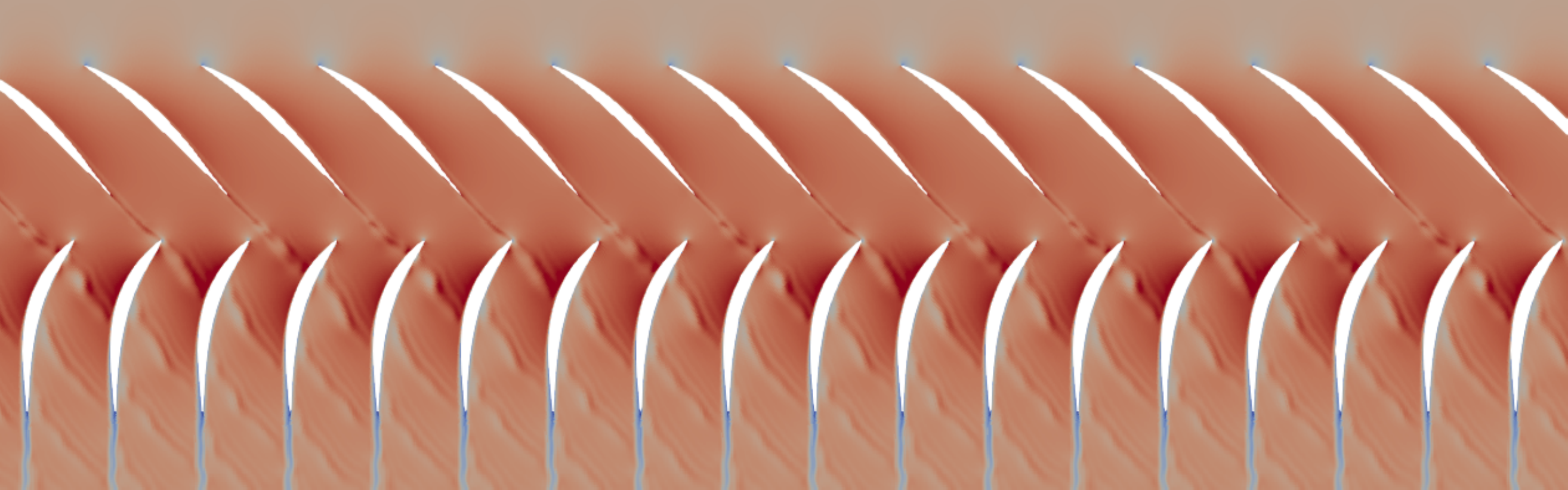} \\
    (a) & (b) \\[0.2cm]
    \end{tabular}
    \includegraphics[width=0.45\textwidth]{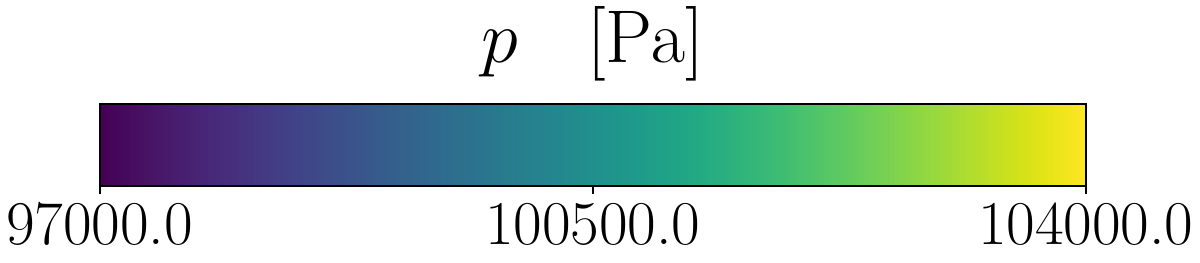}
    \begin{tabular}{cc}
    \includegraphics[width=0.45\textwidth]{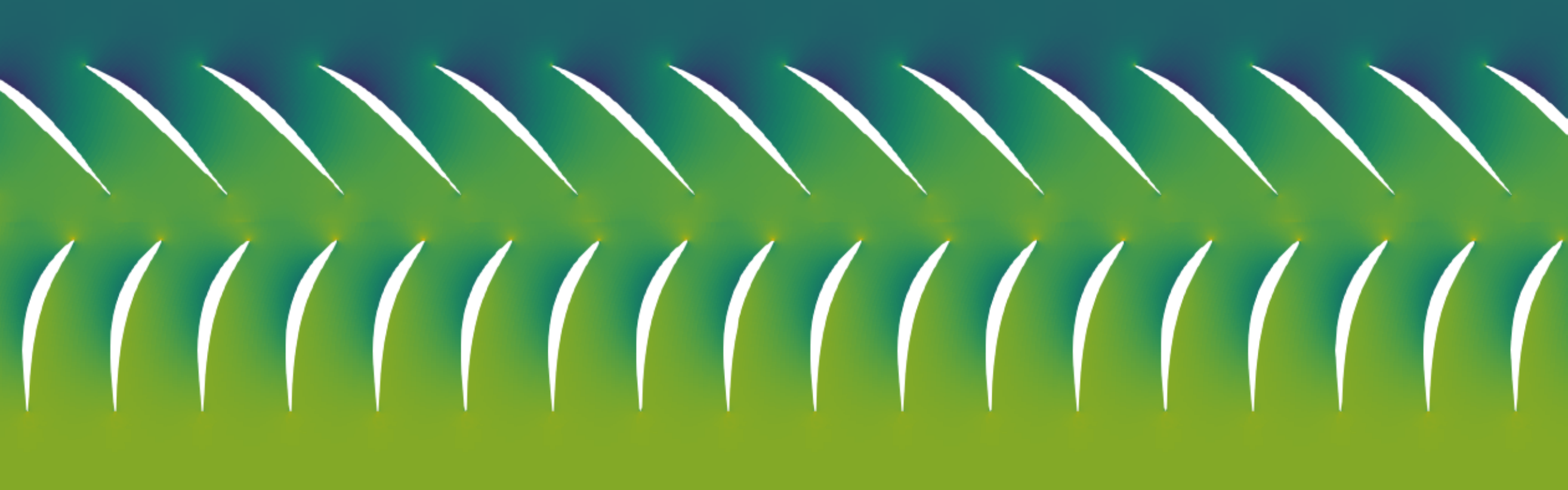} & \includegraphics[width=0.45\textwidth]{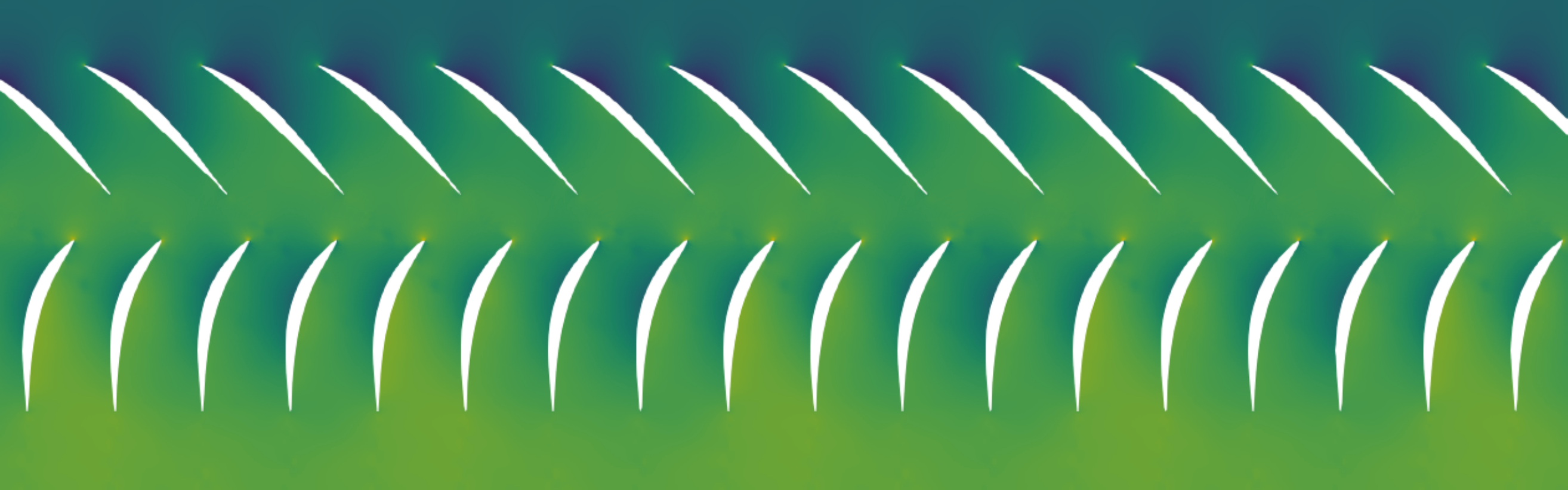} \\
    (c) & (d) \\[0.2cm]
    \end{tabular}
    \includegraphics[width=0.45\textwidth]{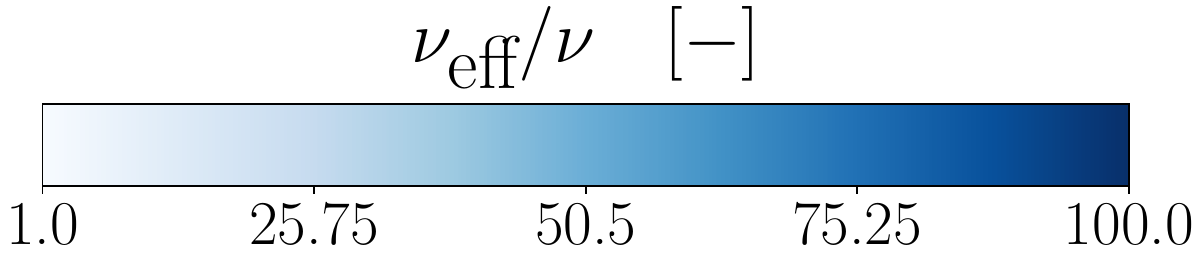}
    \begin{tabular}{cc}
    \includegraphics[width=0.45\textwidth]{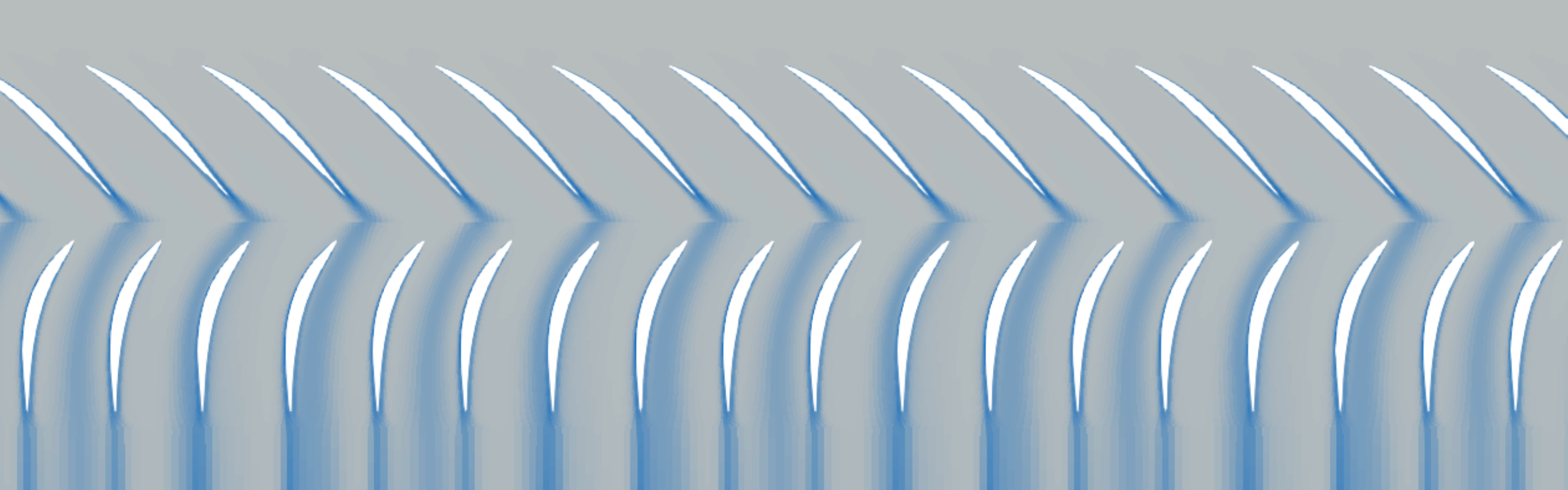} & \includegraphics[width=0.45\textwidth]{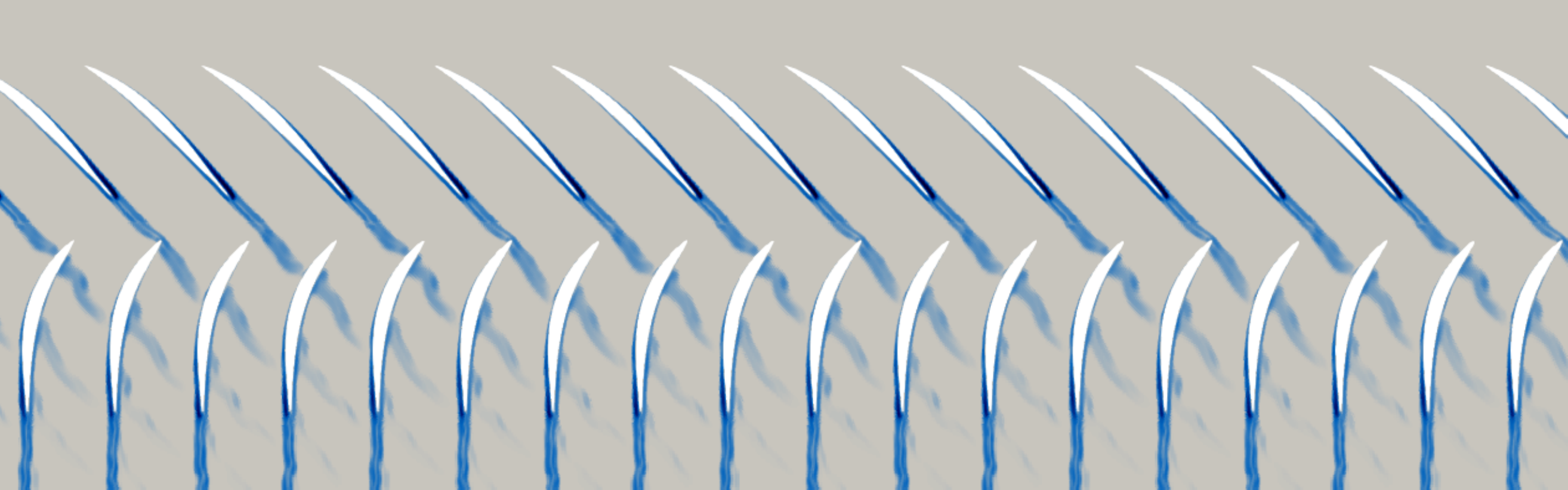} \\
    (e) & (f) \\[0.2cm]
    \end{tabular}
    
    \caption{Results for numerical simulations of the CME2 axial compressor. Top row: isocontours of the axial velocity $u_z$; Middle row: isocontours of the pressure $p$; Bottom row: isocontours of effective viscosity normalized by molecular viscosity $\nu_\textrm{eff}/\nu$. (a), (c) and (e) are reporting RANS solutions and (b), (d) and (f) are reporting DDES solutions. Visualizations are provided for a cylindrical surface for $r=0.2475\,$m. Rotor is displayed at the top and stator at the bottom, flow is going from top to bottom.}
    \label{fig:isocontours_ddes_rans_simulations}
\end{figure}

RANS and DDES simulations have been initially run using the numerical setup previously described to assess their accuracy. First of all, a qualitative visualization of the axial velocity fields is shown in \autoref{fig:isocontours_ddes_rans_simulations}. One can see that both RANS and DDES simulations are able to correctly represent the topology of the flow, which does not exhibit large separations as expected. For the RANS runs, one can see that the wake out of the rotor suddenly changes direction when entering the stator region. This effect is due to the MRF technique here employed, which is consistent with the steady flow representation. On the other hand, the AMI technique implemented for the DDES simulations complies with the scale-resolving features of the CFD solver. One can clearly see that the wake of the rotor is advected through the blades of the stator. Similar physical considerations can be drawn for the pressure $p$ which is shown in \autoref{fig:isocontours_ddes_rans_simulations} (c) and (d), and for the effective viscosity normalized by the molecular viscosity $\nu_\textrm{eff}/\nu$ fields displayed in \autoref{fig:isocontours_ddes_rans_simulations} (e) and (f). 

\begin{figure}[h!]
    \centering
    \includegraphics[width=0.999\linewidth]{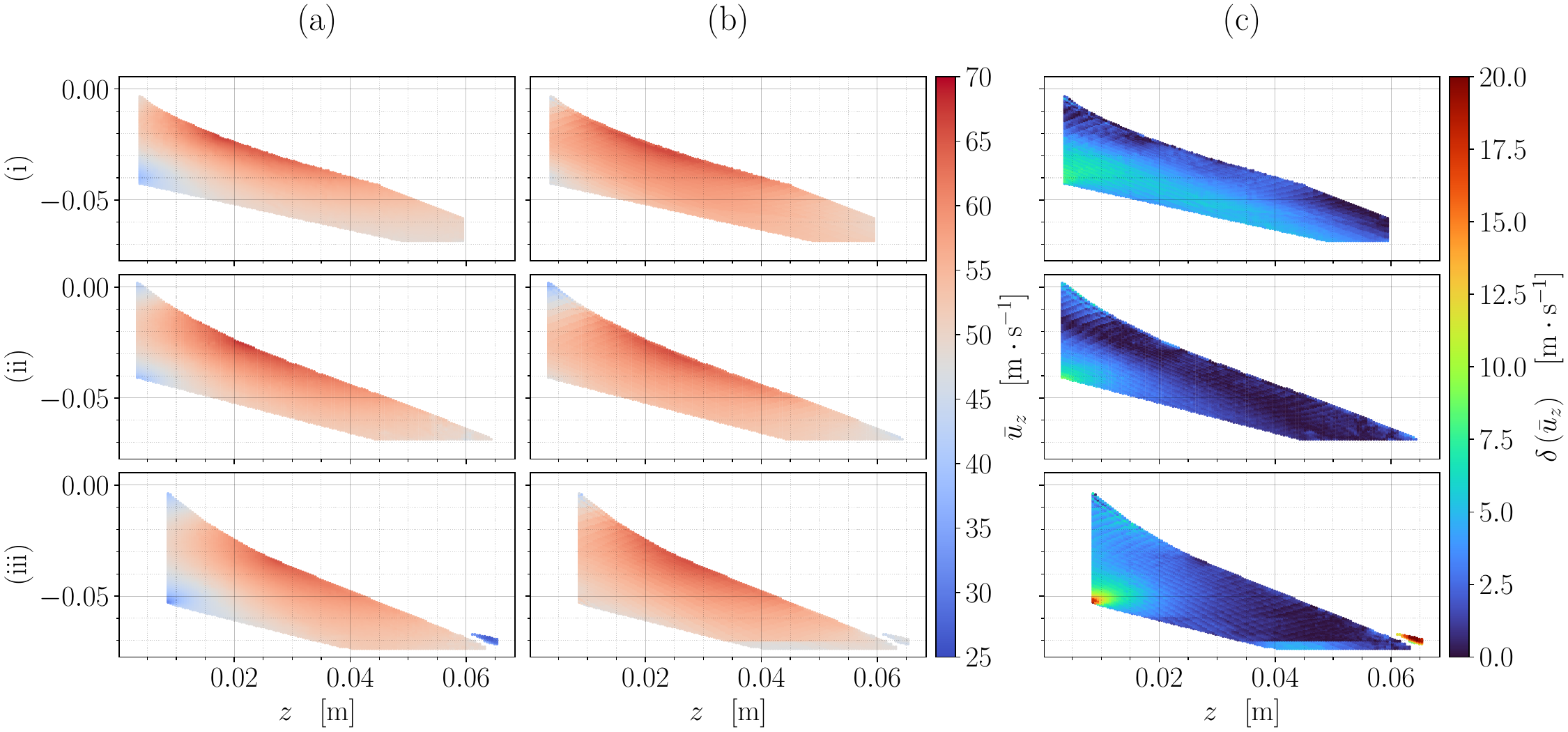}
    \caption{Comparison of the RANS-predicted axial velocity $\bar{u}_z$ with the available experimental data. Column (a): PIV-measured axial velocity $u_z^{\textrm{PIV}}$ data; column (b): RANS-predicted axial velocity $\bar{u}_z^{\textrm{RANS}}$ data; column (c): absolute value of the difference between PIV and RANS axial velocities $\left\vert u_z^\textrm{PIV} - u_z^\textrm{RANS} \right\vert $. Row (i): results for \textbf{P1}; row (ii): results for \textbf{P2}; row (iii): results for \textbf{P3}.}
    \label{fig:RANS_PIV_initial_comparison}
\end{figure}

\begin{figure}[h!]
    \centering
    \includegraphics[width=0.999\linewidth]{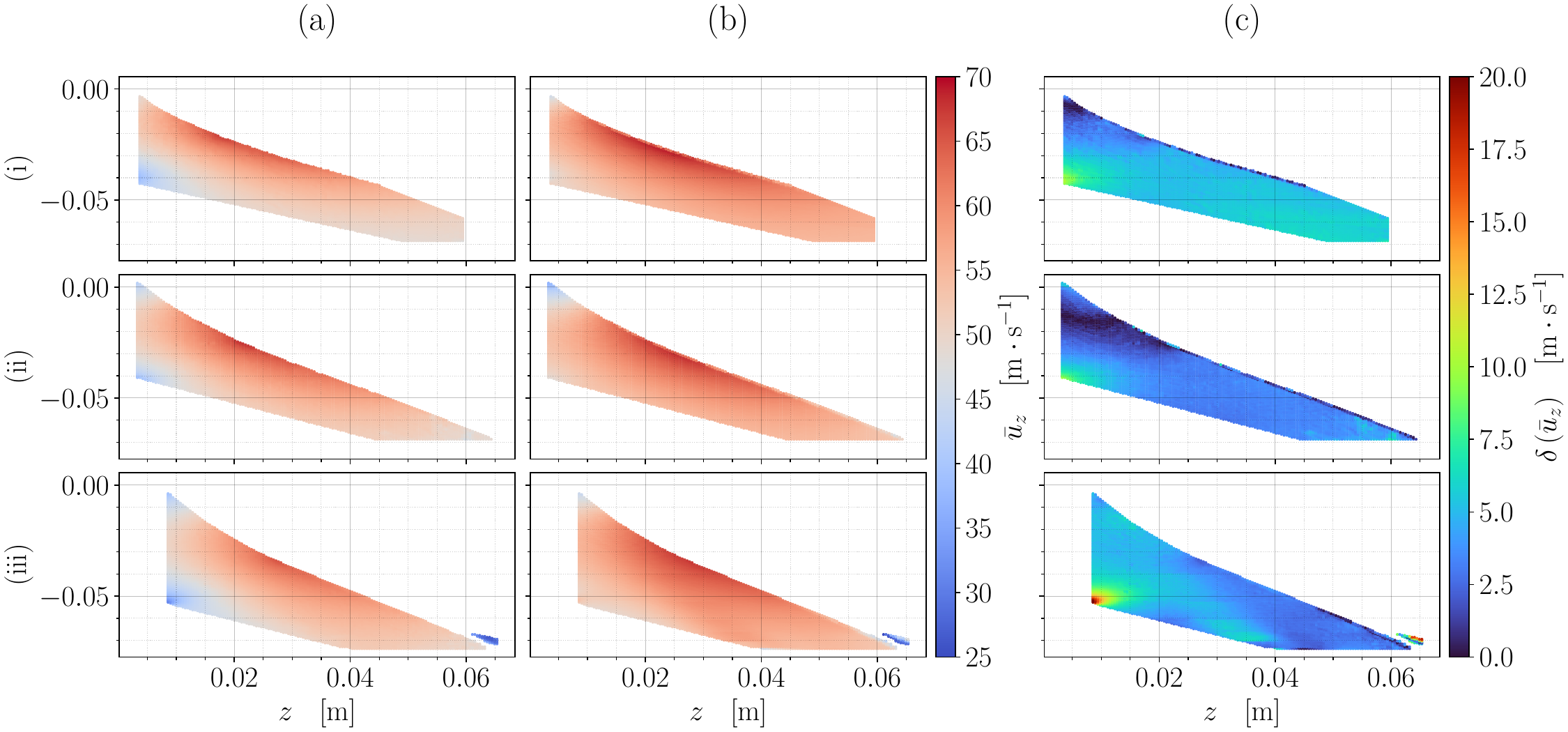}
    \caption{Comparison of the DDES-predicted axial velocity $\bar{u}_z$ with the available experimental data. Column (a): PIV-measured axial velocity $u_z^{\textrm{PIV}}$ data; column (b): time-averaged DDES-predicted axial velocity $\bar{u}_z^{\textrm{DDES}}$ data; column (c): absolute value of the difference between PIV and DDES axial velocities $\left\vert u_z^\textrm{PIV} - u_z^\textrm{DDES} \right\vert $. Row (i): results for \textbf{P1}; row (ii): results for \textbf{P2}; row (iii): results for \textbf{P3}.}
    \label{fig:DDES_PIV_initial_comparison}
\end{figure}

The results obtained via RANS and DDES are now validated via the comparison of the predicted axial velocity $u_z$ with the available experimental data in \autoref{fig:RANS_PIV_initial_comparison} and \autoref{fig:DDES_PIV_initial_comparison}, respectively.

\autoref{fig:RANS_PIV_initial_comparison} compares the time-averaged axial velocity $\bar{u}_z$ predicted by RANS with the PIV measurements. Columns (a), (b) and (c) respectively display the PIV data, the RANS prediction and the absolute value of the difference between the two, while rows (i), (ii) and (iii) correspond to \textbf{P1}, \textbf{P2} and \textbf{P3}, respectively. For the lower plane \textbf{P1} (row (i)), the discrepancy between the RANS prediction and the PIV data is mainly located in the bottom left-hand corner of the plane, which corresponds to the leading edge of the rotor blade located below the PIV plane. The maximum discrepancy reaches $12.65\,\textrm{m}\,\cdot\,\textrm{s}^{-1}$, with a mean absolute discrepancy of $3.52\,\textrm{m}\,\cdot\,\textrm{s}^{-1}$ over the plane. The RANS prediction on the mid-span plane \textbf{P2} (row (ii)) shows the best agreement with the PIV results: the maximum discrepancy, still located at the leading edge of the rotor blade, is $10.90\,\textrm{m}\,\cdot\,\textrm{s}^{-1}$, with a mean absolute discrepancy of $1.54\,\textrm{m}\,\cdot\,\textrm{s}^{-1}$. For the higher plane \textbf{P3} (row (iii)), the RANS/PIV discrepancy is mainly located at the leading edge of the lower rotor blade, but also at the trailing edge of the upper rotor blade, as observed in the right-hand corner of \autoref{fig:RANS_PIV_initial_comparison} (iii,c). On this plane, the maximum discrepancy reaches $20.96\,\textrm{m}\,\cdot\,\textrm{s}^{-1}$, while the mean absolute discrepancy is $2.73\,\textrm{m}\,\cdot\,\textrm{s}^{-1}$. The discontinuity visible in \autoref{fig:RANS_PIV_initial_comparison} (iii,b) and (iii,c) originates from the boundary condition, due to the height at which \textbf{P3} is located in the compressor.

\autoref{fig:DDES_PIV_initial_comparison} presents the comparison between the mean axial velocity fields $\bar{u}_z$ obtained from DDES simulations and PIV measurements. Once again, columns (a), (b), and (c) illustrate the experimental PIV maps, the numerical DDES results, and the absolute difference between the two aforementioned approaches, respectively. Lines (i), (ii), and (iii) correspond to planes \textbf{P1}, \textbf{P2}, and \textbf{P3}. For \textbf{P1} on line (i), the major discrepancies between the DDES model and the experimental data are mainly concentrated in the lower-left area, as visible in sub-figure (i,c), which coincides with the leading edge of the rotor blading positioned below the plane. Locally, the maximum difference rises to $10.79\,\textrm{m}\,\cdot\,\textrm{s}^{-1}$, with an average discrepancy over the entire map (i,c) of $5.32\,\textrm{m}\,\cdot\,\textrm{s}^{-1}$. Regarding \textbf{P2}, which can be visualized on line (ii), the maximum error is once again located near the leading edge of the lower rotor blading with a value of $10.28\,\textrm{m}\,\cdot\,\textrm{s}^{-1}$, while the average error stands at $3.31\,\textrm{m}\,\cdot\,\textrm{s}^{-1}$. Finally, the analysis of \textbf{P3} on line (iii) also reveals differences between the PIV and DDES fields, where less homogeneous disparities are noticeable in sub-figure (iii,c) compared to the other sub-figures in column (c). With this configuration, the maximum discrepancy reaches a value of $21.75\,\textrm{m}\,\cdot\,\textrm{s}^{-1}$ at the leading edge of the lower profile, while maintaining an average discrepancy value of $4.66\,\textrm{m}\,\cdot\,\textrm{s}^{-1}$. This analysis shows that the discrepancy observed for the DDES and RANS results is similar and that on average the RANS run performs better in terms of pure accuracy. This finding is not completely surprising considering that the grid resolution for \textbf{M2} is on average two times more resolved than \textbf{M1} but the DDES simulation implements different techniques for the representation of the moving rotor and is unsteady, therefore introducing an additional level of complexity due to convergence of the time averages of the physical field. One interesting point is also that the cartography of the discrepancy with experimental data shows different patterns for the DDES and the RANS. This point will be exploited in the multifidelity DA application which is presented in the next section.

%% file: sections/data_assimilation_work.tex
\section{Data Assimilation: application of the EnKF to the analysis of the CME2 test case} \label{sec:data_assimilation_work}

The EnKF is here used to combine the flow prediction obtained with the CFD solvers and the available experimental data. More precisely, the multifidelity framework provided by the MFEnKF is used to obtain a state estimation combining results from several RANS runs, a smaller number of DDES simulations and the PIV measurements. The state estimation is here obtained via observation of the time-averaged axial velocity $\bar{u}_z$. This flow field is naturally obtained as a result of RANS simulation and it has been calculated performing phase average from instantaneous flow fields for both the DDES runs and the experiments. In the following, the architecture of the DA procedure is detailed.  

\subsection{Setup of MFEnKF strategy}

The MFEnKF strategy relies on the following elements:
\begin{enumerate}
    \item \textit{Observations} are sampled from the PIV maps of the phase-averaged axial velocity $\bar{u}_z$ obtained during the experimental campaign. From the $8\,180$, $6\,944$, and $8\,090$ points respectively available in planes \textbf{P1}, \textbf{P2}, and \textbf{P3}, $1\,000$ points are randomly selected from each plane, as shown in \autoref{fig:obs_locations}. This selection has been performed to test the capabilities of the DA algorithm to reconstruct an accurate state estimation even with limited available data. For each plane, we thus have:
    \begin{equation}
        \boldsymbol{y} = \left( \bar{u}_{z_0}, \bar{u}_{z_1}, \dots , \bar{u}_{z_{n_o -1}} \right) \in \mathbb{R}^{n_o}
    \end{equation}
    \begin{figure}[h!]
        \centering
        \includegraphics[width=0.99\linewidth]{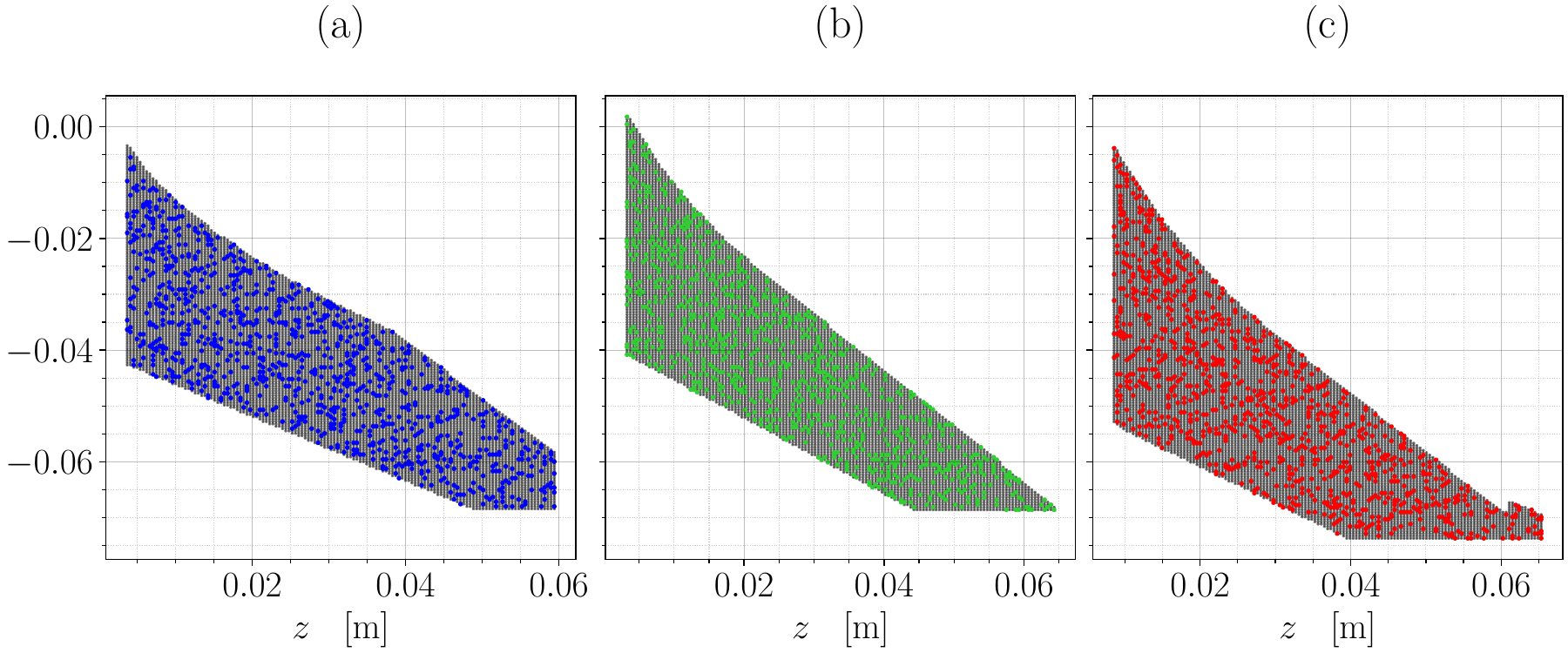}
        \caption{Blue, green and red dots represent the selected observation points for MFEnKF among available data for (a) \textbf{P1}; (b) \textbf{P2}; (c) \textbf{P3}.}
        \label{fig:obs_locations}
    \end{figure}
    
    Considering that a stochastic version of the filter is used, observation is perturbed following a truncated normal law defined as:
    \begin{equation}
        \mathcal{X} \sim \mathcal{N}\left(\mu, \sigma^2\right) : \mathcal{X} \in \left[\mu - 3\sigma, \mu + 3\sigma \right]
    \end{equation}
    with the mean value $\mu = \boldsymbol{y}$ and its standard deviation defined as $\sigma = 0.03\,\boldsymbol{y}$. The selected value corresponds to the global measurement uncertainty experimentally quantified for the PIV samples.
    \item An unsteady, computationally expensive \textit{model} is used, which is based on DDES simulation using the AMI technique with the \texttt{rhoPimpleFoam} unsteady solver. Five DDES ensemble members are performed to this purpose. The variability of the prior states for these simulations is obtained by imposing slightly different outlet conditions. More precisely, the parameter $\alpha$ in \autoref{eq:loi2vanne} is defined as $\alpha \sim \mathcal{N} \left( 0.27, 0.02^2 \right)$. This variability is essential to obtain different physical states to improve the convergence of the filter. Each DDES simulation has been initialized with an already-converged RANS field and run for a total solver time of $0.005\,t$, which corresponds to $50\,000$ iterations using a time step of $\delta t = 1\cdot 10^{-7}\,t$, which represents $8/30$ of a complete rotation of the rotor. The phase averages have been performed after a time of $0.0025\,t$, which was necessary to eliminate the effects of initial conditions.        
    \item A steady, computationally inexpensive \textit{model} is run via RANS simulation with the MRF method. These simulations are performed using the steady-state solver \texttt{rhoSimpleFoam}. Thirty ensemble members are performed with this model. Five are the \textit{control} members (i.e. the ones which are connected with the DDES realizations) while the other 25 are the \textit{ancillary} members. Therefore, the total number of RANS+DDES ensemble members is $n_e = 35$. Each RANS simulation has been initialized with an already-converged RANS field, and run for a total of $6\,250$ iterations. 
    The variability between the ensemble members is obtained via variation of the parameter $\alpha$ in \autoref{eq:loi2vanne}, defined as $\alpha \sim \mathcal{N} \left( 0.27, 0.02^2 \right)$.
    \item The coupling between the model and observation is performed using the numerical library CONES \cite{Villanueva2024_cof, errante_2026_20433499}. The hyperlocalization technique recently proposed by \citet{Villanueva2025_jfm} for the EnKF is here employed. The idea here is that the state update produced by the EnKF is not calculated and applied on the complete physical domain of the simulations, but it is restricted to a determined physical region around the sensors. This procedure allows filtering out sampling errors due to the limited number of ensemble members as well as to reduce the computational costs required. Therefore, the state estimation is performed only for three regions which are created around planes \textbf{P1}, \textbf{P2} and \textbf{P3}. A qualitative representation of such DA zones when compared with the experimental PIV planes is shown in \autoref{fig:colors_plans_boxes}. In the zones determined for DA purposes, the state vector is built considering the three Cartesian components of the velocity field for the selected subset of grid elements:
    \begin{equation}
        \boldsymbol{x} = \left( u_{x_1}, \dots,   u_{x_{n_{\textrm{C}}}}, u_{y_1}, \dots, u_{y_{n_{\textrm{C}}}}, u_{z_1}, \dots, u_{z_{n_{\textrm{C}}}} \right) \in \mathbb{R}^{3 \times n_{\textrm{C}}}
    \end{equation}
    where $n_\textrm{C}$ represents the number of cells in the aforementioned clipping boxes. It must be stressed that, for the grid M2 used for DDES calculations, $n_{\textrm{C,}\textbf{P1}} = 38\,462$, $n_{\textrm{C,}\textbf{P2}} = 44\,072$ and $n_{\textrm{C,}\textbf{P3}} = 59\,113$. In this case, one can see that the total information produced by the DA state estimation is significantly larger than the availability of the experimental data, even if hyperlocalization is used. This point is important because it indicates that the DA approach generates a larger amount of high fidelity data starting from a limited amount of observation data, therefore providing more suitable conditions for the training of machine learning models.

    \begin{figure}[h!]
        \centering
        \begin{tabular}{cc}
            \includegraphics[width=0.45\linewidth]{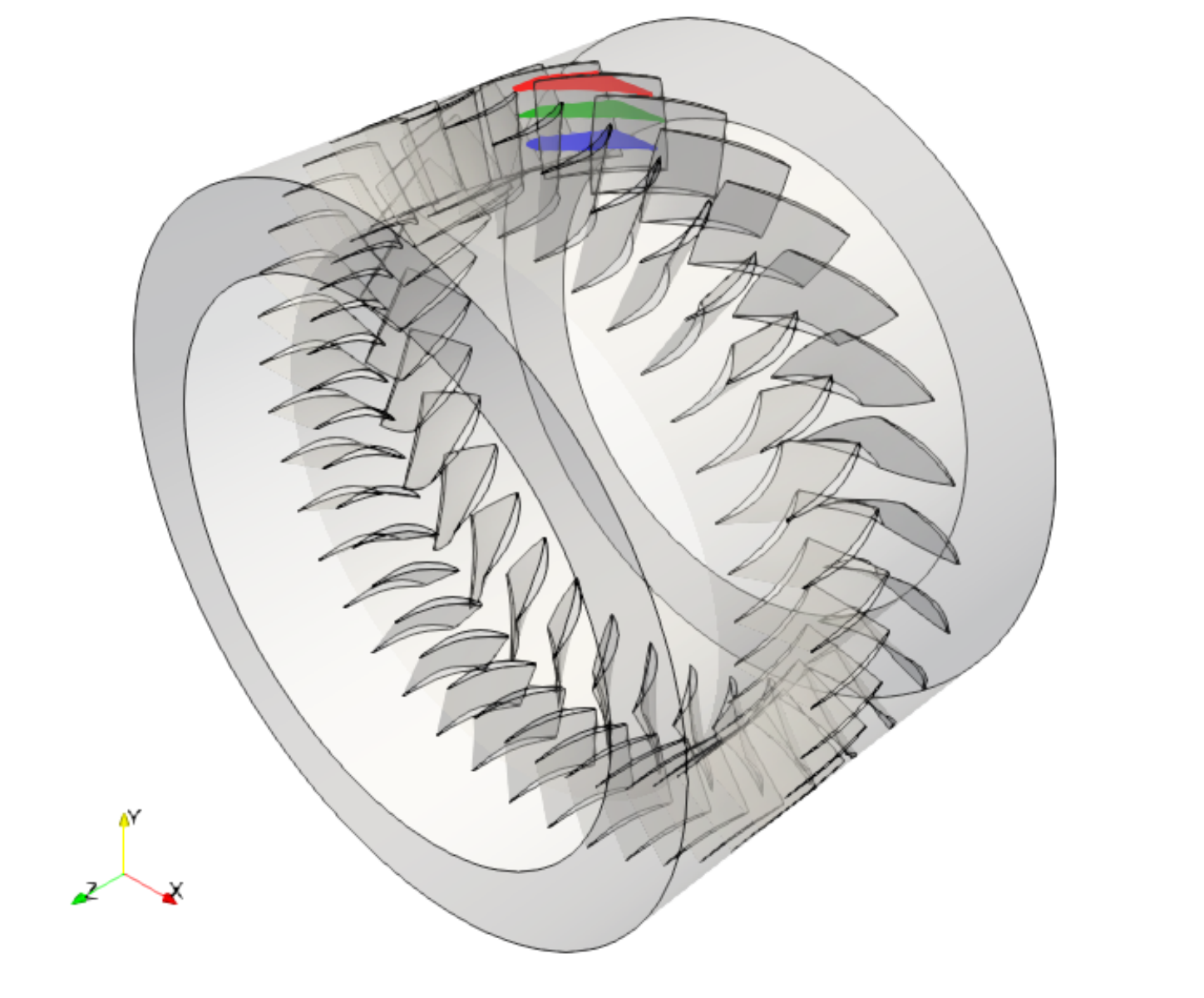} & \includegraphics[width=0.45\linewidth]{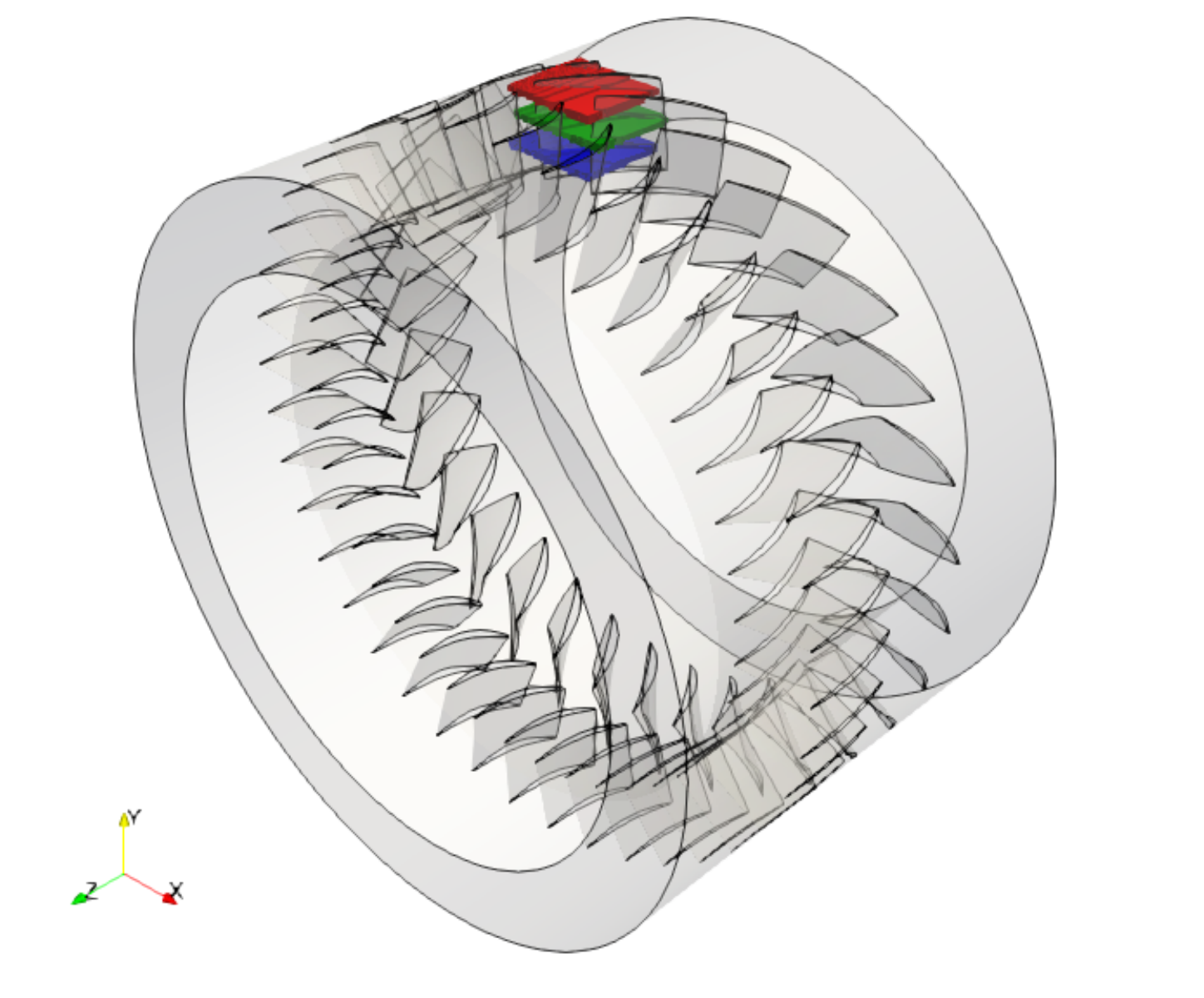} \\
            \includegraphics[width=0.45\linewidth]{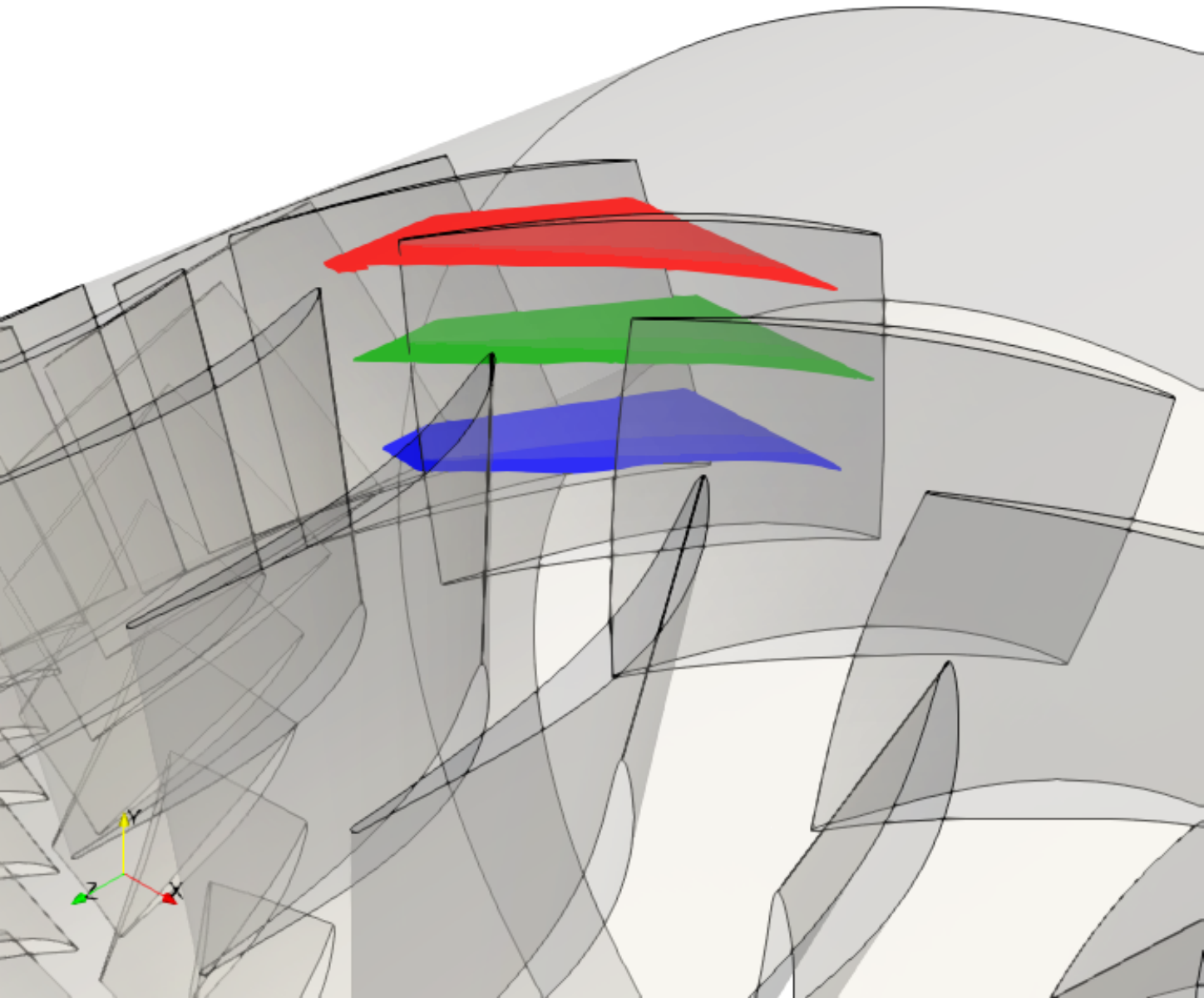} & \includegraphics[width=0.45\linewidth]{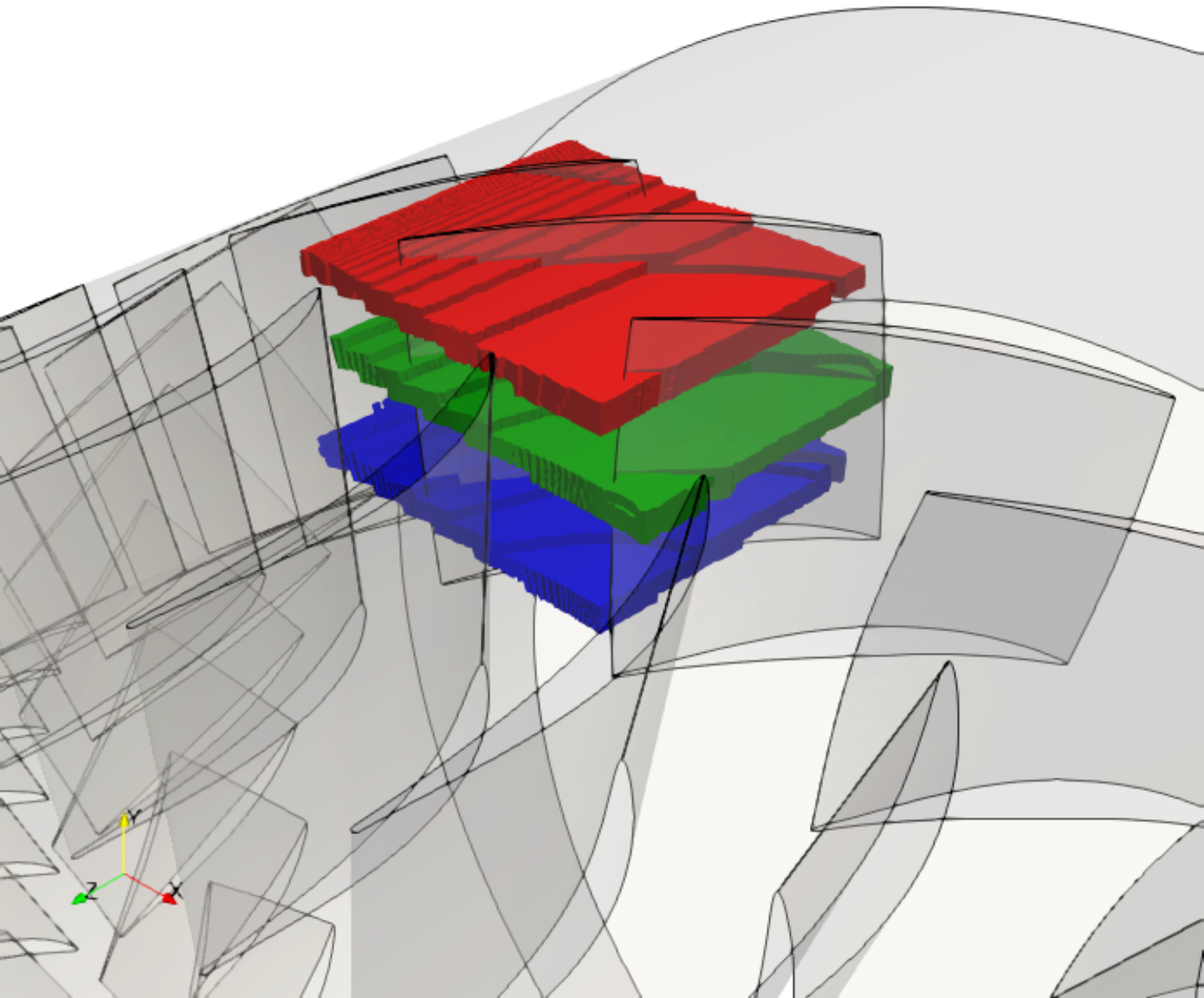} \\
            (a) & (b) \\[0.25cm]
        \end{tabular}
        \caption{(a) column: position of \textbf{P1}, \textbf{P2} and \textbf{P3} PIV planes in the {CME2} compressor, respectively colored in blue, green and red; (b) column: position of the DA hyperlocalization clipping boxes corresponding to the three PIV planes earlier mentioned, with identical color matching.}
        \label{fig:colors_plans_boxes}
    \end{figure}

\end{enumerate}

Finally, the PIV, RANS and DDES data were obtained by phase, iteration and time averaging, respectively. For clarity, all three are hereafter referred to as \textit{mean} data.

\subsection{State estimation using the MFEnKF}

\begin{figure}[h!]
    \centering
    \includegraphics[width=0.99\linewidth]{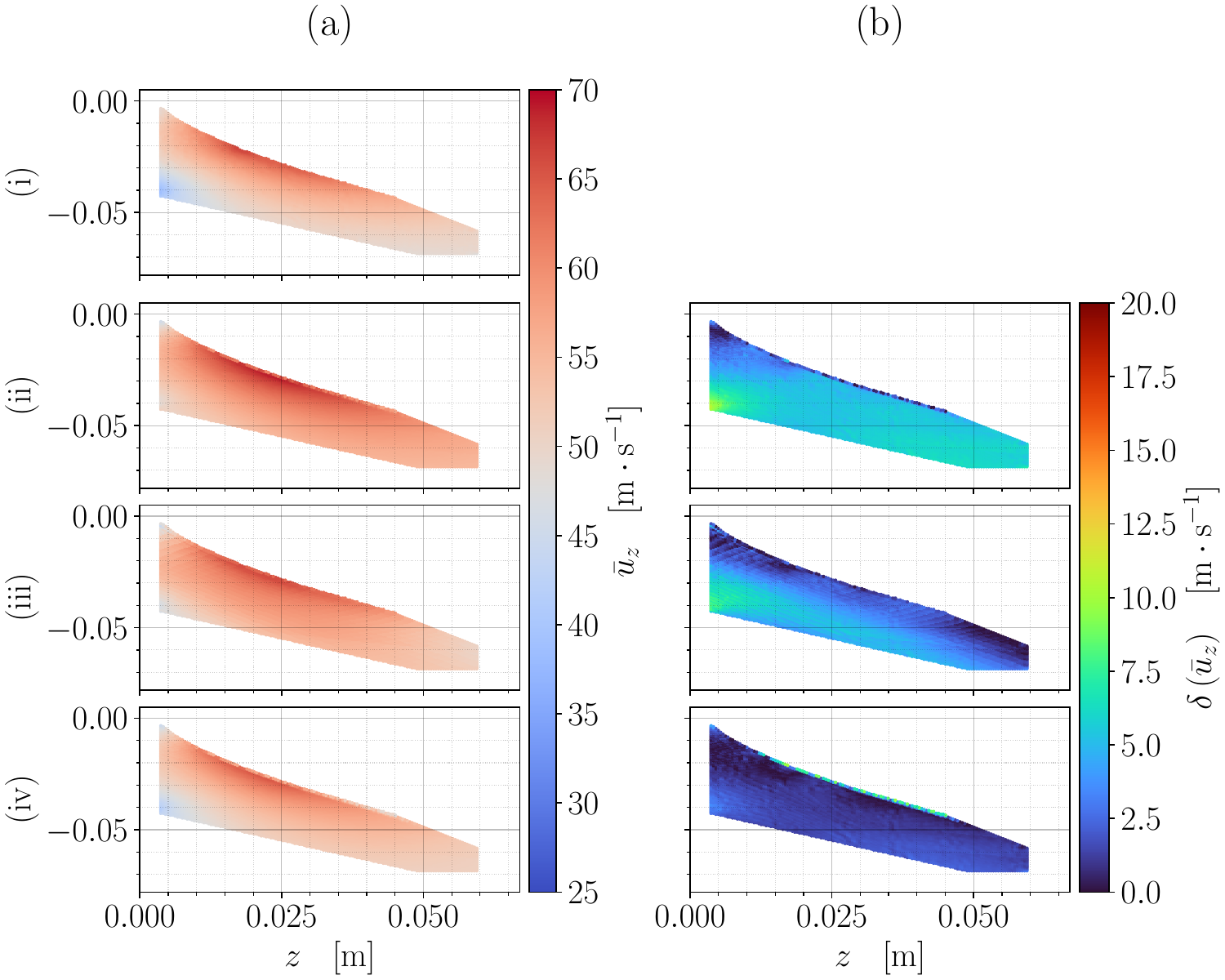}
    \caption{Results obtained via DA using the MFEnKF for \textbf{P1}. Column (a): Maps of the time-averaged axial velocity $\bar{u}_z$. Line (i) Reference experimental data (PIV); line (ii) Initial numerical prediction (DDES); line (iii) Initial numerical prediction (RANS); line (iv) Updated field after data assimilation (ensemble mean of the MFEnKF update $\left \langle \bar{u}_z^a \right \rangle$); column (b): maps of the absolute discrepancy $\delta(\bar{u}_z) = \left \vert u_z^\textrm{PIV} - \bar{u}_z \right \vert$ relative to the reference data. From top to bottom, the image displays the DDES model error $\delta(\bar{u}_z^{\textrm{DDES}})$, the RANS model error $\delta(\bar{u}_z^{\textrm{RANS}})$, and the bottom image displays the residual error of the model corrected by the MFEnKF algorithm ($\left \vert u_z^\textrm{PIV} - \left \langle \bar{u}_z^a \right \rangle \right \vert$).}
    \label{fig:mfenkf_results_low_plan}
\end{figure}

Results obtained via the MFEnKF are now discussed. The results are presented for each plane and box showcased in \autoref{fig:colors_plans_boxes}. After a single analysis phase, the MFEnKF results are displayed in \autoref{fig:mfenkf_results_low_plan} for \textbf{P1}, corresponding to the blue plane and box in \autoref{fig:colors_plans_boxes}. \autoref{fig:mfenkf_results_low_plan} (a,i) shows the axial velocity $u_z^{\textrm{PIV}}$ obtained from the PIV measurements, (a,ii) displays the DDES-predicted time-averaged axial velocity $\bar{u}_z^{\textrm{DDES}}$, (a,iii) displays the RANS-predicted mean axial velocity $\bar{u}_z^{\textrm{RANS}}$, while (a,iv) exhibits the ensemble average of the updated axial velocity from the MFEnKF analysis $\left \langle \bar{u}_z^a \right \rangle$, where the $\left \langle \, \cdot \, \right \rangle$ denotes the average of any quantity over the $n_e$ ensemble members. It is easily observable that the velocity field in \autoref{fig:mfenkf_results_low_plan} (a,i) is in better agreement with \autoref{fig:mfenkf_results_low_plan} (a,iv) than with \autoref{fig:mfenkf_results_low_plan} (a,ii) and (a,iii). More precisely, the improvement is qualitatively observed on the bottom left-hand corner of the (a,iv) field, which corresponds to the leading edge of the rotor blade below \textbf{P1}, where the axial velocity has been reduced to match the PIV data.  \autoref{fig:mfenkf_results_low_plan} (b,ii) displays the magnitude of the difference between the axial velocities obtained by PIV and DDES prediction $\left \vert u_z^\textrm{PIV} - \bar{u}_z^\textrm{DDES} \right \vert$, (b,iii) shows the magnitude of the difference between the axial velocities obtained by PIV and RANS prediction $\left \vert u_z^\textrm{PIV} - \bar{u}_z^\textrm{RANS} \right \vert$, while \autoref{fig:mfenkf_results_low_plan} (b,iv) shows the magnitude of the difference between the axial velocities obtained by PIV and MFEnKF ensemble average  update $\left \vert u_z^\textrm{PIV} - \left \langle \bar{u}_z^a \right \rangle \right \vert$. The DA updated field shows a lower discrepancy with PIV data than the DDES and RANS-predicted fields, which is measured via the L2 norm error $\bar{\varepsilon}$ which is defined as:
\begin{equation}
\label{eq:varepsi}
    \bar{\varepsilon} = \frac{1}{n_{\textrm{p}}} \sum \frac{\vert \boldsymbol{x}_{\textrm{PIV}} - \boldsymbol{x}_{\textrm{CFD}} \vert}{\boldsymbol{x}_{\textrm{PIV}}}
\end{equation}
in which $n_\textrm{p}$ represents the total number of points in the PIV planes, $\boldsymbol{x}_{\textrm{CFD}}$ represents data obtained using a CFD solver and $\boldsymbol{x}_{\textrm{PIV}}$ data from PIV measurements. The results, provided in \autoref{tab:comparaison_results_moe}, show that $\bar{\varepsilon}_{\textrm{DDES}}=10.2\,\%$, $\bar{\varepsilon}_{\textrm{RANS}}=6.87\,\%$ and $\bar{\varepsilon}_{\textrm{MFEnKF}}=2.59\,\%$, therefore indicating that the DA procedure successfully improves flow prediction.

\begin{figure}[h!]
    \centering
    \includegraphics[width=0.99\linewidth]{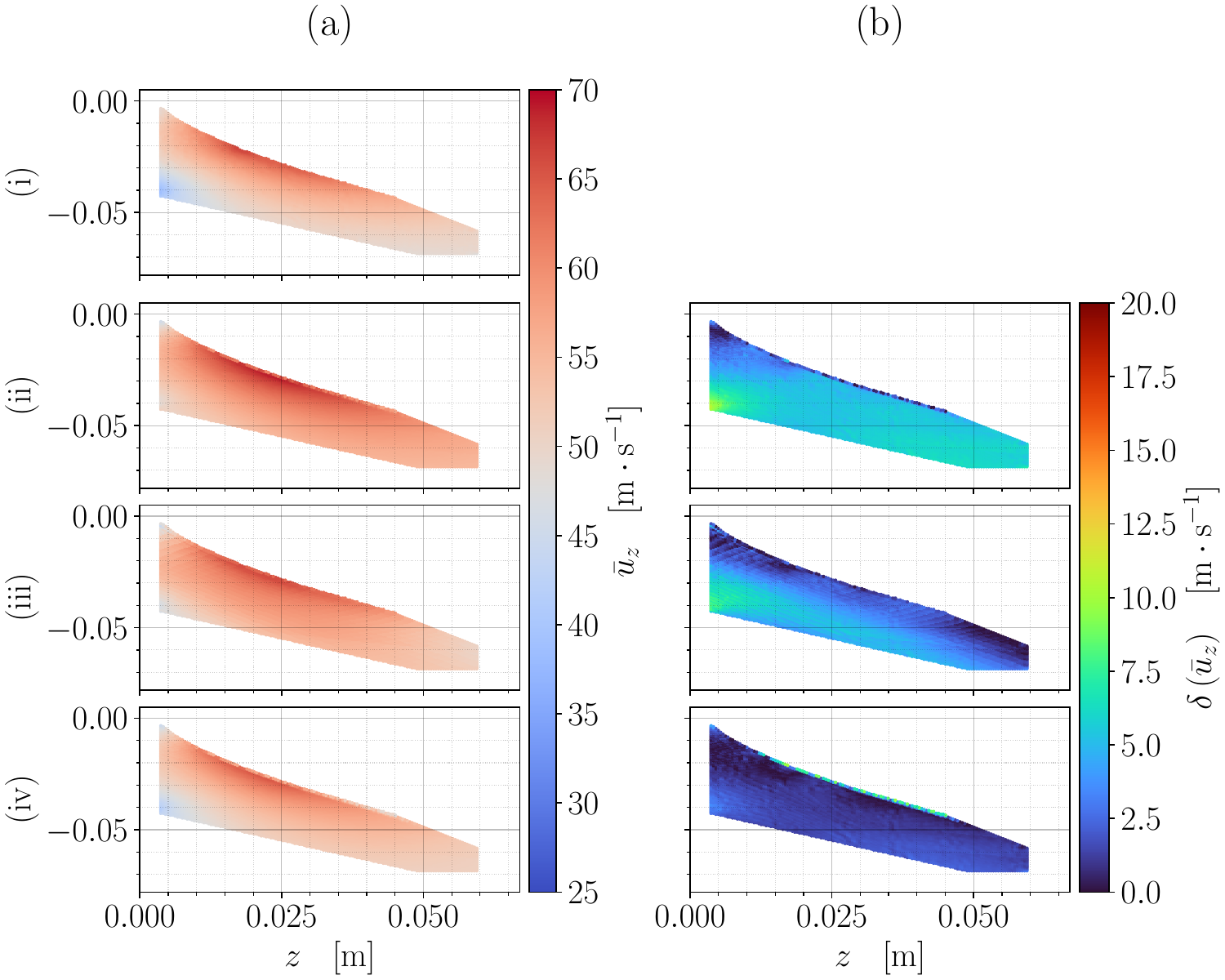}
    \caption{Results obtained via DA using the MFEnKF for \textbf{P2}. Column (a): Maps of the time-averaged axial velocity $\bar{u}_z$. Line (i) Reference experimental data (PIV); line (ii) Initial numerical prediction (DDES); line (iii) Initial numerical prediction (RANS); line (iv) Updated field after data assimilation (ensemble mean of the MFEnKF update $\left \langle \bar{u}_z^a \right \rangle$); column (b): maps of the absolute discrepancy $\delta(\bar{u}_z) = \left \vert u_z^\textrm{PIV} - \bar{u}_z \right \vert$ relative to the reference data. From top to bottom, the image displays the DDES model error $\delta(\bar{u}_z^{\textrm{DDES}})$, the RANS model error $\delta(\bar{u}_z^{\textrm{RANS}})$, and the bottom image displays the residual error of the model corrected by the MFEnKF algorithm ($\left \vert u_z^\textrm{PIV} - \left \langle \bar{u}_z^a \right \rangle \right \vert$).}
    \label{fig:mfenkf_results_mid_plan}
\end{figure}

Similarly, the results of the MFEnKF analysis for \textbf{P2} are presented in \autoref{fig:mfenkf_results_mid_plan}, which corresponds to the green plane and box in \autoref{fig:colors_plans_boxes}. \autoref{fig:mfenkf_results_mid_plan} (a,i) shows the axial velocity $u_z$ obtained from PIV measurements, (a,ii) displays the time-averaged axial velocity predicted by DDES $\bar{u}_z^{\textrm{DDES}}$, (a,iii) displays the mean axial velocity predicted by RANS $\bar{u}_z^{\textrm{RANS}}$, while (a,iv) presents the ensemble mean of the updated axial velocity from the MFEnKF analysis $\left \langle \bar{u}_z^a \right \rangle$. Just as for \textbf{P1}, it is easily observed that the velocity field in \autoref{fig:mfenkf_results_mid_plan} (a,i) corresponding to PIV data is in much better agreement with the results from the MFEnKF in \autoref{fig:mfenkf_results_mid_plan} (a,iv) than with the results from baseline DDES and RANS in \autoref{fig:mfenkf_results_mid_plan} (a,ii) and (a,iii). More precisely, the improvement is qualitatively observable around the leading edge of the lower rotor blade profile, where the inaccuracies of the initial prediction have been corrected to fit the PIV data more accurately. From a quantitative standpoint, \autoref{fig:mfenkf_results_mid_plan} (b,ii) displays the magnitude of the difference between the axial velocities obtained by PIV and by the DDES prediction $\left \vert u_z^\textrm{PIV} - \bar{u}_z^\textrm{DDES} \right \vert$, (b,iii) shows the magnitude of the difference between the axial velocities obtained by PIV and by the RANS prediction $\left \vert u_z^\textrm{PIV} - \bar{u}_z^\textrm{RANS} \right \vert$, while \autoref{fig:mfenkf_results_mid_plan} (b,iv) shows the magnitude of the difference between the axial velocities obtained by PIV and the updated ensemble mean from the MFEnKF analysis $\delta\left( \bar{u}_z \right) = \left \vert u_z^\textrm{PIV} - \left \langle \bar{u}_z^a \right \rangle \right \vert$. It is then clearly noticeable that the updated field exhibits a lower residual error than the fields predicted by DDES and RANS, with the respective average relative difference $\bar{\varepsilon}$ decreasing from $\bar{\varepsilon}_{\textrm{DDES}}=6.21\,\%$ and $\bar{\varepsilon}_{\textrm{RANS}}=2.91\,\%$ down to $\bar{\varepsilon}_{\textrm{MFEnKF}}=1.81\,\%$.

\begin{figure}[h!]
    \centering
    \includegraphics[width=0.99\linewidth]{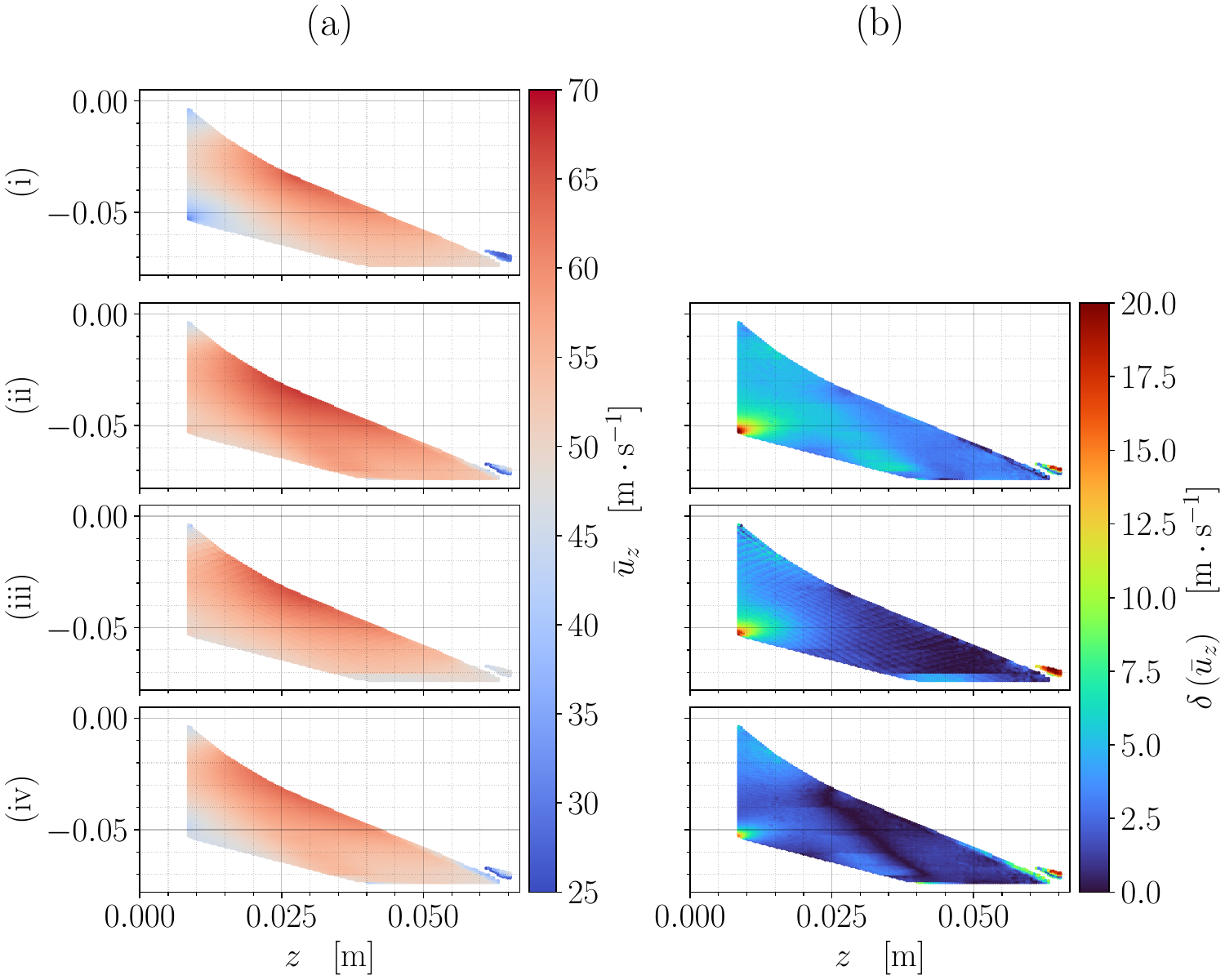}
    \caption{Results obtained via DA using the MFEnKF for \textbf{P3}. Column (a): Maps of the time-averaged axial velocity $\bar{u}_z$. Line (i) Reference experimental data (PIV); line (ii) Initial numerical prediction (DDES); line (iii) Initial numerical prediction (RANS); line (iv) Updated field after data assimilation (ensemble mean of the MFEnKF update $\left \langle \bar{u}_z^a \right \rangle$); column (b): maps of the absolute discrepancy $\delta(\bar{u}_z) = \left \vert u_z^\textrm{PIV} - \bar{u}_z \right \vert$ relative to the reference data. From top to bottom, the image displays the DDES model error $\delta(\bar{u}_z^{\textrm{DDES}})$, the RANS model error $\delta(\bar{u}_z^{\textrm{RANS}})$, and the bottom image displays the residual error of the model corrected by the MFEnKF algorithm ($\left \vert u_z^\textrm{PIV} - \left \langle \bar{u}_z^a \right \rangle \right \vert$).}
    \label{fig:mfenkf_results_high_plan}
\end{figure}

Finally, following the same procedure as for the previous planes, MFEnKF results for \textbf{P3} are presented in \autoref{fig:mfenkf_results_high_plan}, corresponding to the red plane and box in \autoref{fig:colors_plans_boxes}. Here too, it is observed that the velocity field in \autoref{fig:mfenkf_results_high_plan} (a,i) is in better agreement with \autoref{fig:mfenkf_results_high_plan} (a,iv) than with \autoref{fig:mfenkf_results_high_plan} (a,ii) and (a,iii). More precisely, the improvement is qualitatively observable at the leading edge of the lower rotor blading, where the axial velocity has been corrected to adjust to the PIV data. It is clear that the updated field exhibits a lower residual error than the fields initially predicted by DDES and RANS, with the respective average relative difference $\bar{\varepsilon}$ dropping from $9.09\,\%$ and $5.55\,\%$ down to $3.81\,\%$.

To summarize, the application of the MFEnKF data assimilation (DA) algorithm, coupled with the physical hyperlocalization technique, successfully and effectively corrected the errors of the DDES and RANS simulations. The results of the data assimilation procedure are gathered in \autoref{tab:comparaison_results_moe}. By integrating numerical and experimental (PIV) data, a continuous flow field of near-experimental fidelity ($\left \langle \bar{u}_z^a \right \rangle$) was generated. However, the DA procedure requires available data for the calibration. 

\begin{table}[htbp]
\centering
\begin{tabular}{
    l
    >{\centering\arraybackslash}p{2.5cm}
    >{\centering\arraybackslash}p{2.5cm}
    >{\centering\arraybackslash}p{2.5cm}
}
\toprule
 & \textbf{P1} & \textbf{P2} & \textbf{P3} \\
\cmidrule(lr){2-4}
DDES   & $10.2\,\%$ & $6.21\,\%$ & $9.09\,\%$ \\
RANS   & $6.87\,\%$ & $2.91\,\%$ & $5.55\,\%$ \\
MFEnKF & $2.59\,\%$ & $1.81\,\%$ & $3.81\,\%$ \\
\addlinespace[2pt]\hdashline[1pt/2pt]\addlinespace[2pt]
ML-CR1 & $3.08\,\%$ & $4.15\,\%$ & $4.78\,\%$ \\
ML-CR2 & $4.81\,\%$ & $2.65\,\%$ & $5.17\,\%$ \\
ML-CR3 & $9.49\,\%$ & $4.75\,\%$ & $7.66\,\%$ \\
\addlinespace[2pt]\hdashline[1pt/2pt]\addlinespace[2pt]
MoE    & $4.17\,\%$ & $2.78\,\%$ & $3.84\,\%$ \\
\bottomrule
\end{tabular}
\caption{Comparison of average discrepancies with PIV data obtained with the DDES and RANS simulations, the MFEnKF data assimilation, and the four machine learning models investigated. The average discrepancies are measured on the three planes where PIV measurements are available.}
\label{tab:comparaison_results_moe}
\end{table}

The next challenge to be overcome is to obtain turbulence models which can automatically include and use the experimental information that was available for the DA procedure. To this purpose, generative tools from Machine Learning will be used to locally correct the performance of the turbulence models. The expanded data produced by the MFEnKF from the initial PIV measurements provide efficient training conditions for a machine learning approach, therefore setting ideal conditions to transfer the DA improved prediction to black box tools. The following section will thus detail the development of an Artificial Neural Network (ANN) capable of learning and predicting this correction law, allowing it to be applied \textit{a priori} and autonomously to future simulations.

%% file: sections/ann_enhancement.tex
\section{Enhancement of turbulence models via Artificial Neural Networks} \label{sec:ann_enhancement}

\subsection{Introduction}

In this section, the development of an artificial neural network is described, with the objective to mimic the corrections obtained with the Multi-Fidelity Ensemble Kalman Filter. The objective here is to obtain volume forcing source terms to be directly integrated in the Navier--Stokes equations, whose purpose is to locally correct the Spalart-Allmaras model used for turbulence closure.

Usually referred to as ANN, these models are sequences of operations designed to perform various tasks through machine learning \cite{cybenko1989approximation}, leveraging a function approximation $\mathcal{F}$  whose objective is to mathematically connect input and output data. This type of learning is ranked as \textit{supervised learning}, as the input and output data are known \textit{a priori} \cite{alloghani2020systematic}. Supervised learning is generally used for two main purposes: classification problems and regression problems.
\begin{figure}[h!]
    \centering
    \begin{tabular}{cc}
        \includegraphics[width=0.35\linewidth]{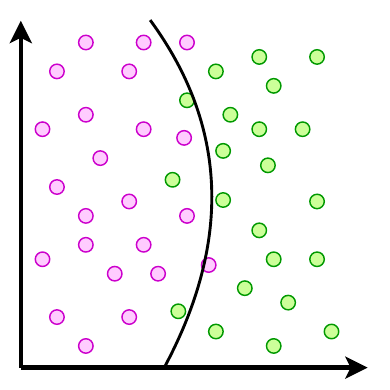} & \includegraphics[width=0.35\linewidth]{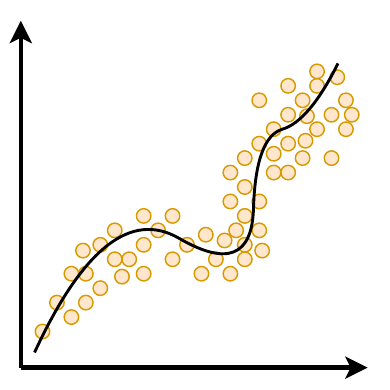} \\
        (a) & (b)
    \end{tabular}
    \caption{Diagram illustrating an example of supervised learning methodology. (a) Classification principle; (b) Regression principle.}
    \label{fig:classification_regression}
\end{figure}

On the one hand, classification involves predicting a category, or a \textit{label}, from a finite set of predefined categories, as illustrated in \autoref{fig:classification_regression} (a). There are many algorithms for solving a classification problem, including Support Vector Machines (SVMs), decision trees and neural networks.

On the other hand, regression involves predicting a continuous numerical value based on a set of input variables, as shown in \autoref{fig:classification_regression} (b). Unlike classification, regression establishes a mathematical relationship between the input variables and the expected output. Being quantitative in nature, this output is often referred to as the \textit{target} and can therefore take on an infinite number of possible numerical values. The applications of regression are also numerous and span a wide range of fields. In industry, it can be used to estimate the lifespan of a mechanical component \citep{alsina2018use} or to predict energy consumption \citep{kapp2023predictingEnergy}.

In the present section, the challenge of mimicking a corrective forcing term implies that ANNs are going to be used as \textbf{regression} methods rather than classification methods. 

\subsection{Basic neural network architecture and training} \label{subsec:neural_networks}

The architecture of the Neural Network is now described. The current network is composed of an input layer, followed by 4 hidden layers and finally an output layer. From \autoref{tab:ann_tab_description}, the number of trainable parameters is equal to $8\,675$.
\begin{table}[h!]
\centering
\begin{tabular}{lccc}
\toprule
\textbf{Layers}    & \textbf{Output shape} & \textbf{Number of parameters} & \textbf{Activation function} \\ \midrule
Input     &   $6$  &    $0$    &  --    \\ 
Hidden 1     &   $32$  &    $224$  &  ReLU      \\
Hidden 2     &   $64$  &    $2\,112$   &   ReLU    \\
Hidden 3     &   $64$  &    $4\,160$  &   ReLU     \\
Hidden 4     &   $32$  &    $2\,080$ &   ReLU      \\
Output     &   $3$  &    $99$  &  Linear      \\ \bottomrule
\end{tabular}
\caption{Description of the structure of the regression ANN model}
\label{tab:ann_tab_description}
\end{table}

Activation functions introduce non-linearity into the network. Without them, stacking multiple linear layers would be only able to represent a linear transformation of the input, regardless of the depth of the network. Thus, the introduction of activation functions allows the development of non-linear relationships between inputs and outputs, which is here essential for CFD applications. The ReLU activation function is defined as $f_\textrm{ReLU}\left( x\right) = \max\left(0,x \right)$ \cite{sharma2017activation}.


The utilized datasets are now described. For each of the clipping boxes displayed in \autoref{fig:colors_plans_boxes}, \textbf{input} data can be described as a vector $\mathbf{z}_{\textrm{input}}$ such that:
\begin{equation}
    \mathbf{z}_{\textrm{input}} = \left(\bar{u}_x,\bar{u}_y,\bar{u}_z,\mathbf{\mathcal{S}}, \mathcal{W}, \tilde{\nu} \right)^\top
\end{equation}
where $\left( \bar{u}_x, \bar{u}_y, \bar{u}_z\right)$ represents the three components of the velocity in a Cartesian framework, $\mathbf{\mathcal{S}}$ is the magnitude of the symmetric part of the velocity gradient tensor defined as $\mathbf{\mathcal{S}} = \left \vert{\frac{1}{2} \left( \nabla \bar{\mathbf{u}} + \nabla \bar{\mathbf{u}}^\top \right)} \right \vert$, $\mathcal{W}$ is the magnitude of the antisymmetric part of the velocity gradient tensor expressed as $\mathcal{W} = \left \vert{\frac{1}{2} \left( \nabla \bar{\mathbf{u}} - \nabla \bar{\mathbf{u}}^\top \right)} \right \vert$ and $\tilde{\nu}$ is the viscosity-like variable used in the Spalart-Allmaras turbulence model.

The \textbf{output} data is described as a vector $\mathbf{z}_{\textrm{output}}$ with three components such that:
\begin{equation}
    \mathbf{z}_{\textrm{output}} = \mathbf{f}^\mathcal{R} =  \left( f^\mathcal{R}_{\bar{u}_x}, f^\mathcal{R}_{\bar{u}_y}, f^\mathcal{R}_{\bar{u}_z} \right)^\top
\end{equation}

The $\mathbf{f}^\mathcal{R}$ vector measures the discrepancy between the flow prediction obtained by the CFD model forecast and the DA state estimation. This term can be written as:
\begin{equation}
    \mathbf{f}^\mathcal{R} = \mathbf{f}^{\mathcal{R}^a} - \mathbf{f}^{\mathcal{R}^f}
\end{equation}
with
\begin{align}
    \mathbf{f}^{\mathcal{R}^f} &= \rho \left(\bar{\mathbf{u}}^f \cdot \nabla \right) \bar{\mathbf{u}}^f  + \nabla p - (\mu + \mu_t) \nabla^2 \bar{\mathbf{u}}^f \\
    \mathbf{f}^{\mathcal{R}^a} &= \rho \left(\bar{\mathbf{u}}^a \cdot \nabla \right) \bar{\mathbf{u}}^a  + \nabla p - (\mu + \mu_t) \nabla^2 \bar{\mathbf{u}}^a
\end{align}
where the $f$ and $a$ superscripts respectively stand for the fields produced during the \textit{forecast} and the \textit{analysis} steps in the MFEnKF algorithm. The term $\mathbf{f}^\mathcal{R}$ can be seen as an additional volume source term to be integrated in the discretized form of the Navier-Stokes equations. 

\begin{equation}
    \pd{(\rho u_i)}{t} + \pd{}{x_j} \left( \rho u_i u_j + p \delta_{ij} - \tau_{ij} \right) = f_i^{\mathcal{R}}
\end{equation}

The idea, initially investigated by Villiers et al. \cite{Villiers2025_ftc}, is that the inclusion of this term can lead the numerical model to naturally produce the flow field obtained via the DA state estimation instead of the forecast result, therefore improving the accuracy of the underlying numerical model.

In order to proceed with an efficient training process and obtain a robust model, pre-processing is performed to eliminate outliers from the dataset. The InterQuartile Range method (IQR) \cite{vinutha2018detectionIQR} is used to prevent the training process from focusing on a small number of data points which could affect the network's parameter optimization. For any $\left(i,j \right)$ entry from dataset $z$.  The IQR method reads:
\begin{equation}
    z_{i,j} \in \text{Clean Set} \iff \forall j : (Q_{1,j} - r \cdot IQR_j) \leq z_{i,j} \leq (Q_{3,j} + r \cdot IQR_j) \label{eq:IQR}
\end{equation}
where the first and third quartiles, $Q_1$ and $Q_3$ , denote the values  below which $25\,\%$ and $75\,\%$ of the target values fall.
$IQR = Q_3 - Q_1$ is the interquartile range and $j$ indexes the features from the dataset. $r$ is the spread to which the outliers will be identified. For this work, \autoref{eq:IQR} is applied to the target vector $\mathbf{z}_\textrm{output}$, i.e. the output vector containing the forcing vector to be regressed, using a spread value equal to $r=10$ to exclude strong outlying data. The IQR enforces target datasets to keep coherent values, ensuring a strong trained model. Data is split in three subsets, namely the train, validation and test set with a random split respectively corresponding to $80\,\%$, $10$\,\% and $10$\,\% of the dataset. The validation set provides feedback during the training process, while the test set offers a final and independent performance assessment after the completed training. The model is compiled using the Mean Square Error (MSE) loss function and the Adam optimizer \cite{jais_adam_2019}. The initial learning rate is set to $10^{-4}$, with an automatic reduction checkpoint down to $10^{-9}$ if the validation loss metric during training is not reducing for 10 consecutive epochs. In that same way, an automatic stop checkpoint is set to prevent training from continuing if relevant metrics do not improve anymore. In particular, the criterion here set to trigger the stop checkpoint is that the validation loss metric does not improve for more than 20 consecutive epochs. 
Finally, the whole procedure is implemented with the Python TensorFlow \cite{tensorflow2015-whitepaper} and Keras \cite{chollet2015keras} packages. 
The training algorithm is repeated three times, using data for each clipping box previously displayed in \autoref{fig:colors_plans_boxes}. The obtained corrective models are referred to as CR1, CR2 and CR3 as they use DA data obtained with samples from the planes \textbf{P1}, \textbf{P2} and \textbf{P3}, respectively. 
\begin{figure}[h!]
    \centering
    \includegraphics[width=0.999\linewidth]{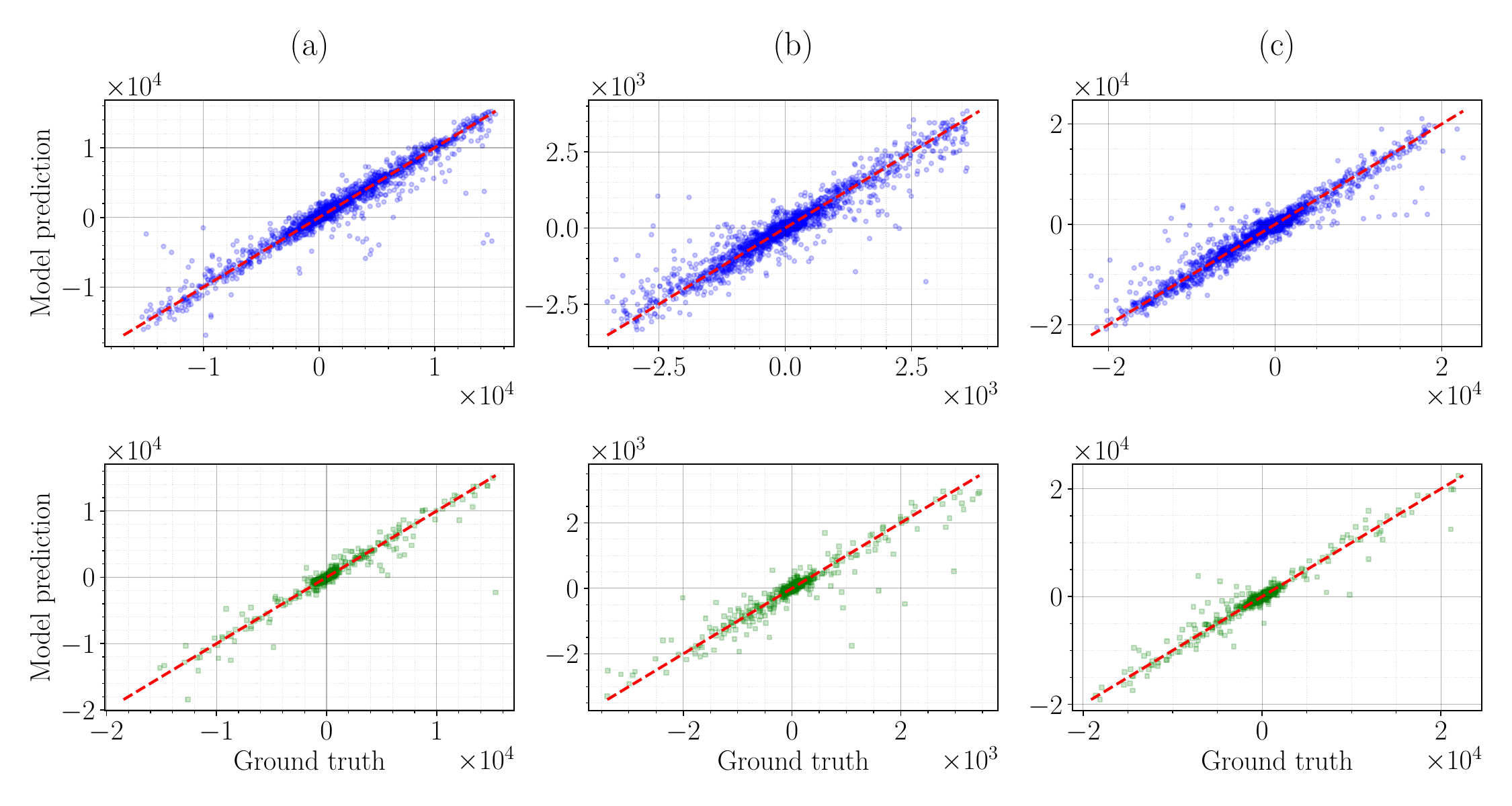}
    \caption{Parity plots corresponding to the CR1 neural network training for \textbf{P1}. Column (a) corresponds to $f_{u_x}^\mathcal{R}$; column (b) corresponds to $f_{u_y}^\mathcal{R}$; column (c) corresponds to $f_{u_z}^\mathcal{R}$. Top row with blue dots represents the validation data; bottom row with green squares represents the test data.}
    \label{fig:parity_plots_Low}
\end{figure}

The accuracy of these models is first assessed via the analysis of the parity plots comparing the predicted forcing components $f_{\bar{u}_x}^\mathcal{R}$, $f_{\bar{u}_y}^\mathcal{R}$ and $f_{\bar{u}_z}^\mathcal{R}$ against their ground truth values. CR1 training results obtained using data from \textbf{P1} are shown in \autoref{fig:parity_plots_Low}. The dashed red lines represent the $1:1$ identity, which indicates a perfect agreement between ground truth and predictions. The top row (blue dots) represents the validation data, while the bottom row (green squares) represents the test data. For each subplot, the coefficient of determination $R^2$ \cite{chicco2021coefficientR2} is calculated to quantify the quality of the regression. This coefficient is defined as:
\begin{equation}
    R^2 = 1 - \frac{\sum_{i=1}^N \left( z_i - \hat{z}_i\right)^2 }{\sum_{i=1}^N \left( z_i - \overline{z}_i\right)^2 }
\end{equation}
where $z_i$ are the ground-truth values, $\hat{z}_i$ are the predicted values and $\overline{z}_i$ is the average of all the ground-truth values. An $R^1$ of $1$ score corresponds to a perfect regression fit, while an $R^2$ of $0$ score indicates that the model is not more accurate than predicting the average value of the dataset. 

The parity plots in \autoref{fig:parity_plots_Low} show an encouraging level of convergence of the regression models for the three forcing components. More precisely, the model for $f_{u_z}^\mathcal{R}$ shown in \autoref{fig:parity_plots_Low} column (c), achieves an $R^2$ score of $0.93$ for the test set, indicating that the main flow features are captured with high accuracy. This component of the volume forcing is integrated in the streamwise component of the discretized equations. The component of the forcing vector $f_{u_x}^\mathcal{R}$ is also well predicted as an $R^2$ score of $0.91$ is reached for the test dataset. A slight performance decrease is observed for the $f_{u_y}^\mathcal{R}$ component of the forcing vector, with an $R^2$ score of $0.87$ for the test set. However, this score still represents a strong correlation between ground-truth and predictions. For all the forcing vector components, the close proximity between validation and test metrics confirms the generalization of the model on unseen data and it indicates that the model does not suffer from overfitting.


\subsection{Integration of the Neural Network based corrections in the RANS solver} \label{subsec::neural_networks_results}

\begin{figure}[h!]
    \centering
    \includegraphics[width=0.75\linewidth]{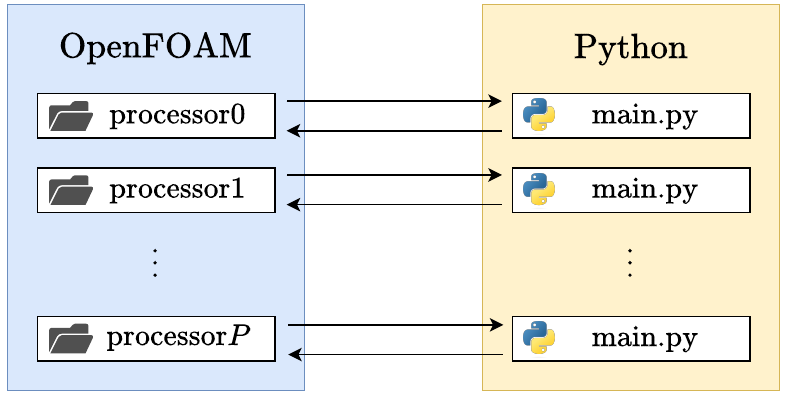}
    \vspace{0.35cm}
    \caption{Scheme of the coupling strategy between the CFD solver and the ML tools.}
    \label{fig:openfoam_python_coupling}
\end{figure}

The usage of black-box correction terms based on neural networks is now described.
The integration of the pre-trained models with OpenFOAM's solvers is realized through a Multiple Program, Multiple Data (MPMD) \cite{gropp1994} architecture, which allows the simultaneous execution of distinct computational tasks. Rather than directly embedding the Machine Learning libraries, which are developed with Python, into the C++ OpenFOAM solvers, the coupling strategy relies on the Message Passing Interface (MPI) \cite{barker2015MPI} routines to bridge the two environments. 
The connection between the two environments is schematized in \autoref{fig:openfoam_python_coupling}.

The machine learning volume forces are integrated within the resolution of the momentum equation of the RANS solvers. The reason is that the models are trained with mean data, which is a constraint associated with the available PIV samples. In order to develop a model suitable for time resolved simulations such as DDES one would have to perform DA state estimation using instantaneous snapshots for both CFD runs and experiments, which were not available for this analysis. However, the procedure would be exactly the same, therefore opening research opportunities for future investigations.   

\begin{figure}[h!]
    \centering
    \includegraphics[width=0.99\linewidth]{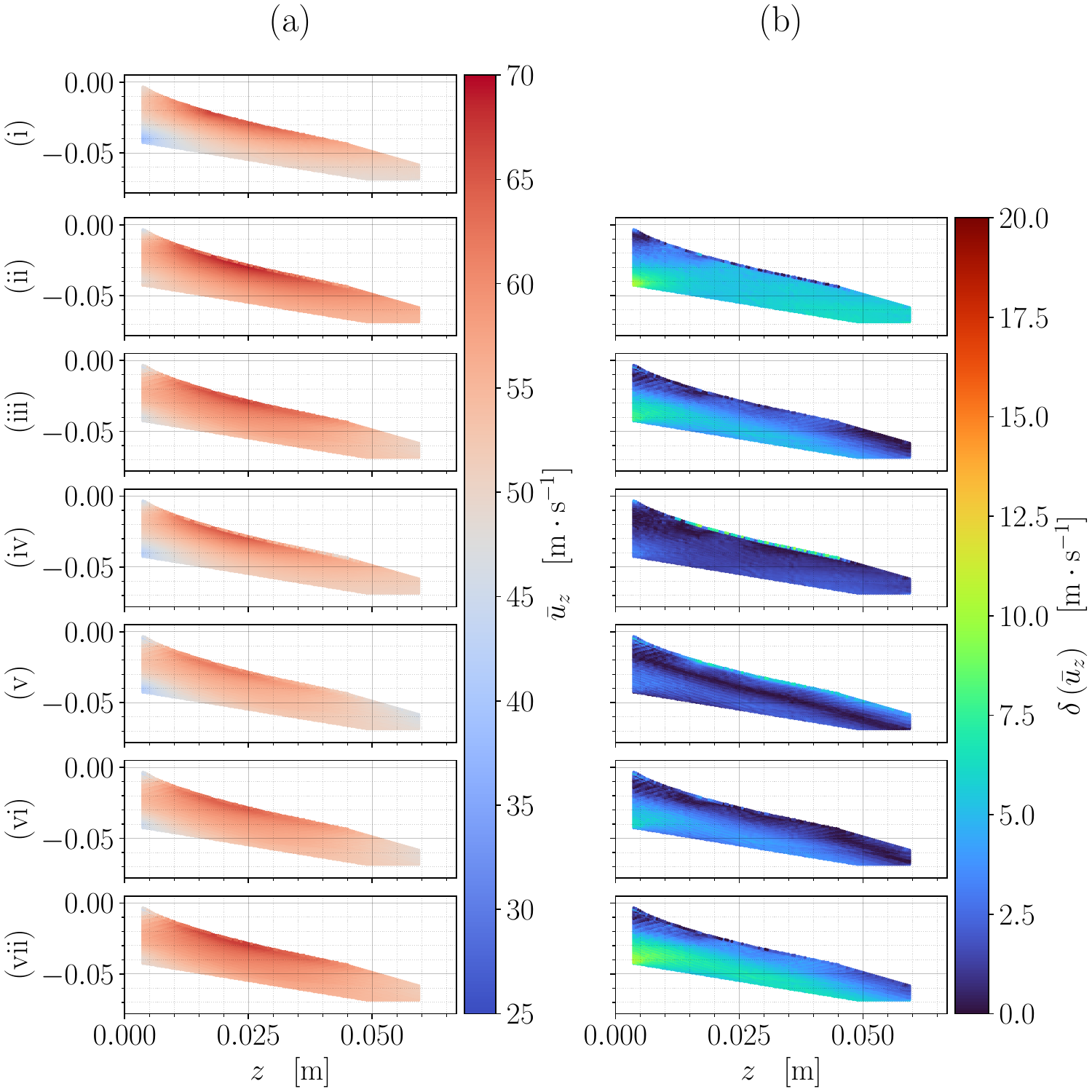}
    \caption{Mean axial velocity $\bar{u}_z$ on \textbf{P1}. Column (a): isocontours of $\bar{u}_z$. Column (b): L2 norm measuring the discrepancy $\delta(\bar{u}_z) = \left \vert u_z^\textrm{PIV} - \bar{u}_z \right \vert$ relative to the reference PIV data. Line (i) Reference experimental data (PIV); line (ii) DDES results; line (iii) RANS results; line (iv) DA state estimation; line (v) RANS + CR1 ML model; line (vi) RANS + CR2 ML model; line (vii) RANS + CR3 ML model.}
    \label{fig:ann_results_low_plan}
\end{figure}

The flow prediction results obtained using RANS combined with ML tools are analyzed in the remainder of this section. The comparisons are performed considering the mean velocity fields on the planes \textbf{P1}, \textbf{P2} and \textbf{P3}, where PIV data is available. First, \autoref{fig:ann_results_low_plan} presents the results obtained on the plane \textbf{P1} (blue plane and box in \autoref{fig:colors_plans_boxes}). As a reminder, \autoref{fig:ann_results_low_plan} (a,i) displays the axial velocity $u_z$ obtained from PIV measurements, (a,ii) the mean axial velocity predicted by the prior DDES, (a,iii) the mean axial velocity $\bar{u}_z$ predicted by the prior RANS, while (a,iv) presents the ensemble mean of the updated axial velocity from the MFEnKF analysis $\left \langle \bar{u}_z^a \right \rangle$. In addition to these previously presented results, \autoref{fig:ann_results_low_plan} also provides the mean velocity fields obtained with the data-driven models RANS + CR1, RANS + CR2 and RANS + CR3. The second column reports the L2 norm of the discrepancy with PIV data $\left \vert u_z^{\textrm{PIV}} - \bar{u}_z^{i} \right \vert$, where $i$ indicates one of the numerical simulations previously listed. 

A first qualitative analysis reveals that the field predicted by the model RANS + CR1, shown in \autoref{fig:ann_results_low_plan} (a,v), shows strong similarities with the field $\left \langle \bar{u}_z^a \right \rangle$ obtained via DA in \autoref{fig:ann_results_low_plan} (a,iv), and therefore with the $u_z^{\textrm{PIV}}$ measurements. Comparison with the RANS data in (a,iii) highlights an improvement in the prediction of the axial velocity at the leading edge of the lower profile, located in the lower-left corner of the plane. However, some discrepancies remain between the $\bar{u}_z^{\textrm{PIV}}$ and $\bar{u}_z^{\textrm{ML}}$ fields, particularly at the trailing edge of the upper profile where the ML approach underestimates the axial velocity. Discrepancies calculated via \autoref{eq:varepsi} are reported in \autoref{tab:comparaison_results_moe}. From a quantitative standpoint, as a reminder, the DA analysis reduced the discrepancy for the velocity field from $6.87\,\%$ (RANS value) to $2.59\,\%$. The integration of the CR1 ML model in the RANS run results in an L2 error norm of $3.08\,\%$ on the plane from which it received training data.  This result proves to be satisfactory, as not only are these errors lower than those of the original RANS simulation, but they are also very similar to those observed for the DA approach. For comparison, the velocity field obtained on \textbf{P1} by the models CR2 and CR3 is shown in lines (vi) and (vii). The L2 error norms are $4.81\,\%$ and $9.49\,\%$ respectively. These results are more in line with the prior RANS prediction, even if one can see that the run RANS + CR3 is the one exhibiting the largest discrepancy. This simulation relies on the model trained with data close to the blade tip.

\begin{figure}[h!]
    \centering
    \includegraphics[width=0.99\linewidth]{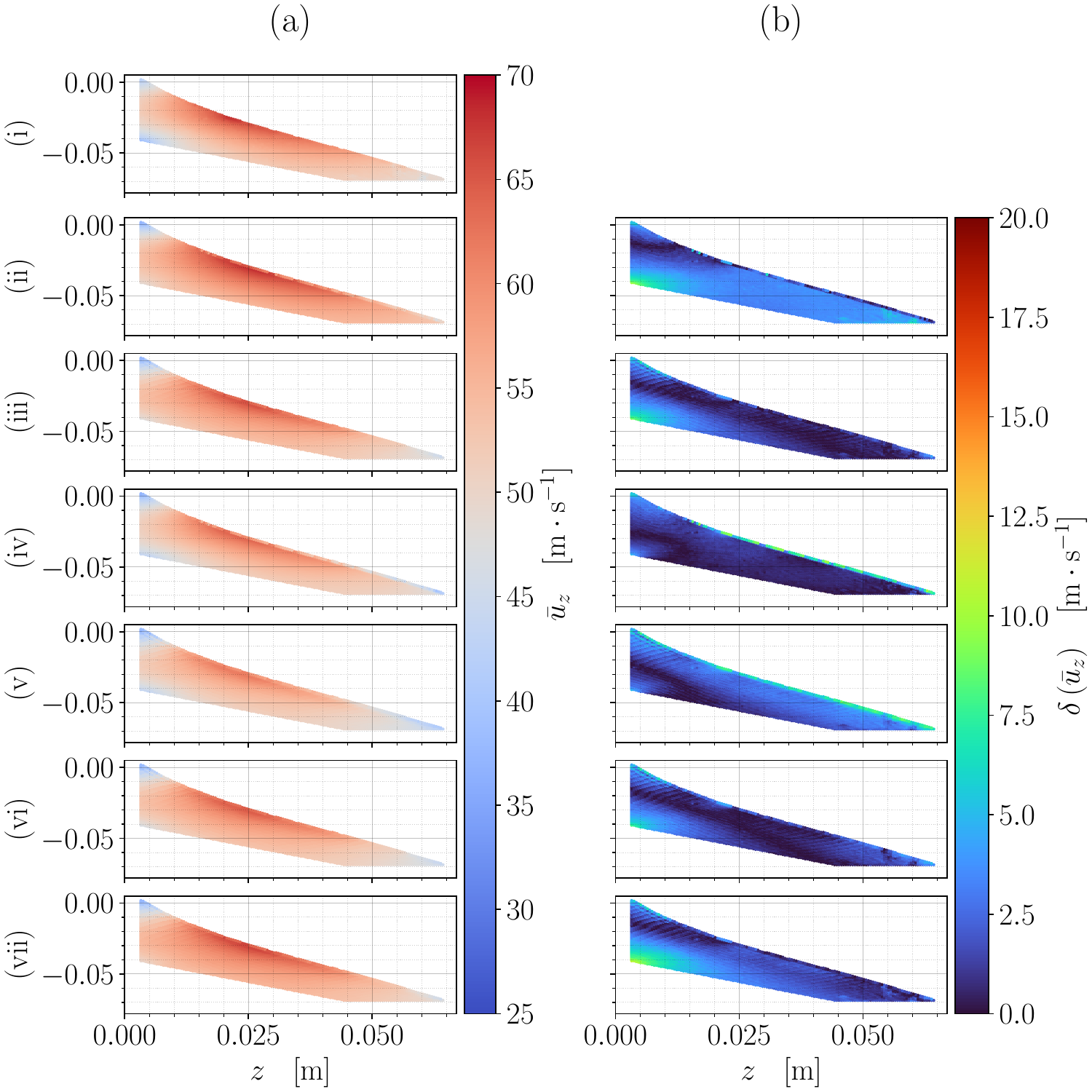}
    \caption{Mean axial velocity $\bar{u}_z$ on \textbf{P2}. Column (a): isocontours of $\bar{u}_z$. Column (b): L2 norm measuring the discrepancy $\delta(\bar{u}_z) = \left \vert u_z^\textrm{PIV} - \bar{u}_z \right \vert$ relative to the reference PIV data. Line (i) Reference experimental data (PIV); line (ii) DDES results; line (iii) RANS results; line (iv) DA state estimation; line (v) RANS + CR1 ML model; line (vi) RANS + CR2 ML model; line (vii) RANS + CR3 ML model.}
    \label{fig:ann_results_mid_plan}
\end{figure}

The velocity fields predicted by the numerical simulations for the plane \textbf{P2} are now investigated. The results are shown in \autoref{fig:ann_results_mid_plan} following the same structure previously used for the plane \textbf{P1} i.e. velocity fields and their discrepancy with respect to PIV data are shown. 
The combination of RANS and the ML model CR2 is first considered. This model is trained using DA data in the proximity of the plane \textbf{P2}, therefore it is expected to perform the best. The analysis of the results in \autoref{fig:ann_results_mid_plan} (vi) confirms the accuracy of the prediction. 

A qualitative analysis of the axial velocity isocontours in column (a) shows that the axial velocity field $\bar{u}_z^{\textrm{ML}}$ predicted using the RANS + CR2 model (a,vi) well reproduces overall the topology of the flow resulting from the DA analysis, thus exhibiting a reduced discrepancy with the reference experimental measurements $\bar{u}_z^{\textrm{PIV}}$. Examination of the L2 discrepancies $\delta \left( \bar{u}_z \right)$ in column (b) (\autoref{tab:comparaison_results_moe}) allows for quantification of these previous observations. The initial error of the RANS model was already relatively low ($2.91\,\%$) but the MFEnKF DA algorithm successfully reduced it to $1.81\,\%$. The application of the RANS + CR2 model leads to an L2 error norm of $2.65\,\%$, which is moderately reduced but still improved when compared with the baseline. The performance of the CR1 and CR3 ML models for the prediction of the flow on the plane \textbf{P2} is shown in \autoref{fig:ann_results_mid_plan} (v) and (vii). Recall that the corrective models are here used far from the region where they were trained. Their average discrepancy shows an L2 error norm of $4.15\,\%$ and $4.75\,\%$, respectively, which are higher than those observed for the solution via the CR2 model in \textbf{P2}, but also higher than the $2.91\,\%$ of the prior RANS solution.

\begin{figure}[h!]
    \centering
    \includegraphics[width=0.99\linewidth]{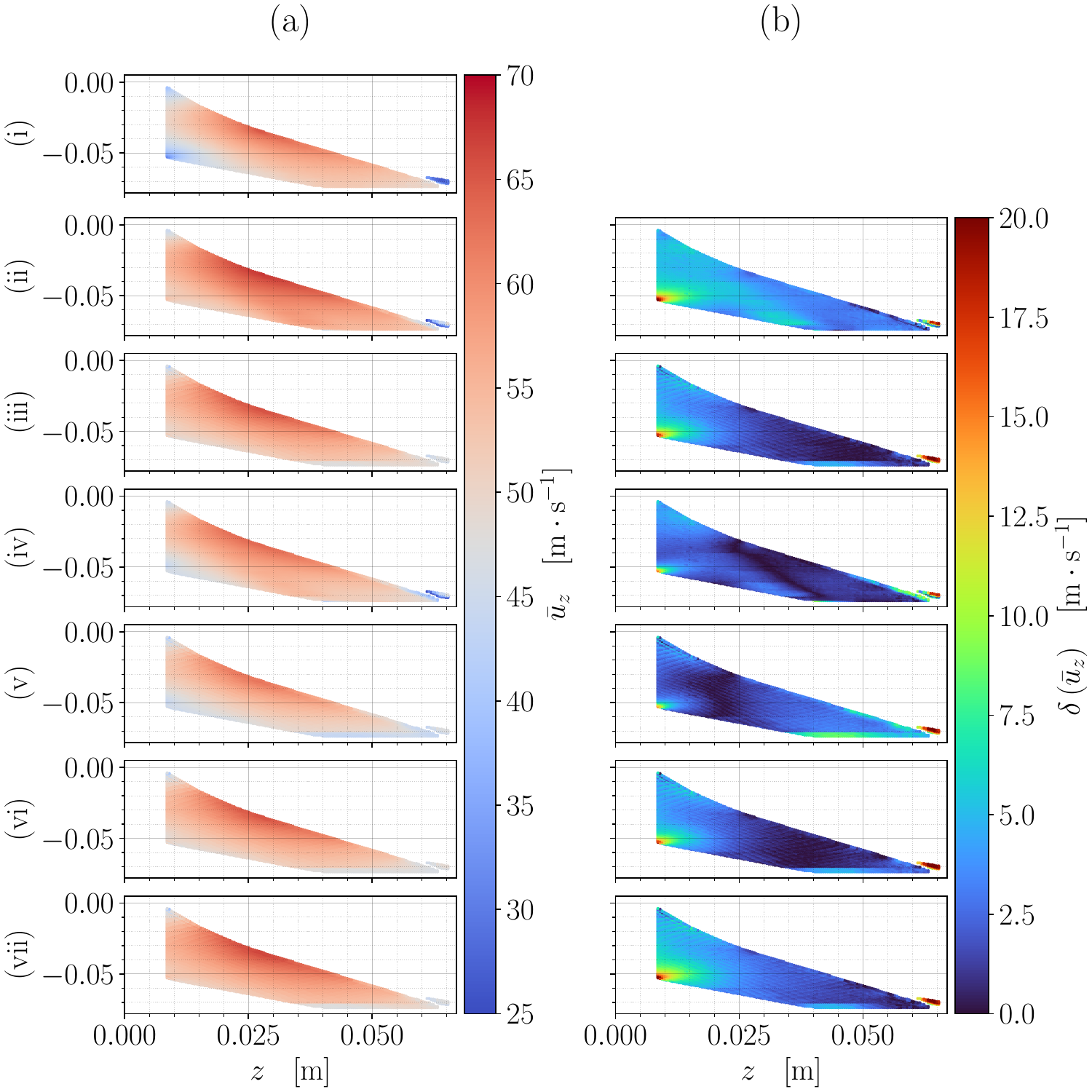}
    \caption{Mean axial velocity $\bar{u}_z$ on \textbf{P3}. Column (a): isocontours of $\bar{u}_z$. Column (b): L2 norm measuring the discrepancy $\delta(\bar{u}_z) = \left \vert u_z^\textrm{PIV} - \bar{u}_z \right \vert$ relative to the reference PIV data. Line (i) Reference experimental data (PIV); line (ii) DDES results; line (iii) RANS results; line (iv) DA state estimation; line (v) RANS + CR1 ML model; line (vi) RANS + CR2 ML model; line (vii) RANS + CR3 ML model.}
    \label{fig:ann_results_high_plan}
\end{figure}

The velocity fields predicted by the numerical simulations for the plane \textbf{P3} are finally investigated. The results are shown in \autoref{fig:ann_results_high_plan}, following the same structure  previously used for the planes \textbf{P1} and \textbf{P2}. As a reminder, \textbf{P3} is located at $79\,\%$ of the rotor blade height, in the proximity of the blade tip where complex and three-dimensional dynamics are observed. 
The combination of RANS and the ML model CR3 is first considered. This model is trained using the DA data in the proximity of the plane \textbf{P3}, so it is expected to perform the best. However, the analysis of the results in \autoref{fig:ann_results_high_plan} (vii) does not confirm that expectation. A qualitative analysis of the mean axial velocity field isocontours in column (a) shows that the axial velocity field $\bar{u}_z^{\textrm{ML}}$ predicted using the RANS+CR3 model shown in (a,vii) is not able to match the accuracy of the MFEnKF prediction, which is shown in \autoref{fig:ann_results_high_plan} (a,iv). As a consequence, it exhibits a large discrepancy with the reference experimental measurements $\bar{u}_z^{\textrm{PIV}}$. 
Using \autoref{eq:varepsi}, the error of the prior RANS is $5.55\,\%$ and the MFEnKF algorithm successfully reduces it to $3.81\,\%$ thanks to a coherent correction of the physical field. Nevertheless, the application of the RANS+CR3 model leads to an L2 error norm of $7.66\,\%$, which represents a significant degradation of the accuracy with respect to the baseline results. This loss of accuracy can be explained by the local aerodynamic complexity of this region in the compressor, where the flow physics is governed by strong gradients, tip clearance flows and other vortical structures. Therefore, the accumulation of the physical phenomena involved represents a major hurdle for the application of ML tools, even in the proximity of the region where the CR3 training data was sampled. The performance of the CR1 and CR2 ML models for the prediction of the flow on the plane \textbf{P3} is shown in \autoref{fig:ann_results_high_plan} (v) and (vi) respectively. It is reminded that these corrective models are here evaluated far from the location where they received data for training. Their average discrepancy shows an L2 error norm of $4.78\,\%$ and $5.17\,\%$. These values are higher than the one obtained with DA, but they are slightly lower than the $5.55\,\%$ of the prior RANS prediction. Therefore, on \textbf{P3}, the RANS+CR3 run not only fails to approach the MFEnKF results, but is also the least accurate of all the performed RANS simulations. 


The analysis of the corrective models CR1, CR2 and CR3 has shown that, while an improvement of the prediction can be obtained, the results may be unreliable and also be responsible for a decrease in accuracy with respect to the baseline RANS prediction. This is the case, for example, for the CR1 model applied on \textbf{P2}, as well as the model CR3 on all three planes. The next research question is whether this sensitivity of the models can be reduced by combining them in a common numerical framework, which is the mixture of experts model.

\subsection{Mixture of experts development}
In this section, the three corrective models CR1, CR2, and CR3 are blended with a single RANS simulation. This procedure exploits their higher accuracy when applied in physical regions which show similarities with their training data, therefore producing a more efficient global model. This procedure is performed with a mixture of experts (MoE) technique \citep{masoudnia2014mixture} which has been recently applied in fluid mechanics studies \cite{oulghelou2025machine}. The procedure trains a master model which, via classification, decides which combination of the available model corrections is going to be employed. In the present work, for the sake of simplicity, the master model works with the geometric features of the compressor. While an algebraic law  depending on the geometry could have been used in its place, the present work prepares a platform to include more sophisticated features in the selection of the model combination in future analyses. These features will include for example the local characteristics of the flow in terms of velocity, pressure and temperature.

\subsubsection{Architecture setup}
\begin{figure}[h!]
    \centering
    \includegraphics[width=0.99\linewidth]{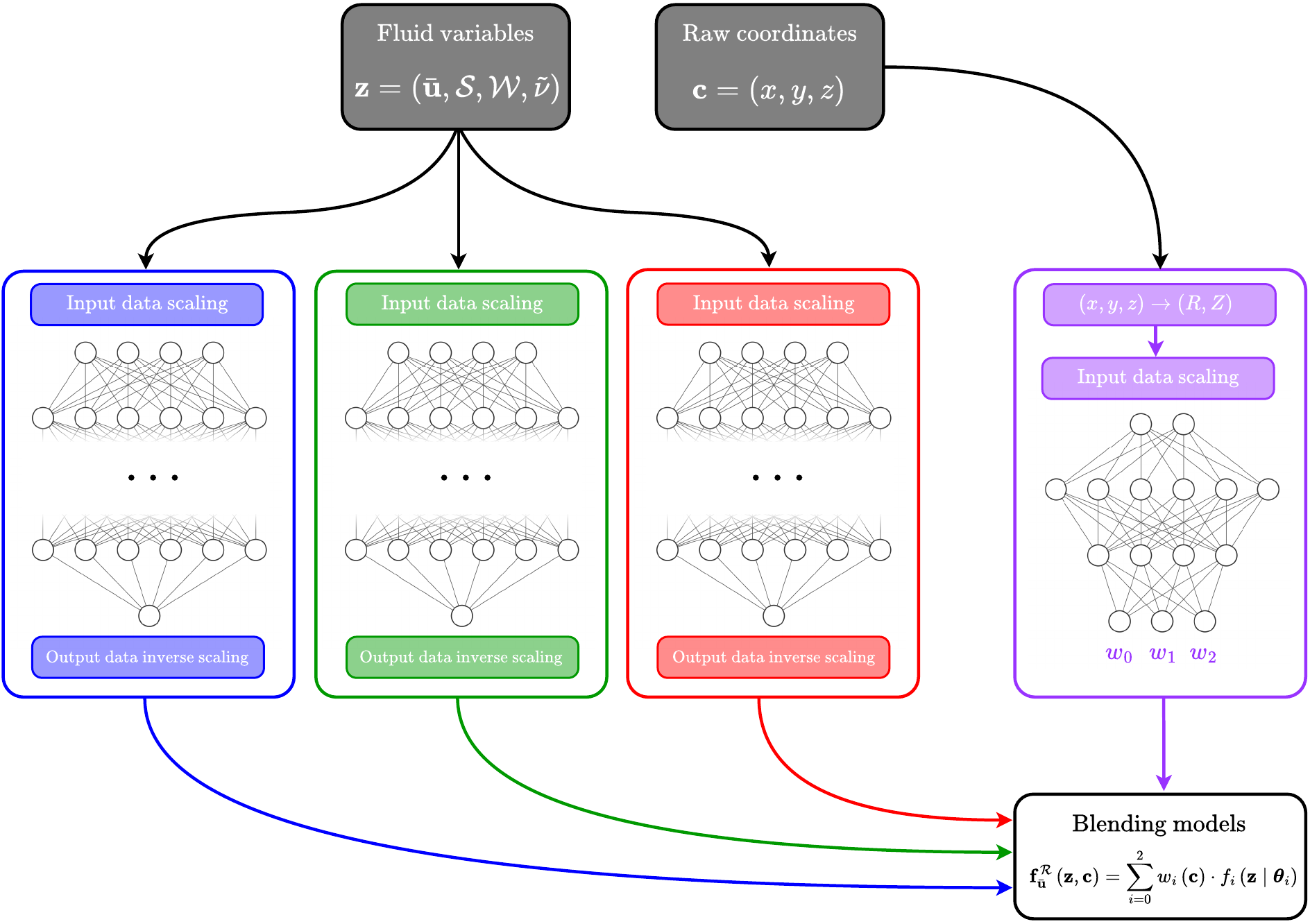}
    \caption{Mixture of experts architecture.}
    \label{fig:MoE_arch}
\end{figure}

The architecture used for the present analysis is schematized in \autoref{fig:MoE_arch}. The three models CR1, CR2 and CR3, which are trained independently using different datasets, are referred to as \textit{the experts}. They are now blended together via a classification network which routes each grid element to the appropriate combination of experts.

The training process of the classification network is now described. First, the \textbf{input} data is stored in the  $\mathbf{c}_{\textrm{input}}$ vector with:
\begin{equation}
    \mathbf{c}_{\textrm{input}} = \left( R,Z \right)
\end{equation}
where $\left( R,Z\right)$ corresponds to the cylindrical coordinates computed from the cartesian coordinates $\left( x,y,z\right)$ as:
\begin{equation}
    \begin{cases}
        R = \sqrt{x^2 + y^2} \\
        Z = z
    \end{cases}
\end{equation}

\begin{table}[h!]
\centering
\begin{tabular}{lccc}
\toprule
\textbf{Layers}    & \textbf{Output shape} & \textbf{Number of parameters} & \textbf{Activation function} \\ \midrule
Input     &   $2$  &    $0$    &  --    \\ 
Hidden 1     &   $32$  &    $96$  &  ReLU      \\
Hidden 2     &   $16$  &    $528$   &   ReLU    \\
Output     &   $3$  &    $51$  &  Softmax      \\ \bottomrule
\end{tabular}
\caption{Hyperparameters used for the classification ANN used by the MoE}
\label{tab:class_ann_tab_description}
\end{table}

The architecture of the classification network, composed of 675 trainable parameters, is described in \autoref{tab:class_ann_tab_description}. The softmax activation function $\bm{\sigma}$ is used on the output layer, for its ability to convert a tuple of real values $\mathbf{c} = \left( c_1, \dots, c_K \right) \in \mathbb{R}^K$ to a probability distribution $\bm{\sigma} \left( \mathbf{c} \right) \in \left( 0,1\right)^K$, thanks to the relation:
\begin{equation}
    \bm{\sigma} \left( \mathbf{c} \right)_i = \frac{\exp{\left( c_i \right)}}{\sum_{j=1}^K \exp{\left( c_j \right)}}
\end{equation}

The \textbf{output} data is extracted from the state estimation boxes produced via MFEnKF in \autoref{fig:colors_plans_boxes} (b), where grid elements receive a label depending on their geometric position and whether they are inside a state estimation region. 
We have the following encoding:
\begin{equation}
    \begin{cases}
        \textrm{Elements in Box 1 } \longrightarrow \textrm{Label }0 \longrightarrow \texttt{(1,0,0)} \\
        \textrm{Elements in Box 2 } \longrightarrow \textrm{Label }1 \longrightarrow \texttt{(0,1,0)} \\
        \textrm{Elements in Box 3 } \longrightarrow \textrm{Label }2 \longrightarrow \texttt{(0,0,1)} 
    \end{cases}
\end{equation}

These labels, which act as weights in the blending procedure, are stored in the $\mathbf{c}_{\textrm{output}} = \left(w_0, w_1, w_2 \right)$ probability vector.

Data are split to a training and a validation set, respectively representing $80\, \%$ and $20\,\%$ of the full dataset. This classification model is compiled using the Categorical Cross-Entropy (CCE) loss function $\mathcal{L}_{\textrm{CCE}}$, defined as:
\begin{equation}
    \mathcal{L}_{\textrm{CCE}} = - \sum_{i=1}^K y_i \, \log{\left( \hat{y}_i \right)}
\end{equation}
where $y_i$ is the ground truth and $\hat{y}_i$ is the output predicted by the network.


Finally, the training process uses the Adam optimizer with a learning rate set to $10^{-3}$ and an automatic reduction based on the validation set accuracy if the metric does not improve for 10 consecutive epochs, down to a learning rate of $10^{-9}$. An early stopping checkpoint is likewise implemented to halt training when performance stagnates. Specifically, training stops if the validation loss shows no improvement for $20$ consecutive epochs.

\begin{figure}[h!]
    \centering
    \includegraphics[width=0.45\linewidth]{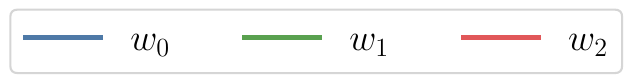}
    \includegraphics[width=0.95\linewidth]{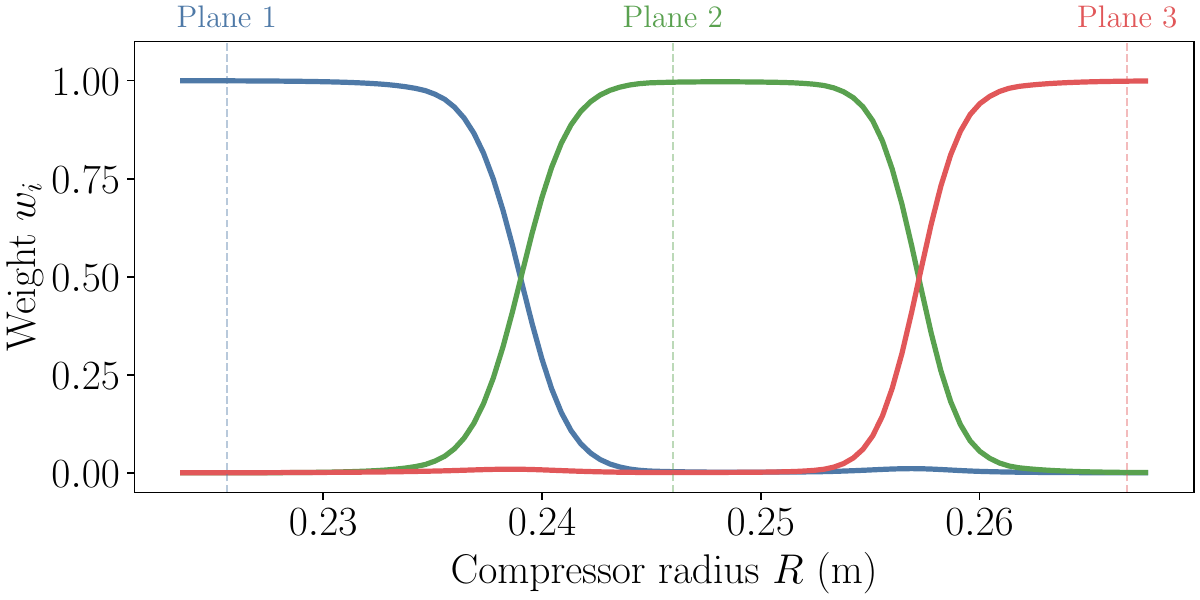}\\
    \caption{Spatial distribution of the expert weighting coefficients across the computational domain.}
    \label{fig:weight_distribution}
\end{figure}

Once the three experts and the classifier are trained, the models are blended into a single prediction $\mathbf{f}_{\bar{\mathbf{u}}}^{\mathcal{R}} \left( \mathbf{z},\mathbf{c} \right)$:
\begin{equation}
    \mathbf{f}_{\bar{\mathbf{u}}}^{\mathcal{R}} \left( \mathbf{z},\mathbf{c} \right) = \sum_{i=0}^2 w_i \left( \mathbf{c} \right) \cdot f_i \left( \mathbf{z} \mid \bm{\theta}_i \right)
\end{equation}

where $w_i \left( \mathbf{c} \right)$ represents the weights from the classification network and $f_i \left( \mathbf{z} \mid \bm{\theta}_i \right)$ the prediction from each expert CR$i$. It is important to stress that the weights may vary for each grid element, therefore tuning the correction performed. The behavior of the trained MoE model is assessed in \autoref{fig:weight_distribution}, which presents the variation of the expert weighting coefficients $w_{i \in [0,2]}$ along the channel height, illustrating how the classification network attributes each expert to its corresponding flow region. It can be seen that these distributions are smooth and they only depend on the radius of the compressor stage. Again, future analyses will substitute geometric information with local features of the flow field.

\subsubsection{MoE results}

\begin{figure}[h!]
    \centering
    \includegraphics[width=0.99\linewidth]{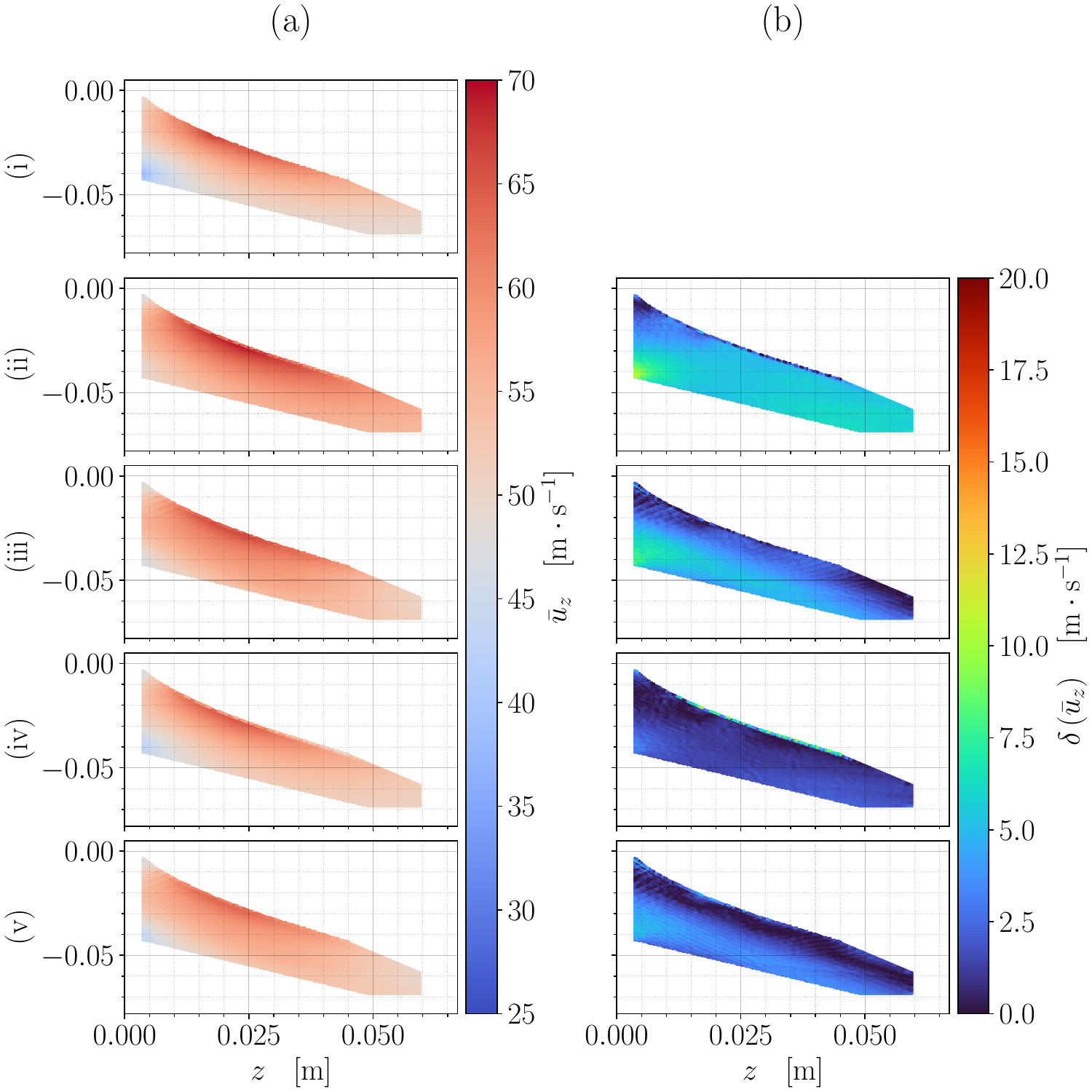}
    \caption{Mean axial velocity $\bar{u}_z$ on \textbf{P1}. Column (a): isocontours of $\bar{u}_z$. Column (b): L2 norm measuring the discrepancy $\delta(\bar{u}_z) = \left \vert u_z^\textrm{PIV} - \bar{u}_z \right \vert$ relative to the reference PIV data. Line (i) Reference experimental data (PIV); line (ii) DDES results; line (iii) RANS results; line (iv) DA state estimation; line (v) RANS + MoE ML model.}
    \label{fig:moe_ann_results_low_plan}
\end{figure}

The mixture of experts architecture is integrated within a RANS simulation of the CME2 axial compressor and results are now discussed.
Similarly to what was proposed in \autoref{subsec::neural_networks_results}, the results are discussed for each plane where PIV data are available. \autoref{fig:moe_ann_results_low_plan} shows for \textbf{P1} that the axial velocity field $\bar{u}_z^{\textrm{MoE}}$ obtained with the new model, which is displayed in (a,v), exhibits strong similarities with the field $\left \langle \bar{u}_z^a \right \rangle$ obtained by DA in (a,iv), and therefore the experimental measurements $\bar{u}_z^{\textrm{PIV}}$ as well. The MFEnKF results on this plane exhibited an average discrepancy with PIV of $2.59\,\%$, which is significantly lower than the $6.87\,\%$ value observed for RANS. The combination of RANS with the correction model CR1 provided an average discrepancy of $3.08\,\%$, very close to the MFEnKF result. The RANS+MoE model yields a higher average L2 error norm of $4.17\,\%$. This outcome can be considered as satisfactory, as it remains significantly lower than the prior RANS field prediction and it is of the same order of magnitude observed for the DA results and the RANS + CR1 model results. Looking at the distribution of the L2 discrepancy norm $\delta \left( \bar{u}_z \right)$ over the plane \textbf{P1}, one can see that the topology of the RANS + MoE model is similar to the one of the initial RANS simulation, but the discrepancy is reduced everywhere. This suggests that the MoE correction acts as a global attenuation of the RANS error rather than a local correction of specific flow features. 

\begin{figure}[h!]
    \centering
    \includegraphics[width=0.99\linewidth]{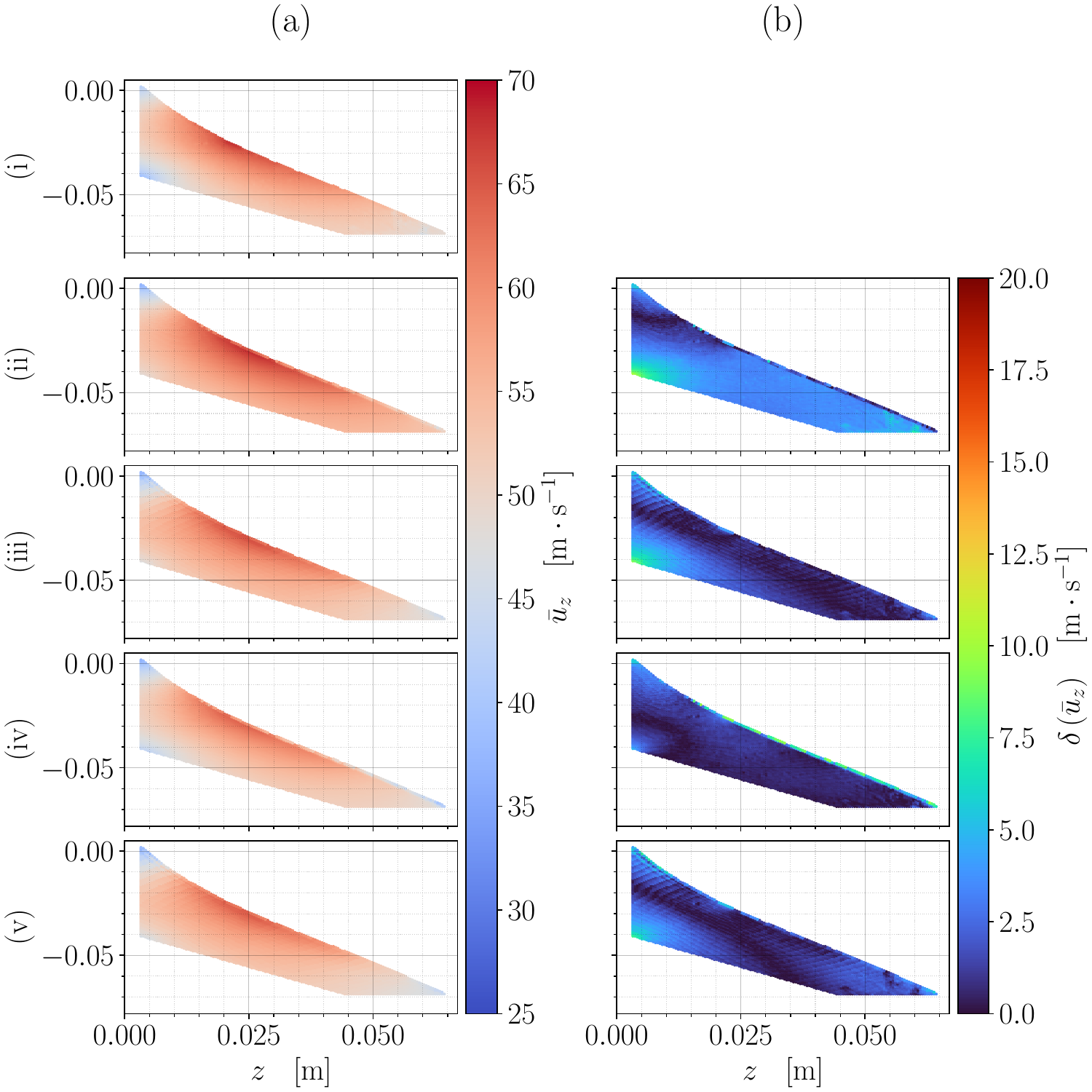}
    \caption{Mean axial velocity $\bar{u}_z$ on \textbf{P2}. Column (a): isocontours of $\bar{u}_z$. Column (b): L2 norm measuring the discrepancy $\delta(\bar{u}_z) = \left \vert u_z^\textrm{PIV} - \bar{u}_z \right \vert$ relative to the reference PIV data. Line (i) Reference experimental data (PIV); line (ii) DDES results; line (iii) RANS results; line (iv) DA state estimation; line (v) RANS + MoE ML model.}
    \label{fig:moe_ann_results_mid_plan}
\end{figure}

\autoref{fig:moe_ann_results_mid_plan} shows the mean axial velocity field on \textbf{P2}. As a reminder, the average discrepancies with respect to the PIV data of the DDES and RANS simulations were $6.21\,\%$ and $2.91\,\%$, respectively. The MFEnKF analysis in (a,iv) reduced this error to $1.81\,\%$, and the combination of RANS and CR2 model produces an average L2 error norm of $2.65\,\%$. 
The application of the MoE architecture shown in \autoref{fig:moe_ann_results_mid_plan} (a,v) yields an average L2 discrepancy norm of $2.78\,\%$, which is very similar to what is observed for the RANS+CR2 model combination and slightly lower than the results obtained for the baseline RANS prediction. This observation suggests that the ML training was able to provide a robust behavior of both the experts and the classification network, as the accuracy of the RANS prediction is preserved and even slightly improved. The examination of the distribution of the L2 discrepancy norm $\delta \left( \bar{u}_z \right)$ over the plane \textbf{P2} shows that the baseline RANS solution and the RANS + MoE model exhibit very similar distributions, highlighting the robustness of the model and the consistency observed on the plane \textbf{P1}.

\begin{figure}[h!]
    \centering
    \includegraphics[width=0.99\linewidth]{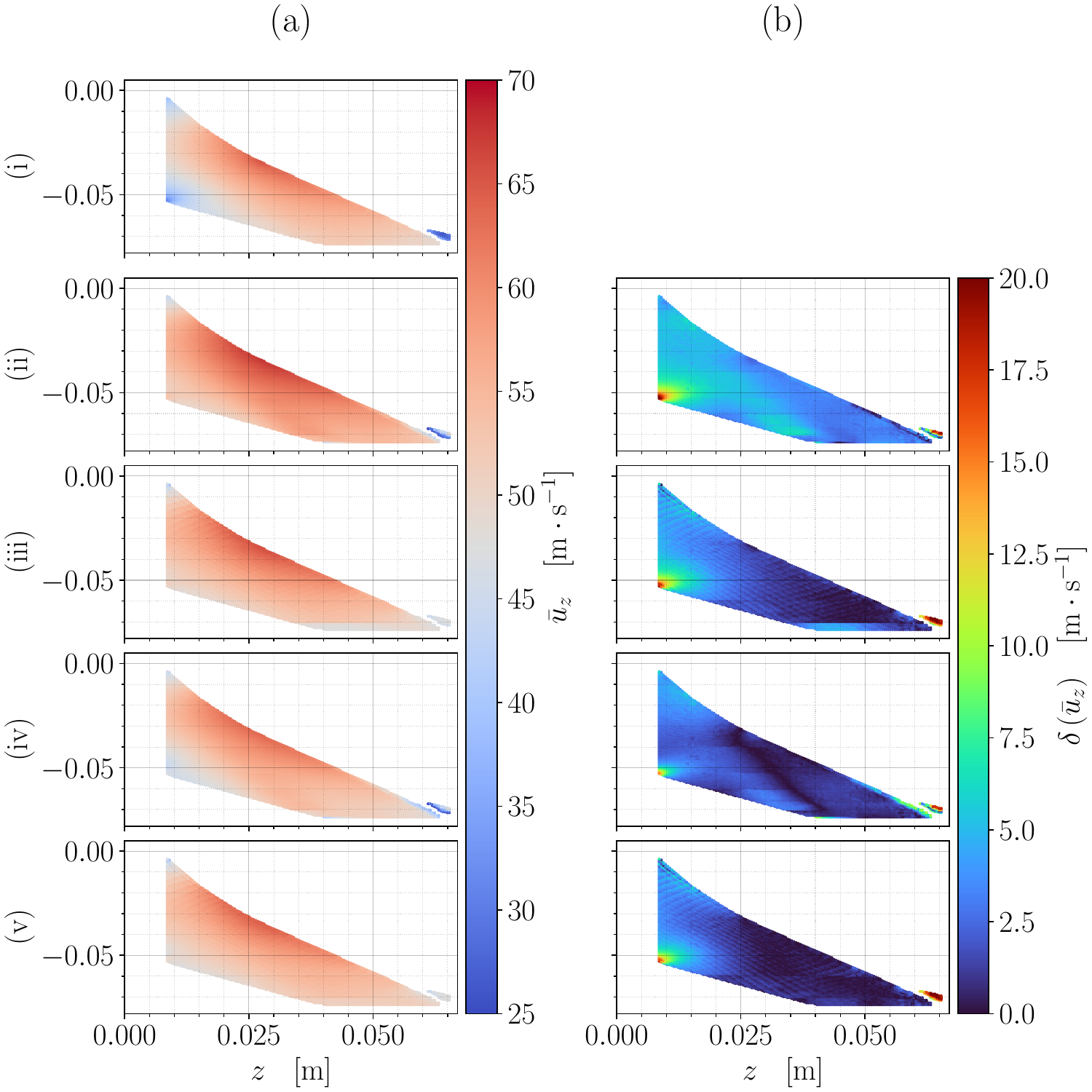}
    \caption{Mean axial velocity $\bar{u}_z$ on \textbf{P3}. Column (a): isocontours of $\bar{u}_z$. Column (b): L2 norm measuring the discrepancy $\delta(\bar{u}_z) = \left \vert u_z^\textrm{PIV} - \bar{u}_z \right \vert$ relative to the reference PIV data. Line (i) Reference experimental data (PIV); line (ii) DDES results; line (iii) RANS results; line (iv) DA state estimation; line (v) RANS + MoE ML model.}
    \label{fig:moe_ann_results_high_plan}
\end{figure}

Finally, \autoref{fig:moe_ann_results_high_plan} shows the analysis of data sampled on \textbf{P3}. It is reminded that this plane is located at $79\,\%$ of the rotor blade height, and in this region three-dimensional effects due to tip vortices are present. On this plane, the average L2 discrepancy norm for the RANS baseline model was $5.55\,\%$ while the MFEnKF analysis reduced the L2 error norm to $3.81\,\%$. The combination of RANS with the CR3 model failed to provide accurate predictions, with an L2 error norm rising to $7.66\,\%$, thus degrading the  results with respect to the RANS baseline prediction. The RANS + MoE model provides a significant improvement in accuracy in this region, with an average L2 discrepancy norm of $3.84\,\%$ which is in close agreement with the MFEnKF result. This can be particularly appreciated by analyzing the topology of the L2 error distribution in \autoref{fig:moe_ann_results_high_plan} (b,v). This result suggests that the specialization of each expert allows the MoE model to correctly identify the local physical phenomena acting in this plane, rather than attempting to understand the full complexity of the flow with a single-network model.



%% file: sections/conclusion.tex
\section{Conclusions}
\label{sec::conclusions}

This work presents the development and assessment of a hybrid methodology coupling numerical simulation, data assimilation (DA), and machine learning (ML) for the analysis of the flow within the CME2 compressor. First, the DA algorithm used for the present analysis, the Multi-Fidelity Ensemble Kalman Filter (MFEnKF), was implemented combining PIV data and model realizations from RANS and DDES runs. These two modeling techniques used different representation strategies for the movement of the rotor. Hyperlocalization of the state estimation was performed in three regions in the rotor region of the compressor surrounding the planes where experimental data was available. 
The MFEnKF was able to significantly reduce the discrepancy of the base model with the available observations while ensuring an accurate  prediction of the governing physical features of the flow. In addition, the MFEnKF was able to expand the size of the initial PIV dataset, therefore producing a larger number of high fidelity samples which can be used for machine learning training purposes. 
ML correction tools were then developed using Neural Networks in the form of source terms to be integrated within the RANS momentum equation. 
More precisely, the developed architecture retrieves various physical data, such as the velocity, the velocity gradients and  turbulent viscosity, to predict a volume forcing with the objective to reproduce the DA solution. The assessment of this methodology shows a global satisfactory accuracy across most of the application domain. More precisely, results using the ML models CR1, and CR2, which are trained using data in the proximity of planes \textbf{P1} and \textbf{P2}, yield average errors close to the ones observed for the DA prediction when used on \textbf{P1} and \textbf{P2} respectively. Nevertheless, this study also highlighted the robustness limits inherent to this ML approach. In regions with high aerodynamic complexity, the lack of complete physical information prevents the generalization of the ANN model, which then can degrade the prediction of the base RANS model. This has been observed for example for the model CR3, which was trained using data in the proximity of the blade tip. 

These findings prompted further development to consolidate the hybrid DA/ML approach. For this reason, a mixture of experts (MoE) approach has been developed to combine the performance of the three models CR1, CR2 and CR3 in a single model. More precisely, the MoE recognizes which blend of the base ML models should be used in each grid element of the domain, trying to improve the global accuracy of the prediction. The results of such method for the flow prediction in planes \textbf{P1}, \textbf{P2} and \textbf{P3} shows that the model globally produces results similar to the DA state estimation, therefore opening perspectives of applications in several fields where sparse data may be used for training of ML models.  

%% file: sections/appendix.tex
\appendix

\section{Spalart-Allmaras modeling configuration} \label{sec:appendix_spalart_allmaras_config}

Here is the complete set of equations used for the Spalart-Allmaras model as it is coded in the numerical solver of OpenFOAM:

\begin{align}
    \begin{split}
        \tilde{S} \equiv S+{\frac  {{\tilde  {\nu }}}{\kappa ^{2}d^{2}}}f_{{v2}},&\qquad f_{{v2}}=1-{\frac  {\chi }{1+\chi f_{{v1}}}} \\
        S={\sqrt  {2 W _{{ij}} W _{{ij}}}}, & \qquad W _{ij}={\frac{1}{2}} \left( \pd{u_i}{x_j} - \pd{u_j}{x_i} \right) \\
        f_{{t2}}=C_{{t3}}\exp \left(-C_{{t4}}\chi ^{2}\right) , & \qquad f_{w}=g\left[{\frac  {1+C_{{w3}}^{6}}{g^{6}+C_{{w3}}^{6}}}\right]^{{1/6}} \\
        g=r+C_{{w2}}(r^{6}-r),&\qquad r\equiv {\frac  {{\tilde  {\nu }}}{{\tilde{S}}\kappa ^{2}d^{2}}} \label{eq:SA_equation_parameters}
    \end{split}
\end{align}
\begin{table}[h!]
\centering
\begin{tabular}{cccccccccccc} \toprule[1.5pt]
$\sigma$ & $C_{{b1}}$ & $C_{{b2}}$ & $\kappa$ & $C_{{w1}}$ & $C_{{w2}}$ & $C_{{w3}}$ & $C_{{v1}}$ & $C_{{t1}}$ & $C_{{t2}}$ & $ C_{{t3}}$ & $C_{{t4}}$ \\
2/3 & 0.1355 & 0.622 & 0.41 & 3.2391 & 0.3 & 2 & 7.1 & 1 & 2 & 1.1 & 2 \\
\bottomrule[1.5pt]
\end{tabular}
\caption{Values chosen for the model constants for the turbulence closure used in the present analysis. These values are the ones provided as default in the implementation of the code.}
\label{tab:SA_constants}
\end{table}

\section{Ensemble Kalman Filter} \label{sec:appendix_EnKF}

The stochastic Ensemble Kalman Filter (EnKF) builds upon the fundamental principle of the classic Kalman Filter (KF) \citep{kalman_new_1960}, but tailored for high-dimensional and non-linear systems. Initially proposed by \cite{evensen_sequential_1994} for state estimation and later generalized for parameter optimization \cite{Evensen2009_ieee}, this methodology utilizes a population of $n_e$ distinct realizations commonly referred to as \textit{ensemble members}. The primary features of the EnKF algorithm include a \textit{forecast} and an \textit{analysis}, and are now detailed:
\begin{enumerate}
    \item \textbf{Forecast: } Every state vector $i \in [1,n_e]$ is advanced from time $k-1$ to $k$, using the model $\mathcal{M}$:
    \begin{equation}
        \state{i,k}{f} = \model{k-1:k}{}{\state{i,k-1}{a}, \bm{\varpi}_{i,k-1}^a }
    \end{equation}
    in which $f$ and $a$ superscripts refer to \textit{forecast} and \textit{analysis} states respectively. 
    
    In computational fluid dynamics, the model usually represents a discretized formulation of the governing equations, namely the compressible Navier-Stokes equations in that case. The numerical model $\model{}{}{}$ advances the state vector which encompasses the physical quantities of interest, such as velocity, pressure or any other relevant quantities defined at time $k-1$. Simultaneously, the time propagation can include parameter vectors $\bm{\varpi}_{i,k-1}$ which characterize underlying physical features and laws, or boundary conditions. The EnKF algorithm is then solved using the so called \textit{extended state} \cite{Asch2016_SIAM}. The ensemble representation gives an estimate of the model's uncertainty. Indeed, the ensemble average is assumed to be an approximation of the state ground truth $\state{k-1}{\textrm{truth}}$, which is used to define the error covariance matrix $\boldsymbol{P}_{k-1}$:
    \begin{equation}
        \boldsymbol{P}_{k-1}= \mathbb{E} \left(\left( \state{k-1}{} -\state{k-1}{\textrm{truth}} \right) \left(\state{k-1}{}- \state{k-1}{\textrm{truth}}  \right)^\top\right)
    \end{equation}
    \item \textbf{Observation data construction: } A number of $n_o$ observation data available at time $k-1$ is set in a vector $\obsvec{k-1}{} \in \mathbb{R}^{n_o}$. This observation vector is then arranged into a matrix $\obsvec{i,k-1}{} \in \mathbb{R}^{n_o, n_e}$ such that:
    \begin{equation}
        \obsvec{i,k-1}{} \equiv \obsvec{k-1}{} + \bm{\varepsilon}_i
    \end{equation}

    where $\bm{\varepsilon}_i \sim \mathcal{N}\left( 0, \bm{\varsigma}_k \right)$. $ \bm{\varsigma}_k$ is a diagonal matrix dimensioned as $\mathbb{R}^{n_o, n_o}$ and representing the unbiased and uncorrelated uncertainty of the given observation data. The $\obsop{\cdot}$ operator is introduced as a projection operator from the model space to the observation space. Thus, $\obsop{\state{}{}}$ is used to obtain the interpolation of the model's data onto the position of the observation sensors. \\
    \item \textbf{Ensemble averaging: } $\state{i,k}{f}$ and $\obsop{\state{i,k}{f}}$ are averaged over the $n_e$ ensemble members.
        \begin{equation}
            \enkfmean{k}{\state{}{f}} = \frac{1}{n_e} \sum_{i=1}^{n_e} \state{i,k}{f} \quad ; \quad \enkfmean{k}{\obsop{\state{}{f}}} = \frac{1}{n_e} \sum_{i=1}^{n_e} \obsop{\state{i,k}{f}}
        \end{equation}
        \item \textbf{Anomaly matrices: } Using the hypothesis of statistical independence between the $n_e$ ensemble members, the error covariance matrix is propagated from time $k-1$ to time $k$ via some anomaly matrices $\enkfanomaly{k}{\state{}{f}}$, through the expression $\boldsymbol{P}_k^f \approx \enkfanomaly{k}{\state{}{f}}\left(\enkfanomaly{k}{\state{}{f}}\right)^\top$, for which each column describes the deviation of the $i^{\textrm{th}}$ ensemble member from the ensemble mean. Analogously, the projection of the model onto the observation space can be expressed by means of another anomaly matrix $\enkfanomaly{k}{\obsop{\state{}{f}}}$. Therefore:
    \begin{equation}
        \enkfanomaly{k}{\left(\bm{\cdot} \right)} = \frac{1}{\sqrt{n_e-1}} \left((\bm{\cdot})_{1,k} - \enkfmean{k}{\left( \bm{\cdot} \right)}, \dots, (\bm{\cdot})_{n_e,k} - \enkfmean{k}{\left( \bm{\cdot} \right)} \right)
    \end{equation}
    where $(\bm{\cdot})$ may be replaced with $\state{}{f}$ and $\obsop{\state{}{f}}$ to respectively obtain $\enkfanomaly{k}{\state{}{f}}$ and $\enkfanomaly{k}{\obsop{\state{}{f}}}$. \\
    \item \textbf{Covariance matrices: } Covariance matrices $\enkfcovariance{}{\left(\bm{\cdot}\right)}{\left(\bm{\cdot}\right)}$ are calculated from the previously computed anomaly matrices.
    \begin{align}
        \enkfcovariance{k}{\state{}{f}}{\obsop{\state{}{f}}} &= \enkfanomaly{k}{\state{}{f}} \left( \enkfanomaly{k}{\obsop{\state{}{f}}} \right)^\top \\
        \enkfcovariance{k}{\obsop{\state{}{f}}}{\obsop{\state{}{f}}} &= \enkfanomaly{k}{\obsop{\state{}{f}}} \left( \enkfanomaly{k}{\obsop{\state{}{f}}} \right)^\top
    \end{align}

    \item \textbf{Kalman Gain matrix: } The Kalman Gain matrix $\boldsymbol{K}$ determines the optimal weighting when combining the model prediction and the observations, based on their respective error statistics, to minimize the posterior estimation error.
    \begin{equation}
        \kalmangain_k = \enkfcovariance{k}{\state{}{f}}{\obsop{\state{}{f}}}\left( \enkfcovariance{k}{\obsop{\state{}{f}}}{\obsop{\state{}{f}}} + \bm{\varsigma}_k \right)^{-1}
    \end{equation}

    \item \textbf{Analysis: } This step computes the updated state vector $\boldsymbol{x}^a_{i,k}$ and model parameters $\bm{\varpi}^a_{i,k}$ by correcting the forecast state vector $\boldsymbol{x}_{i,k}^f$ and the previous model parameters $\bm{\varpi}_{i,k-1}^f$ using the difference between the observation vector $\boldsymbol{y}_k$ and the sampled data $\mathcal{H} \left( \boldsymbol{x}^f_{i,k} \right)$. This difference, known as the \textit{innovation term}, is weighted by the Kalman Gain matrix $\kalmangain$. 
    \begin{align}
        \state{i,k}{a} &= \state{i,k}{f} + \boldsymbol{K}_k \left( \obsvec{k}{} - \obsop{\state{i,k}{f}}\right) \label{eq:state_correction} \\
        \bm{\varpi}_{i,k}^a &= \bm{\varpi}_{i,k}^f + \boldsymbol{K}_k \left( \obsvec{k}{} - \obsop{\state{i,k}{f}}\right) \label{eq:param_correction}
    \end{align}
\end{enumerate}

Both corrections given in \autoref{eq:state_correction} and \autoref{eq:param_correction} are proportional to the discrepancy between the observations $\obsvec{}{}$ and the model equivalent values $\obsop{\state{}{}}$ weighted by the Kalman gain $\kalmangain$.

The performance of the EnKF heavily depends on how well the model and observation uncertainties are characterized. The update estimation is the most effective when the uncertainty regions of both sources of information show significant intersection. However, if the observations are treated with low confidence, the analysis will remain biased toward the forecast state of the model, resulting in a limited increase in accuracy. Despite its great accuracy, the EnKF relies on a relatively high number of numerical simulations, which can rapidly become computationally expensive when dealing with high dimensional systems. Constraining the ensemble size to mitigate this cost often introduces sampling errors and a systematic underestimation of state variance, which can ultimately lead to filter divergence and a failure to converge. \\

\section{Multi-Fidelity Ensemble Kalman Filter} \label{sec:appendix_mfenkf}

First introduced by~\citep{popov_multifidelity_2021}, the Multi-Fidelity Ensemble Kalman Filter is a dedicated Data Assimilation procedure combining full order models, referred to as FOM, along which an ensemble of reduced order stand-in models (ROM) are utilized to increase the computational efficiency while increasing physical accuracy of Data Assimilation. This multi-level method leverages the approach of the linear control variate, which is a variance reduction technique often used in Monte Carlo methodologies \citep{rubinstein_efficiency_1985}. 

We consider a multi-fidelity framework comprising a high-fidelity model, denoted by $\princ$, and a computationally efficient low-fidelity surrogate, represented by $\cont$. $\cont$ is a control variate of $\princ$ if $\cont$ shows a strong correlation with the $\princ$ variate and its expectation $\expectcont$ is \textit{a priori} known.

The main objective of the control variate methodology is to build estimations of statistics for the high fidelity state, thanks to information coming from the lower-fidelity states, which becomes \textit{de facto} computationally less expensive. This goal is reached with the usage of the total variate, expressed with $\tot$ and defined with a linear combination involving $\expectprinc$, $\expectcont$, respectively the expectations of $\princ$ and $\cont$:
\begin{equation}
    \tot = \princ - \lambda \left( \cont - \expectcont \right) \label{eq:total_variate_1}
\end{equation}

The random variable $\princ$ is called the \textit{principal variate}, $\cont$ the \textit{control variate} and $\tot$ the \textit{total variate}. In \autoref{eq:total_variate_1}, variable $\tot$ has a lower variance compared to the principal variable $\princ$, while their expectations exhibit $\expecttot = \expectprinc$. When working in a practical framework, the $\expectcont$ is not known and the MFEnKF framework needs a new variable. Therefore, the variable $\anc$, referred to as the \textit{ancillary variate}, is introduced and selected to obtain $\expectcont = \expectanc$. The ancillary variate is also computed on an ensemble framework and runs on the same low-resolution model as the control variate $\cont$. Thanks to the introduction of the ancillary variate $\anc$, \autoref{eq:total_variate_1} can now be written as:
\begin{equation}
    \tot = \princ - \lambda \left( \cont - \anc \right) \label{eq:total_variate_2}
\end{equation}

\autoref{eq:total_variate_2} is now transposed to a multi-level approach. Two models are now defined. On the one hand, the CFD model $\finemodel$ is meant to be the \textit{fine} model, leveraging a fine grid and accurate turbulence modeling. On the other hand, the CFD model $\coarsemodel$ uses a coarser grid and lower-accuracy turbulence modeling. Analogously to the control variates method presented earlier, three sub-ensembles are built to fit the ensemblist methodology of the MFEnKF. The first sub-ensemble corresponds to the principal variate $\princ$, composed of $N_{\princ}$ members and run using the fine model $\finemodel$. In a similar manner, the control variate $\cont$ is represented by $N_{\cont} = N_{\princ}$ ensemble members run with the coarse model $\coarsemodel$. Finally, the ancillary variate $\anc$ is represented with $N_{\anc}$ ensemble members, also run on the coarse model $\coarsemodel$. The total number of ensemble members $N_e$ simultaneously run is equal to:
\begin{equation}
    N_e = 2N_{\princ} + N_{\anc}
\end{equation}

One can then notice that $N_{\cont} + N_{\anc}$ ensemble members are run with the coarse model $\coarsemodel$, while $N_{\princ}$ are run using the fine model $\finemodel$. Thus, the control variate theorem implies that a state estimation can be realized with a limited number of principal ensemble members $N_{\princ}$, while using a much larger amount of coarse-model simulations $N_{\cont} + N_{\anc}$. Steps of the algorithm are now detailed:
\begin{enumerate}
    \item \textbf{Time Forecast: } Members from the three sub-ensembles are advanced in time thanks to their corresponding non-linear models $\mathcal{M}$:
    \begin{equation}
        \begin{aligned}
            \principalstate{i,k}{f} = \princmodel \left( \principalstate{i,k-1}{a} \right) \\
            \controlstate{i,k}{f} = \contmodel \left( \controlstate{i,k-1}{a} \right) \\
            \ancillarystate{i,k}{f} = \ancmodel \left( \ancillarystate{i,k-1}{a} \right)
        \end{aligned}
    \end{equation}

    \item \textbf{Projection from principal space $\princ$ to control space $\cont$: } As $\princmodel \equiv \finemodel$ and $\contmodel \equiv \coarsemodel$ are defined with different fidelities, physical quantities obtained on the principal and the control members become decorrelated along time advancement. Then, physical fields from the principal ensemble members are projected to the control ensemble members with the projection operator $\projMinus$, which produces a mapping of the fine model to the coarse model:
    \begin{equation}
        \controlstate{i,k}{f} = \projMinus \left( \principalstate{i,k}{f} \right)
    \end{equation}
    \item \textbf{Observation operator for the total variate: } The observation operator $\obsop{\bm{\cdot}}$ may be used for the principal, as well as for the control and ancillary members using the projection operator $\projPlus$, contrary to the total variate ensemble members which are not directly observable. Nevertheless, a twin operator defined as $\widetilde{\mathcal{H}\left({\bm{\cdot}}\right)}$ is introduced and equal to:
    \begin{equation}
        \totalobsop = \obsop{\principalstate{i,k}{f}} - \lambda \left( \obsop{\projPlus \left( \controlstate{i,k}{f} \right)} \right) + \lambda \left( \obsop{\projPlus \left( \ancillarystate{i,k}{f} \right)} \right)
    \end{equation}
    where $\lambda$ is here equal to $1/2$.
    \item \textbf{Computation of the covariance matrices: } The approximated covariance matrices are built using the anomaly matrices from the three sub-ensembles. Assuming the principal and the ancillary are defined as random and independent, we have:
    \begin{equation}
        \begin{aligned}
            \mfenkfcov{\totalstate{i,k}{f}}{\widetilde{\mathcal{H}\left(\totalstate{i,k}{f} \right)}} &= \mfenkfcov{\principalstate{i,k}{f}}{\obsop{\principalstate{i,k}{f}}} \\ 
            &+ \frac{1}{4} \left( \projPlus \left( \mfenkfcov{\controlstate{i,k}{f}}{\obsop{\projPlus\left( \controlstate{i,k}{f} \right)}} \right)  + \projPlus \left( \mfenkfcov{\ancillarystate{i,k}{f}}{\obsop{\projPlus\left( \ancillarystate{i,k}{f} \right)}} \right) \right)\\
            &- \frac{1}{2} \left( \mfenkfcov{\principalstate{i,k}{f}}{\obsop{\projPlus\left( \controlstate{i,k}{f} \right)}} + \projPlus\left( \mfenkfcov{\controlstate{i,k}{f}}{\obsop{\principalstate{i,k}{f}}} \right)\right)
        \end{aligned}
    \end{equation}
    \begin{equation}
        \begin{aligned}
            \mfenkfcov{\totalobsop}{\totalobsop} &= \mfenkfcov{\obsop{\principalstate{i,k}{f}}}{\obsop{\principalstate{i,k}{f}}} \\
            &+\frac{1}{4} \left( \mfenkfcov{\obsop{\projPlus\left( \controlstate{i,k}{f} \right)}}{\obsop{\projPlus\left(  \controlstate{i,k}{f}\right)}} + \mfenkfcov{\obsop{\projPlus\left( \ancillarystate{i,k}{f} \right)}}{\obsop{\projPlus\left( \ancillarystate{i,k}{f} \right)}} \right) \\
            &- \frac{1}{2} \left( \mfenkfcov{\obsop{\principalstate{i,k}{f}}}{\obsop{\projPlus\left( \controlstate{i,k}{f} \right)}} + \mfenkfcov{\obsop{\projPlus\left( \controlstate{i,k}{f} \right)}}{\obsop{\principalstate{i,k}{f}}} \right)
        \end{aligned}
    \end{equation}
    The observation error covariance matrices are constructed using independent and random variables such that:
    \begin{align}
        \bm{\vartheta}^{\princ} &\sim \mathcal{N} \left(0, \bm{\varsigma}^{\princ} \right) \\
        \bm{\vartheta}^{\cont}& = \bm{\vartheta}^{\princ} \\
        \bm{\vartheta}^{\anc} &\sim \mathcal{N}\left(0,\bm{\varsigma}^{\anc}  \right)
    \end{align}
    which leads to:
    \begin{equation}
        \mfenkfcov{\bm{\vartheta}^{\princ}}{\bm{\vartheta}^{\princ}} = 
        \mfenkfcov{\bm{\vartheta}^{\cont}}{\bm{\vartheta}^{\cont}} = 
        \mfenkfcov{\bm{\vartheta}^{\anc}}{\bm{\vartheta}^{\anc}} = 
        2\mfenkfcov{\bm{\vartheta}^{\tot}}{\bm{\vartheta}^{\tot}}
    \end{equation}
    and the variables $\bm{\vartheta}^{\princ}$, $\bm{\vartheta}^{\cont}$ and $\bm{\vartheta}^{\anc}$ are used to obtain the noised observation data $\obsvec{}{\princ}$, $\obsvec{}{\cont}$ and $\obsvec{}{\anc}$ respectively.

    \item \textbf{Kalman Gain computation: } The Kalman Gain matrix is computed to estimate the updated state:
    \begin{equation}
        \widetilde{\kalmangain} = \mfenkfcov{\totalstate{i,k}{f}}{\widetilde{\mathcal{H}\left(\totalstate{i,k}{f} \right)}} \left( \mfenkfcov{\totalobsop}{\totalobsop} + \mfenkfcov{\bm{\vartheta}^{\tot}}{\bm{\vartheta}^{\tot}} \right)^{-1}
    \end{equation}

    \item \textbf{Ensemble updates: } Principal, control and ancillary variables are updated:
    \begin{align}
        \principalstate{i,k}{a} &= \principalstate{i,k}{f} + \widetilde{\kalmangain} \left( \obsvec{i,k}{\princ} - \obsop{\principalstate{i,k}{f}} \right) \\
        \controlstate{i,k}{a} &= \controlstate{i,k}{f} + \projMinus \left(\widetilde{\kalmangain} \left( \obsvec{i,k}{\cont} - \obsop{\controlstate{i,k}{f}} \right) \right) \\
        \ancillarystate{i,k}{a} &= \ancillarystate{i,k}{f} + \projMinus\left( \widetilde{\kalmangain} \left( \obsvec{i,k}{\anc} - \obsop{\ancillarystate{i,k}{f}} \right) \right)
    \end{align}
\end{enumerate}

\begin{figure}[h!]
    \centering
    \includegraphics[width=0.75\textwidth]{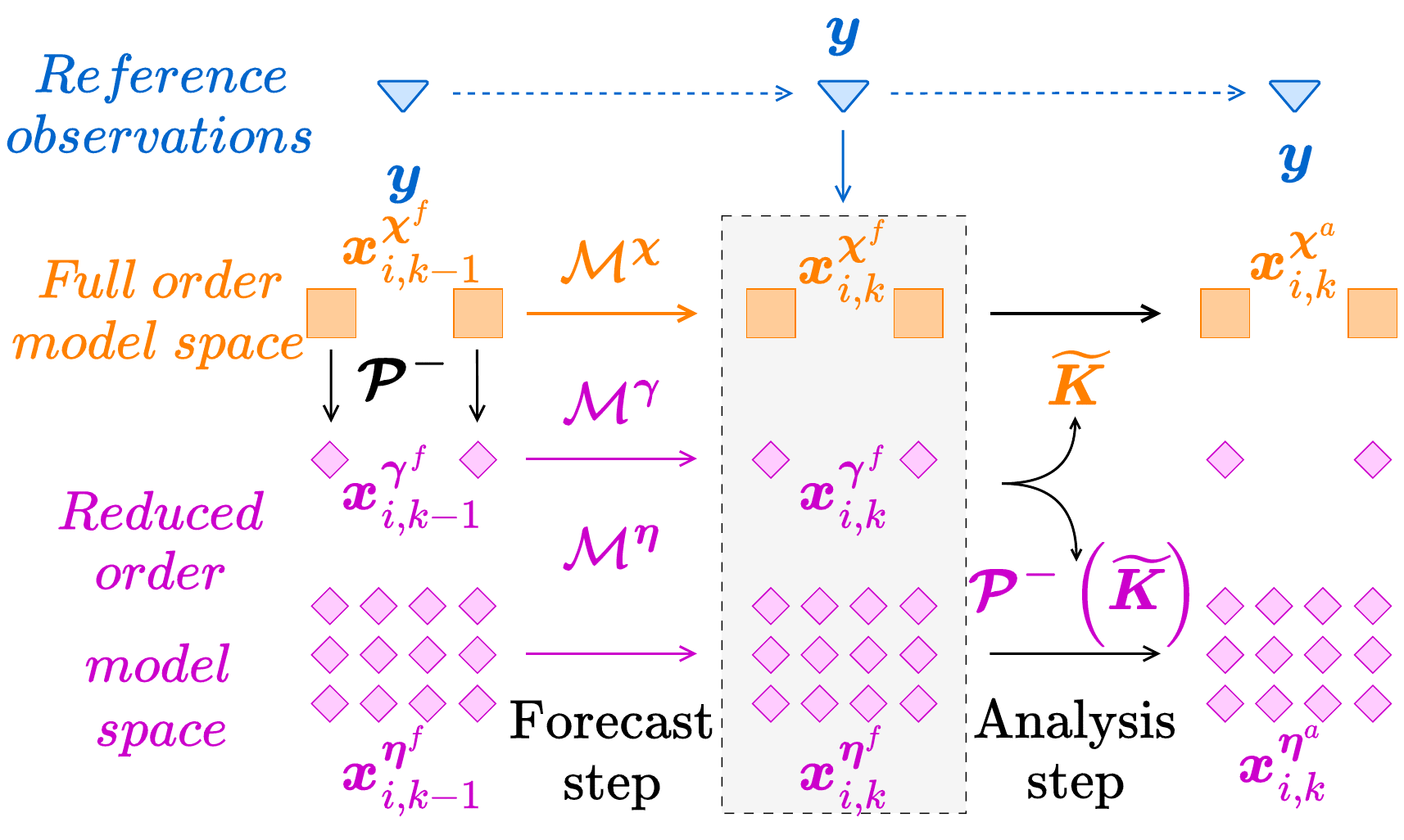}
    \caption{Schematic representation of the MFEnKF algorithm}
    \label{fig:mfenkf_scheme}
\end{figure}

%% file: main.bbl

\begin{thebibliography}{82}
\ifx \bisbn   \undefined \def \bisbn  #1{ISBN #1}\fi
\ifx \binits  \undefined \def \binits#1{#1}\fi
\ifx \bauthor  \undefined \def \bauthor#1{#1}\fi
\ifx \batitle  \undefined \def \batitle#1{#1}\fi
\ifx \bjtitle  \undefined \def \bjtitle#1{#1}\fi
\ifx \bvolume  \undefined \def \bvolume#1{\textbf{#1}}\fi
\ifx \byear  \undefined \def \byear#1{#1}\fi
\ifx \bissue  \undefined \def \bissue#1{#1}\fi
\ifx \bfpage  \undefined \def \bfpage#1{#1}\fi
\ifx \blpage  \undefined \def \blpage #1{#1}\fi
\ifx \burl  \undefined \def \burl#1{\textsf{#1}}\fi
\ifx \doiurl  \undefined \def \doiurl#1{\url{https://doi.org/#1}}\fi
\ifx \betal  \undefined \def \betal{\textit{et al.}}\fi
\ifx \binstitute  \undefined \def \binstitute#1{#1}\fi
\ifx \binstitutionaled  \undefined \def \binstitutionaled#1{#1}\fi
\ifx \bctitle  \undefined \def \bctitle#1{#1}\fi
\ifx \beditor  \undefined \def \beditor#1{#1}\fi
\ifx \bpublisher  \undefined \def \bpublisher#1{#1}\fi
\ifx \bbtitle  \undefined \def \bbtitle#1{#1}\fi
\ifx \bedition  \undefined \def \bedition#1{#1}\fi
\ifx \bseriesno  \undefined \def \bseriesno#1{#1}\fi
\ifx \blocation  \undefined \def \blocation#1{#1}\fi
\ifx \bsertitle  \undefined \def \bsertitle#1{#1}\fi
\ifx \bsnm \undefined \def \bsnm#1{#1}\fi
\ifx \bsuffix \undefined \def \bsuffix#1{#1}\fi
\ifx \bparticle \undefined \def \bparticle#1{#1}\fi
\ifx \barticle \undefined \def \barticle#1{#1}\fi
\bibcommenthead
\ifx \bconfdate \undefined \def \bconfdate #1{#1}\fi
\ifx \botherref \undefined \def \botherref #1{#1}\fi
\ifx \url \undefined \def \url#1{\textsf{#1}}\fi
\ifx \bchapter \undefined \def \bchapter#1{#1}\fi
\ifx \bbook \undefined \def \bbook#1{#1}\fi
\ifx \bcomment \undefined \def \bcomment#1{#1}\fi
\ifx \oauthor \undefined \def \oauthor#1{#1}\fi
\ifx \citeauthoryear \undefined \def \citeauthoryear#1{#1}\fi
\ifx \endbibitem  \undefined \def \endbibitem {}\fi
\ifx \bconflocation  \undefined \def \bconflocation#1{#1}\fi
\ifx \arxivurl  \undefined \def \arxivurl#1{\textsf{#1}}\fi
\csname PreBibitemsHook\endcsname

\bibitem[\protect\citeauthoryear{Day}{2016}]{Day2015}
\begin{barticle}
\bauthor{\bsnm{Day}, \binits{I.J.}}:
\batitle{Stall, surge, and 75 years of research}.
\bjtitle{Journal of Turbomachinery}
\bvolume{138}(\bissue{1}),
\bfpage{011001}
(\byear{2016})
\doiurl{10.1115/1.4031473}
\end{barticle}
\endbibitem

\bibitem[\protect\citeauthoryear{Lord and Suciu}{2016}]{Nasa_Short_Core}
\begin{barticle}
\bauthor{\bsnm{Lord}, \binits{W.K.}},
\bauthor{\bsnm{Suciu}, \binits{G.L.}}:
\batitle{Engine architecture for high efficiency at small core size}.
\bjtitle{Mechanical Engineering}
\bvolume{138}(\bissue{03}),
\bfpage{52}--\blpage{53}
(\byear{2016})
\doiurl{10.1115/1.2016-Mar-4} .
\bcomment{Pratt \& Whitney}
\end{barticle}
\endbibitem

\bibitem[\protect\citeauthoryear{Liu et~al.}{2025}]{Liu2025}
\begin{barticle}
\bauthor{\bsnm{Liu}, \binits{E.}},
\bauthor{\bsnm{Zhou}, \binits{Z.}},
\bauthor{\bsnm{Zheng}, \binits{H.}}:
\batitle{Control of high-load compressor stator corner flow by blade slotting}.
\bjtitle{International Journal of Aerospace Engineering}
\bvolume{2025}(\bissue{1}),
\bfpage{3743172}
(\byear{2025})
\doiurl{10.1155/ijae/3743172}
\end{barticle}
\endbibitem

\bibitem[\protect\citeauthoryear{Kumar et~al.}{2024}]{Tandem2023}
\begin{barticle}
\bauthor{\bsnm{Kumar}, \binits{A.}},
\bauthor{\bsnm{Kumar}, \binits{A.}},
\bauthor{\bsnm{Chhugani}, \binits{H.}},
\bauthor{\bsnm{More}, \binits{S.}},
\bauthor{\bsnm{Pradeep}, \binits{A.M.}}:
\batitle{Understanding the effect of three-dimensional design in tandem blade}.
\bjtitle{Journal of Turbomachinery}
\bvolume{146}(\bissue{3}),
\bfpage{031002}
(\byear{2024})
\doiurl{10.1115/1.4063924}
\end{barticle}
\endbibitem

\bibitem[\protect\citeauthoryear{Li et~al.}{2019}]{LiReview}
\begin{barticle}
\bauthor{\bsnm{Li}, \binits{J.}},
\bauthor{\bsnm{Du}, \binits{J.}},
\bauthor{\bsnm{Nie}, \binits{C.}},
\bauthor{\bsnm{Zhang}, \binits{H.}}:
\batitle{Review of tip air injection to improve stall margin in axial
  compressors}.
\bjtitle{Progress in Aerospace Sciences}
\bvolume{106},
\bfpage{15}--\blpage{31}
(\byear{2019})
\doiurl{10.1016/j.paerosci.2019.01.005}
\end{barticle}
\endbibitem

\bibitem[\protect\citeauthoryear{Rao and Mahulikar}{2002}]{Rao2002}
\begin{barticle}
\bauthor{\bsnm{Rao}, \binits{G.A.}},
\bauthor{\bsnm{Mahulikar}, \binits{S.P.}}:
\batitle{Integrated review of stealth technology and its role in airpower}.
\bjtitle{The Aeronautical Journal}
\bvolume{106}(\bissue{1066}),
\bfpage{629}--\blpage{642}
(\byear{2002})
\doiurl{10.1017/S0001924000011702}
\end{barticle}
\endbibitem

\bibitem[\protect\citeauthoryear{Yildirim et~al.}{2022}]{Yildirim2022}
\begin{barticle}
\bauthor{\bsnm{Yildirim}, \binits{A.}},
\bauthor{\bsnm{Gray}, \binits{J.S.}},
\bauthor{\bsnm{Mader}, \binits{C.A.}},
\bauthor{\bsnm{Martins}, \binits{J.R.R.A.}}:
\batitle{Boundary-layer ingestion benefit for the starc-abl concept}.
\bjtitle{Journal of Aircraft}
\bvolume{59}(\bissue{4}),
\bfpage{896}--\blpage{911}
(\byear{2022})
\doiurl{10.2514/1.C036103}
\end{barticle}
\endbibitem

\bibitem[\protect\citeauthoryear{Cousins}{2004}]{Cousins2004}
\begin{bchapter}
\bauthor{\bsnm{Cousins}, \binits{W.T.}}:
\bctitle{History, philosophy, physics, and future directions of aircraft
  propulsion system/inlet integration}.
\bsertitle{Turbo Expo},
vol. \bseriesno{Volume 2: Turbo Expo 2004},
pp. \bfpage{305}--\blpage{320}
(\byear{2004}).
\doiurl{10.1115/GT2004-54210}
\end{bchapter}
\endbibitem

\bibitem[\protect\citeauthoryear{Camp and Day}{1998}]{Camp1997}
\begin{barticle}
\bauthor{\bsnm{Camp}, \binits{T.R.}},
\bauthor{\bsnm{Day}, \binits{I.J.}}:
\batitle{A study of spike and modal stall phenomena in a {Low-Speed} axial
  compressor}.
\bjtitle{Journal of Turbomachinery}
\bvolume{120}(\bissue{3}),
\bfpage{393}--\blpage{401}
(\byear{1998})
\doiurl{10.1115/1.2841730}
\end{barticle}
\endbibitem

\bibitem[\protect\citeauthoryear{Wilson et~al.}{2025}]{Wilson2025}
\begin{botherref}
\oauthor{\bsnm{Wilson}, \binits{K.N.}},
\oauthor{\bsnm{Jaman}, \binits{G.G.}},
\oauthor{\bsnm{Thapa}, \binits{A.}},
\oauthor{\bsnm{Vivekananda}, \binits{A.}},
\oauthor{\bsnm{Lowe}, \binits{M.}},
\oauthor{\bsnm{Grima}, \binits{Z.}},
\oauthor{\bsnm{Schoen}, \binits{M.P.}}:
A review of stall detection in subsonic axial compressors.
Machines
\textbf{13}(1)
(2025)
\doiurl{10.3390/machines13010013}
\end{botherref}
\endbibitem

\bibitem[\protect\citeauthoryear{Pope}{2000}]{Pope2000_cambridge}
\begin{bbook}
\bauthor{\bsnm{Pope}, \binits{S.B.}}:
\bbtitle{{Turbulent Flows}}.
\bpublisher{Cambridge University Press},
\blocation{Cambridge}
(\byear{2000}).
\doiurl{10.1017/CBO9780511840531}
\end{bbook}
\endbibitem

\bibitem[\protect\citeauthoryear{Sagaut}{2006}]{Sagaut2006_springer}
\begin{bbook}
\bauthor{\bsnm{Sagaut}, \binits{P.}}:
\bbtitle{{Large Eddy Simulation for Incompressible Flows: An Introduction}},
\bedition{3rd} edn.
\bsertitle{Scientific Computation}.
\bpublisher{Springer},
\blocation{Berlin, Heidelberg}
(\byear{2006}).
\doiurl{10.1007/b137536}
\end{bbook}
\endbibitem

\bibitem[\protect\citeauthoryear{Lumley}{1970}]{Lumley1970_AP}
\begin{bbook}
\bauthor{\bsnm{Lumley}, \binits{J.L.}}:
\bbtitle{{Stochastic Tools in Turbulence}}.
\bsertitle{Applied Mathematics and Mechanics},
vol. \bseriesno{12}.
\bpublisher{Academic Press},
\blocation{New York}
(\byear{1970})
\end{bbook}
\endbibitem

\bibitem[\protect\citeauthoryear{Wilcox}{2006}]{Wilcox2006_DCW}
\begin{bbook}
\bauthor{\bsnm{Wilcox}, \binits{D.C.}}:
\bbtitle{{Turbulence Modeling for CFD}},
\bedition{3rd} edn.
\bpublisher{DCW Industries},
\blocation{La Ca{\~n}ada, CA}
(\byear{2006})
\end{bbook}
\endbibitem

\bibitem[\protect\citeauthoryear{Duraisamy et~al.}{2019}]{Duraisamy2019_arfm}
\begin{barticle}
\bauthor{\bsnm{Duraisamy}, \binits{K.}},
\bauthor{\bsnm{Xiao}, \binits{H.}},
\bauthor{\bsnm{Iaccarino}, \binits{G.}}:
\batitle{Turbulence modeling in the age of data}.
\bjtitle{Annual Review of Fluid Mechanics}
\bvolume{51},
\bfpage{1}--\blpage{23}
(\byear{2019})
\end{barticle}
\endbibitem

\bibitem[\protect\citeauthoryear{Schmelzer et~al.}{2020}]{Schmelzer2020_ftc}
\begin{barticle}
\bauthor{\bsnm{Schmelzer}, \binits{M.}},
\bauthor{\bsnm{Dwight}, \binits{R.}},
\bauthor{\bsnm{Cinnella}, \binits{P.}}:
\batitle{{Discovery of algebraic Reynolds-stress models using sparse symbolic
  regression}}.
\bjtitle{Flow, Turbulence and Combustion}
\bvolume{104},
\bfpage{579}--\blpage{603}
(\byear{2020})
\end{barticle}
\endbibitem

\bibitem[\protect\citeauthoryear{Volpiani et~al.}{2021}]{Volpiani2021_prf}
\begin{barticle}
\bauthor{\bsnm{Volpiani}, \binits{P.S.}},
\bauthor{\bsnm{Meyer}, \binits{M.}},
\bauthor{\bsnm{Franceschini}, \binits{L.}},
\bauthor{\bsnm{Dandois}, \binits{J.}},
\bauthor{\bsnm{Renac}, \binits{F.}},
\bauthor{\bsnm{Martin}, \binits{E.}},
\bauthor{\bsnm{Marquet}, \binits{O.}},
\bauthor{\bsnm{Sipp}, \binits{D.}}:
\batitle{{Machine learning-augmented turbulence modeling for RANS simulations
  of massively separated flows}}.
\bjtitle{Physical Review Fluids}
\bvolume{6},
\bfpage{064607}
(\byear{2021})
\end{barticle}
\endbibitem

\bibitem[\protect\citeauthoryear{Daley}{1991}]{Daley1991_cambridge}
\begin{bbook}
\bauthor{\bsnm{Daley}, \binits{R.}}:
\bbtitle{{Atmospheric Data Analysis}}.
\bsertitle{Cambridge Atmospheric and Space Science Series}.
\bpublisher{Cambridge University Press},
\blocation{Cambridge}
(\byear{1991})
\end{bbook}
\endbibitem

\bibitem[\protect\citeauthoryear{Asch et~al.}{2016}]{Asch2016_SIAM}
\begin{bbook}
\bauthor{\bsnm{Asch}, \binits{M.}},
\bauthor{\bsnm{Bocquet}, \binits{M.}},
\bauthor{\bsnm{Nodet}, \binits{M.}}:
\bbtitle{{Data Assimilation: Methods, Algorithms, and Applications}}.
\bsertitle{Fundamentals of Algorithms},
vol. \bseriesno{11}.
\bpublisher{SIAM},
\blocation{Philadelphia, PA}
(\byear{2016}).
\doiurl{10.1137/1.9781611974546}
\end{bbook}
\endbibitem

\bibitem[\protect\citeauthoryear{Evensen et~al.}{2022}]{Evensen2022_springer}
\begin{bbook}
\bauthor{\bsnm{Evensen}, \binits{G.}},
\bauthor{\bsnm{Vossepoel}, \binits{F.C.}},
\bauthor{\bsnm{{van Leeuwen}}, \binits{P.J.}}:
\bbtitle{{Data Assimilation Fundamentals}}.
\bpublisher{Springer},
\blocation{Cham}
(\byear{2022}).
\doiurl{10.1007/978-3-030-96709-3}
\end{bbook}
\endbibitem

\bibitem[\protect\citeauthoryear{Simon}{2006}]{Simon2006_wiley}
\begin{bbook}
\bauthor{\bsnm{Simon}, \binits{D.}}:
\bbtitle{{Optimal State Estimation: Kalman, $H_\infty$, and Nonlinear
  Approaches}}.
\bpublisher{Wiley},
\blocation{Hoboken, NJ}
(\byear{2006}).
\doiurl{10.1002/0470045345}
\end{bbook}
\endbibitem

\bibitem[\protect\citeauthoryear{Evensen}{1994}]{evensen_sequential_1994}
\begin{barticle}
\bauthor{\bsnm{Evensen}, \binits{G.}}:
\batitle{Sequential data assimilation with a nonlinear quasi-geostrophic model
  using {Monte} {Carlo} methods to forecast error statistics}.
\bjtitle{Journal of Geophysical Research}
\bvolume{99}(\bissue{C5}),
\bfpage{10143}
(\byear{1994})
\doiurl{10.1029/94JC00572}
\end{barticle}
\endbibitem

\bibitem[\protect\citeauthoryear{Evensen}{2009}]{Evensen2009_ieee}
\begin{barticle}
\bauthor{\bsnm{Evensen}, \binits{G.}}:
\batitle{The ensemble kalman filter for combined state and parameter estimation
  - monte carlo techniques for data assimilation in large systems}.
\bjtitle{IEEE Control Systems}
\bvolume{29},
\bfpage{83}--\blpage{104}
(\byear{2009})
\end{barticle}
\endbibitem

\bibitem[\protect\citeauthoryear{Popov et~al.}{2021}]{popov_multifidelity_2021}
\begin{barticle}
\bauthor{\bsnm{Popov}, \binits{A.A.}},
\bauthor{\bsnm{Mou}, \binits{C.}},
\bauthor{\bsnm{Sandu}, \binits{A.}},
\bauthor{\bsnm{Iliescu}, \binits{T.}}:
\batitle{A {Multifidelity} {Ensemble} {Kalman} {Filter} with {Reduced} {Order}
  {Control} {Variates}}.
\bjtitle{SIAM Journal on Scientific Computing}
\bvolume{43}(\bissue{2}),
\bfpage{1134}--\blpage{1162}
(\byear{2021})
\doiurl{10.1137/20M1349965}
\end{barticle}
\endbibitem

\bibitem[\protect\citeauthoryear{Villiers et~al.}{2025}]{Villiers2025_ftc}
\begin{barticle}
\bauthor{\bsnm{Villiers}, \binits{R.}},
\bauthor{\bsnm{Mons}, \binits{V.}},
\bauthor{\bsnm{Sipp}, \binits{D.}},
\bauthor{\bsnm{Lamballais}, \binits{E.}},
\bauthor{\bsnm{Meldi}, \binits{M.}}:
\batitle{Enhancing unsteady reynolds-averaged navier–stokes modelling from
  sparse data through sequential data assimilation and machine learning}.
\bjtitle{Flow, Turbulence and Combustion}
\bvolume{115},
\bfpage{991}--\blpage{1029}
(\byear{2025})
\end{barticle}
\endbibitem

\bibitem[\protect\citeauthoryear{Valero and Meldi}{2025}]{Valero2025_cf}
\begin{barticle}
\bauthor{\bsnm{Valero}, \binits{M.M.}},
\bauthor{\bsnm{Meldi}, \binits{M.}}:
\batitle{Improved state estimation for turbulent flows combining ensemble data
  assimilation and machine learning}.
\bjtitle{Computers \& Fluids}
\bvolume{300},
\bfpage{106757}
(\byear{2025})
\end{barticle}
\endbibitem

\bibitem[\protect\citeauthoryear{Moussie et~al.}{2026}]{Moussie026_cf}
\begin{barticle}
\bauthor{\bsnm{Moussie}, \binits{T.}},
\bauthor{\bsnm{Errante}, \binits{P.}},
\bauthor{\bsnm{Meldi}, \binits{M.}}:
\batitle{{Multi-level ensemble Kalman Filter algorithms enhanced by
  Convolutional Neural Networks}}.
\bjtitle{Computers and Fluids}
\bvolume{314},
\bfpage{107082}
(\byear{2026})
\end{barticle}
\endbibitem

\bibitem[\protect\citeauthoryear{Moubogha et~al.}{2022}]{MouboghaASME}
\begin{bchapter}
\bauthor{\bsnm{Moubogha}, \binits{J.M.}},
\bauthor{\bsnm{Joseph}, \binits{P.}},
\bauthor{\bsnm{Roussette}, \binits{O.}},
\bauthor{\bsnm{Dazin}, \binits{A.}}:
\bctitle{Experimental flow investigation inside an axial compressor with active
  flow control}.
In: \bbtitle{Turbo Expo: Power for Land, Sea, and Air}
(\byear{2022}).
\doiurl{10.1115/GT2022-82126} .
\bcomment{American Society of Mechanical Engineers}
\end{bchapter}
\endbibitem

\bibitem[\protect\citeauthoryear{Baretter et~al.}{2025}]{BaretterJPP}
\begin{barticle}
\bauthor{\bsnm{Baretter}, \binits{A.}},
\bauthor{\bsnm{Joseph}, \binits{P.}},
\bauthor{\bsnm{Roussette}, \binits{O.}},
\bauthor{\bsnm{Dazin}, \binits{A.}},
\bauthor{\bsnm{Roman\`{o}}, \binits{F.}}:
\batitle{Scaling laws at stall in an axial compressor with an upstream
  perturbation}.
\bjtitle{Journal of Propulsion and Power}
\bvolume{41}(\bissue{1}),
\bfpage{82}--\blpage{98}
(\byear{2025})
\doiurl{10.2514/1.B39407}
\end{barticle}
\endbibitem

\bibitem[\protect\citeauthoryear{Alekseik et~al.}{2025}]{Alekseik2025}
\begin{botherref}
\oauthor{\bsnm{Alekseik}, \binits{O.}},
\oauthor{\bsnm{Joseph}, \binits{P.}},
\oauthor{\bsnm{Roussette}, \binits{O.}},
\oauthor{\bsnm{Dazin}, \binits{A.}}:
An experimental investigation by particle image velocimetry of the active flow
  control of the stall inception of an axial compressor.
International Journal of Turbomachinery, Propulsion and Power
\textbf{10}(4)
(2025)
\doiurl{10.3390/ijtpp10040040}
\end{botherref}
\endbibitem

\bibitem[\protect\citeauthoryear{Ferziger and
  Peri{\'c}}{1996}]{Ferziger1996_springer}
\begin{bbook}
\bauthor{\bsnm{Ferziger}, \binits{J.H.}},
\bauthor{\bsnm{Peri{\'c}}, \binits{M.}}:
\bbtitle{{Computational Methods for Fluid Dynamics}}.
\bpublisher{Springer},
\blocation{Berlin, Heidelberg}
(\byear{1996}).
\doiurl{10.1007/978-3-642-97651-3}
\end{bbook}
\endbibitem

\bibitem[\protect\citeauthoryear{Spalart and Allmaras}{1992}]{spalart1992one}
\begin{bchapter}
\bauthor{\bsnm{Spalart}, \binits{P.}},
\bauthor{\bsnm{Allmaras}, \binits{S.}}:
\bctitle{A one-equation turbulence model for aerodynamic flows}.
In: \bbtitle{30th Aerospace Sciences Meeting and Exhibit},
p. \bfpage{439}
(\byear{1992})
\end{bchapter}
\endbibitem

\bibitem[\protect\citeauthoryear{Wilhelm}{2015}]{wilhelm2015rotating}
\begin{barticle}
\bauthor{\bsnm{Wilhelm}, \binits{D.}}:
\batitle{Rotating flow simulations with openfoam}.
\bjtitle{Int J Aeronautics Aerospace Res. S}
\bvolume{1},
\bfpage{1}--\blpage{7}
(\byear{2015})
\end{barticle}
\endbibitem

\bibitem[\protect\citeauthoryear{Spalart et~al.}{2006}]{spalart_new_2006}
\begin{barticle}
\bauthor{\bsnm{Spalart}, \binits{P.R.}},
\bauthor{\bsnm{Deck}, \binits{S.}},
\bauthor{\bsnm{Shur}, \binits{M.L.}},
\bauthor{\bsnm{Squires}, \binits{K.D.}},
\bauthor{\bsnm{Strelets}, \binits{M.K.}},
\bauthor{\bsnm{Travin}, \binits{A.}}:
\batitle{A {New} {Version} of {Detached}-eddy {Simulation}, {Resistant} to
  {Ambiguous} {Grid} {Densities}}.
\bjtitle{Theoretical and Computational Fluid Dynamics}
\bvolume{20}(\bissue{3}),
\bfpage{181}--\blpage{195}
(\byear{2006})
\doiurl{10.1007/s00162-006-0015-0}
\end{barticle}
\endbibitem

\bibitem[\protect\citeauthoryear{Carneiro et~al.}{2019}]{carneiro2019AMI}
\begin{barticle}
\bauthor{\bsnm{Carneiro}, \binits{F.}},
\bauthor{\bsnm{Moura}, \binits{L.}},
\bauthor{\bsnm{Rocha}, \binits{P.C.}},
\bauthor{\bsnm{Lima}, \binits{R.P.}},
\bauthor{\bsnm{Ismail}, \binits{K.}}:
\batitle{Application and analysis of the moving mesh algorithm ami in a small
  scale hawt: Validation with field test's results against the frozen rotor
  approach}.
\bjtitle{Energy}
\bvolume{171},
\bfpage{819}--\blpage{829}
(\byear{2019})
\end{barticle}
\endbibitem

\bibitem[\protect\citeauthoryear{Greenshields and
  Weller}{2022}]{greenshieldsweller2022}
\begin{bbook}
\bauthor{\bsnm{Greenshields}, \binits{C.}},
\bauthor{\bsnm{Weller}, \binits{H.}}:
\bbtitle{Notes on Computational Fluid Dynamics: General Principles}.
\bpublisher{CFD Direct Ltd},
\blocation{Reading, UK}
(\byear{2022})
\end{bbook}
\endbibitem

\bibitem[\protect\citeauthoryear{Caretto et~al.}{1973}]{caretto_simple_1973}
\begin{bchapter}
\bauthor{\bsnm{Caretto}, \binits{L.S.}},
\bauthor{\bsnm{Gosman}, \binits{A.D.}},
\bauthor{\bsnm{Patankar}, \binits{S.V.}},
\bauthor{\bsnm{Spalding}, \binits{D.B.}}:
\bctitle{Two calculation procedures for steady, three-dimensional flows with
  recirculation}.
In: \beditor{\bsnm{Cabannes}, \binits{H.}},
\beditor{\bsnm{Temam}, \binits{R.}} (eds.)
\bbtitle{Proceedings of the Third International Conference on Numerical Methods
  in Fluid Mechanics},
pp. \bfpage{60}--\blpage{68}.
\bpublisher{Springer},
\blocation{Berlin, Heidelberg}
(\byear{1973})
\end{bchapter}
\endbibitem

\bibitem[\protect\citeauthoryear{Issa et~al.}{1986}]{ISSA198666}
\begin{barticle}
\bauthor{\bsnm{Issa}, \binits{R.I.}},
\bauthor{\bsnm{Gosman}, \binits{A.D.}},
\bauthor{\bsnm{Watkins}, \binits{A.P.}}:
\batitle{The computation of compressible and incompressible recirculating flows
  by a non-iterative implicit scheme}.
\bjtitle{Journal of Computational Physics}
\bvolume{62}(\bissue{1}),
\bfpage{66}--\blpage{82}
(\byear{1986})
\doiurl{10.1016/0021-9991(86)90100-2}
\end{barticle}
\endbibitem

\bibitem[\protect\citeauthoryear{Jeanney et~al.}{2026}]{jeanney2026ensemble}
\begin{barticle}
\bauthor{\bsnm{Jeanney}, \binits{P.}},
\bauthor{\bsnm{Hetherington}, \binits{A.}},
\bauthor{\bsnm{Ahmed}, \binits{S.E.}},
\bauthor{\bsnm{Lanceta}, \binits{D.}},
\bauthor{\bsnm{Saiz}, \binits{S.}},
\bauthor{\bsnm{Perez}, \binits{J.M.}},
\bauthor{\bsnm{Le~Clainche}, \binits{S.}}:
\batitle{Ensemble kalman filter for data assimilation coupled with
  low-resolution computations techniques applied in fluid dynamics}.
\bjtitle{Applied Mathematics and Computation}
\bvolume{531},
\bfpage{130184}
(\byear{2026})
\end{barticle}
\endbibitem

\bibitem[\protect\citeauthoryear{Donato et~al.}{2024}]{donato2024integrating}
\begin{barticle}
\bauthor{\bsnm{Donato}, \binits{L.}},
\bauthor{\bsnm{Kamal}, \binits{M.M.}},
\bauthor{\bsnm{Procacci}, \binits{A.}},
\bauthor{\bsnm{Cafiero}, \binits{M.}},
\bauthor{\bsnm{Sharma}, \binits{S.}},
\bauthor{\bsnm{Galletti}, \binits{C.}},
\bauthor{\bsnm{Coussement}, \binits{A.}},
\bauthor{\bsnm{Parente}, \binits{A.}}:
\batitle{Integrating data assimilation and sparse sensing for updating a
  digital twin of a semi-industrial furnace}.
\bjtitle{Proceedings of the Combustion Institute}
\bvolume{40}(\bissue{1-4}),
\bfpage{105284}
(\byear{2024})
\end{barticle}
\endbibitem

\bibitem[\protect\citeauthoryear{He and Liu}{2020}]{he2020time}
\begin{botherref}
\oauthor{\bsnm{He}, \binits{C.}},
\oauthor{\bsnm{Liu}, \binits{Y.}}:
Time-resolved reconstruction of turbulent flows using linear stochastic
  estimation and sequential data assimilation.
Physics of Fluids
\textbf{32}(7)
(2020)
\end{botherref}
\endbibitem

\bibitem[\protect\citeauthoryear{Artana et~al.}{2012}]{Artana2012_jcp}
\begin{barticle}
\bauthor{\bsnm{Artana}, \binits{G.}},
\bauthor{\bsnm{Camilleri}, \binits{A.}},
\bauthor{\bsnm{Charlier}, \binits{J.}},
\bauthor{\bsnm{Memin}, \binits{E.}}:
\batitle{{Strong and weak constraint variational assimilations for reduced
  order fluid flow modeling}}.
\bjtitle{Journal of Computational Physics}
\bvolume{231},
\bfpage{3264}--\blpage{3288}
(\byear{2012})
\end{barticle}
\endbibitem

\bibitem[\protect\citeauthoryear{Foures et~al.}{2014}]{Foures2014_jfm}
\begin{barticle}
\bauthor{\bsnm{Foures}, \binits{D.P.G.}},
\bauthor{\bsnm{Dovetta}, \binits{N.}},
\bauthor{\bsnm{Sipp}, \binits{D.}},
\bauthor{\bsnm{Schmid}, \binits{P.J.}}:
\batitle{{A data-assimilation method for Reynolds-averaged Navier-Stokes-driven
  mean flow reconstruction}}.
\bjtitle{Journal of Fluid Mechanics}
\bvolume{759},
\bfpage{404}--\blpage{431}
(\byear{2014})
\end{barticle}
\endbibitem

\bibitem[\protect\citeauthoryear{Onder and Meyers}{2016}]{Onder2016_cf}
\begin{barticle}
\bauthor{\bsnm{Onder}, \binits{A.}},
\bauthor{\bsnm{Meyers}, \binits{J.}}:
\batitle{{Optimal control of a transitional jet using a continuous adjoint
  method}}.
\bjtitle{Computers \& Fluids}
\bvolume{126},
\bfpage{12}--\blpage{24}
(\byear{2016})
\end{barticle}
\endbibitem

\bibitem[\protect\citeauthoryear{Mons et~al.}{2016}]{Mons2016_jcp}
\begin{barticle}
\bauthor{\bsnm{Mons}, \binits{V.}},
\bauthor{\bsnm{Chassaing}, \binits{J.C.}},
\bauthor{\bsnm{Gomez}, \binits{T.}},
\bauthor{\bsnm{Sagaut}, \binits{P.}}:
\batitle{{Reconstruction of unsteady viscous flows using data assimilation
  schemes}}.
\bjtitle{Journal of Computational Physics}
\bvolume{316},
\bfpage{255}--\blpage{280}
(\byear{2016})
\end{barticle}
\endbibitem

\bibitem[\protect\citeauthoryear{Chandramouli
  et~al.}{2020}]{Chandramouli2020_jcp}
\begin{barticle}
\bauthor{\bsnm{Chandramouli}, \binits{P.}},
\bauthor{\bsnm{Memin}, \binits{E.}},
\bauthor{\bsnm{Heitz}, \binits{D.}}:
\batitle{{4D large scale variational data assimilation of a turbulent flow with
  a dynamics error model}}.
\bjtitle{Journal of Computational Physics}
\bvolume{412},
\bfpage{109446}
(\byear{2020})
\end{barticle}
\endbibitem

\bibitem[\protect\citeauthoryear{Padmanaban et~al.}{2026}]{Padmanaban2026_prf}
\begin{barticle}
\bauthor{\bsnm{Padmanaban}, \binits{U.C.}},
\bauthor{\bsnm{Ganapathisubramani}, \binits{B.}},
\bauthor{\bsnm{Symon}, \binits{S.}}:
\batitle{{Three-dimensional variational data assimilation of separated flows
  using time-averaged experimental data}}.
\bjtitle{Physical Review Fluids}
\bvolume{11},
\bfpage{014904}
(\byear{2026})
\end{barticle}
\endbibitem

\bibitem[\protect\citeauthoryear{Rochoux et~al.}{2014}]{Rochoux2014_nhess}
\begin{barticle}
\bauthor{\bsnm{Rochoux}, \binits{M.C.}},
\bauthor{\bsnm{Ricci}, \binits{S.}},
\bauthor{\bsnm{Lucor}, \binits{D.}},
\bauthor{\bsnm{Cuenot}, \binits{B.}},
\bauthor{\bsnm{Trouve}, \binits{A.}}:
\batitle{{Towards predictive data-driven simulations of wildfire spread - Part
  I: Reduced-cost Ensemble Kalman Filter based on a Polynomial Chaos surrogate
  model for parameter estimation}}.
\bjtitle{Natural Hazards and Earth System Sciences}
\bvolume{14},
\bfpage{2951}--\blpage{2973}
(\byear{2014})
\end{barticle}
\endbibitem

\bibitem[\protect\citeauthoryear{Leroux et~al.}{2014}]{Leroux2014_ep}
\begin{barticle}
\bauthor{\bsnm{Leroux}, \binits{R.}},
\bauthor{\bsnm{Chatellier}, \binits{L.}},
\bauthor{\bsnm{David}, \binits{L.}}:
\batitle{{Bayesian inference applied to spatio-temporal reconstruction of flows
  around a NACA0012 airfoil}}.
\bjtitle{Experiments in Fluids}
\bvolume{55},
\bfpage{1699}
(\byear{2014})
\end{barticle}
\endbibitem

\bibitem[\protect\citeauthoryear{Kato et~al.}{2015}]{Kato2015_jcp}
\begin{barticle}
\bauthor{\bsnm{Kato}, \binits{H.}},
\bauthor{\bsnm{Yoshizawa}, \binits{A.}},
\bauthor{\bsnm{Ueno}, \binits{G.}},
\bauthor{\bsnm{Obayashi}, \binits{S.}}:
\batitle{{A data assimilation methodology for reconstructing turbulent flows
  around aircraft}}.
\bjtitle{Journal of Computational Physics}
\bvolume{283},
\bfpage{559}--\blpage{581}
(\byear{2015})
\end{barticle}
\endbibitem

\bibitem[\protect\citeauthoryear{Meldi and Poux}{2017}]{Meldi2017_jcp}
\begin{barticle}
\bauthor{\bsnm{Meldi}, \binits{M.}},
\bauthor{\bsnm{Poux}, \binits{A.}}:
\batitle{{A reduced order model based on Kalman Filtering for sequential Data
  Assimilation of turbulent flows}}.
\bjtitle{Journal of Computational Physics}
\bvolume{347},
\bfpage{207}--\blpage{234}
(\byear{2017})
\end{barticle}
\endbibitem

\bibitem[\protect\citeauthoryear{Labahn et~al.}{2019}]{Labahn2019_pci}
\begin{barticle}
\bauthor{\bsnm{Labahn}, \binits{J.W.}},
\bauthor{\bsnm{Wu}, \binits{H.}},
\bauthor{\bsnm{Coriton}, \binits{B.}},
\bauthor{\bsnm{Frank}, \binits{J.H.}},
\bauthor{\bsnm{Ihme}, \binits{M.}}:
\batitle{{Data assimilation using high-speed measurements and LES to examine
  local extinction events in turbulent flames}}.
\bjtitle{Proceedings of the Combustion Institute}
\bvolume{37},
\bfpage{2259}--\blpage{2266}
(\byear{2019})
\end{barticle}
\endbibitem

\bibitem[\protect\citeauthoryear{Zhang et~al.}{2020}]{Zhang2020_jcp}
\begin{barticle}
\bauthor{\bsnm{Zhang}, \binits{X.}},
\bauthor{\bsnm{Michelin-Ströfer}, \binits{C.}},
\bauthor{\bsnm{Xiao}, \binits{H.}}:
\batitle{{Regularized ensemble Kalman methods for inverse problems}}.
\bjtitle{Journal of Computational Physics}
\bvolume{416},
\bfpage{109517}
(\byear{2020})
\end{barticle}
\endbibitem

\bibitem[\protect\citeauthoryear{Villanueva et~al.}{2024}]{Villanueva2024_cof}
\begin{barticle}
\bauthor{\bsnm{Villanueva}, \binits{L.}},
\bauthor{\bsnm{Valero}, \binits{M.M.}},
\bauthor{\bsnm{\v{S}arki\'{c}-Glumac}, \binits{A.}},
\bauthor{\bsnm{Meldi}, \binits{M.}}:
\batitle{Augmented state estimation of urban settings using on-the-fly
  sequential data assimilation}.
\bjtitle{Computers \& Fluids}
\bvolume{269},
\bfpage{106118}
(\byear{2024})
\end{barticle}
\endbibitem

\bibitem[\protect\citeauthoryear{Moussie et~al.}{2024}]{Moussie2024_ftc}
\begin{barticle}
\bauthor{\bsnm{Moussie}, \binits{T.}},
\bauthor{\bsnm{Errante}, \binits{P.}},
\bauthor{\bsnm{Meldi}, \binits{M.}}:
\batitle{Statistical inference of upstream turbulence intensity for the flow
  around a bluff body with massive separation}.
\bjtitle{Flow, Turbulence and Combustion}
\bvolume{113},
\bfpage{853}--\blpage{889}
(\byear{2024})
\end{barticle}
\endbibitem

\bibitem[\protect\citeauthoryear{Valero and Meldi}{2025}]{Valero2025_jcp}
\begin{barticle}
\bauthor{\bsnm{Valero}, \binits{M.M.}},
\bauthor{\bsnm{Meldi}, \binits{M.}}:
\batitle{An immersed boundary method using online sequential data
  assimilation}.
\bjtitle{Journal of Computational Physics}
\bvolume{524},
\bfpage{113697}
(\byear{2025})
\end{barticle}
\endbibitem

\bibitem[\protect\citeauthoryear{Villanueva et~al.}{2025}]{Villanueva2025_jfm}
\begin{barticle}
\bauthor{\bsnm{Villanueva}, \binits{L.}},
\bauthor{\bsnm{Truffin}, \binits{K.}},
\bauthor{\bsnm{Borée}, \binits{J.}},
\bauthor{\bsnm{Meldi}, \binits{M.}}:
\batitle{Enhancement of large-eddy simulations for the prediction of an intake
  flow rig using sequential data assimilation}.
\bjtitle{Journal of Fluid Mechanics}
\bvolume{1023},
\bfpage{47}
(\byear{2025})
\end{barticle}
\endbibitem

\bibitem[\protect\citeauthoryear{Özalp et~al.}{2026}]{Ozalp2026_cmame}
\begin{barticle}
\bauthor{\bsnm{Özalp}, \binits{E.}},
\bauthor{\bsnm{Nóvoa}, \binits{A.}},
\bauthor{\bsnm{Magri}, \binits{L.}}:
\batitle{Real-time forecasting of chaotic dynamics from sparse data and
  autoencoders}.
\bjtitle{Computer Methods in Applied Mechanics and Engineering}
\bvolume{450},
\bfpage{118600}
(\byear{2026})
\end{barticle}
\endbibitem

\bibitem[\protect\citeauthoryear{Villanueva
  et~al.}{2024}]{Villanueva2024_ijhff}
\begin{barticle}
\bauthor{\bsnm{Villanueva}, \binits{L.}},
\bauthor{\bsnm{Truffin}, \binits{K.}},
\bauthor{\bsnm{Meldi}, \binits{M.}}:
\batitle{Synchronization and optimization of large eddy simulation using an
  online ensemble kalman filter}.
\bjtitle{International Journal of Heat and Fluid Flow}
\bvolume{110},
\bfpage{109597}
(\byear{2024})
\end{barticle}
\endbibitem

\bibitem[\protect\citeauthoryear{Moldovan et~al.}{2024}]{Moldovan2024_cf}
\begin{barticle}
\bauthor{\bsnm{Moldovan}, \binits{G.}},
\bauthor{\bsnm{Mariotti}, \binits{A.}},
\bauthor{\bsnm{Lehnasch}, \binits{G.}},
\bauthor{\bsnm{Cordier}, \binits{L.}},
\bauthor{\bsnm{Salvetti}, \binits{M.V.}},
\bauthor{\bsnm{Meldi}, \binits{M.}}:
\batitle{{Multigrid sequential data assimilation for the large-eddy simulation
  of a massively separated bluff-body flow}}.
\bjtitle{Computers and Fluids}
\bvolume{281},
\bfpage{106385}
(\byear{2024})
\end{barticle}
\endbibitem

\bibitem[\protect\citeauthoryear{Kalman}{1960}]{kalman_new_1960}
\begin{barticle}
\bauthor{\bsnm{Kalman}, \binits{R.E.}}:
\batitle{A {New} {Approach} to {Linear} {Filtering} and {Prediction}
  {Problems}}.
\bjtitle{Journal of Basic Engineering}
\bvolume{82}(\bissue{1}),
\bfpage{35}--\blpage{45}
(\byear{1960})
\doiurl{10.1115/1.3662552}
\end{barticle}
\endbibitem

\bibitem[\protect\citeauthoryear{Moldovan et~al.}{2021}]{Moldovan2021_jcp}
\begin{barticle}
\bauthor{\bsnm{Moldovan}, \binits{G.}},
\bauthor{\bsnm{Lehnasch}, \binits{G.}},
\bauthor{\bsnm{Cordier}, \binits{L.}},
\bauthor{\bsnm{Meldi}, \binits{M.}}:
\batitle{{A multigrid/ensemble Kalman filter strategy for assimilation of
  unsteady flows}}.
\bjtitle{Journal of Computational Physics}
\bvolume{443},
\bfpage{110481}
(\byear{2021})
\end{barticle}
\endbibitem

\bibitem[\protect\citeauthoryear{Moldovan et~al.}{2022}]{Moldovan2022_jcp}
\begin{barticle}
\bauthor{\bsnm{Moldovan}, \binits{G.}},
\bauthor{\bsnm{Lehnasch}, \binits{G.}},
\bauthor{\bsnm{Cordier}, \binits{L.}},
\bauthor{\bsnm{Meldi}, \binits{M.}}:
\batitle{{Optimized parametric inference for the inner loop of the Multigrid
  Ensemble Kalman Filter}}.
\bjtitle{Journal of Computational Physics}
\bvolume{471},
\bfpage{111621}
(\byear{2022})
\end{barticle}
\endbibitem

\bibitem[\protect\citeauthoryear{Rannou et~al.}{2022}]{rannou2022effect}
\begin{bchapter}
\bauthor{\bsnm{Rannou}, \binits{C.}},
\bauthor{\bsnm{Dazin}, \binits{A.}},
\bauthor{\bsnm{Marty}, \binits{J.}},
\bauthor{\bsnm{Tanguy}, \binits{G.}},
\bauthor{\bsnm{Castillon}, \binits{L.}},
\bauthor{\bsnm{Moubogha}, \binits{J.}}:
\bctitle{{Effect of the Axial Compressor Tip Clearance Size: Performance and
  Transition to Rotating Stall}}.
In: \bbtitle{Proceedings of the ASME Turbo Expo 2022: Turbomachinery Technical
  Conference and Exposition},
vol. \bseriesno{10A: Turbomachinery - Axial Flow Fan and Compressor
  Aerodynamics}.
\bpublisher{American Society of Mechanical Engineers},
\blocation{Rotterdam, Netherlands}
(\byear{2022}).
\bcomment{V10AT29A011}
\end{bchapter}
\endbibitem

\bibitem[\protect\citeauthoryear{Marty et~al.}{2023}]{marty2023numerical}
\begin{barticle}
\bauthor{\bsnm{Marty}, \binits{J.}},
\bauthor{\bsnm{Castillon}, \binits{L.}},
\bauthor{\bsnm{Joseph}, \binits{P.}}:
\batitle{Numerical investigations on the rotating stall in an axial compressor
  and its control by flow injection at casing}.
\bjtitle{Journal of Turbomachinery}
\bvolume{145}(\bissue{5}),
\bfpage{051009}
(\byear{2023})
\end{barticle}
\endbibitem

\bibitem[\protect\citeauthoryear{Rannou}{2023}]{rannou2023methodologie}
\begin{botherref}
\oauthor{\bsnm{Rannou}, \binits{C.}}:
M{\'e}thodologie pour la mod{\'e}lisation du contr{\^o}le actif
  d’{\'e}coulement dans un compresseur axial.
PhD thesis,
HESAM Universit{\'e}
(2023)
\end{botherref}
\endbibitem

\bibitem[\protect\citeauthoryear{Errante et~al.}{2026}]{errante_2026_20433499}
\begin{botherref}
\oauthor{\bsnm{Errante}, \binits{P.}},
\oauthor{\bsnm{Meldi}, \binits{M.}},
\oauthor{\bsnm{Moussie}, \binits{T.}},
\oauthor{\bsnm{Martinez-Valero}, \binits{M.}},
\oauthor{\bsnm{Er}, \binits{S.}},
\oauthor{\bsnm{Villanueva}, \binits{L.}}:
CONES.
Zenodo
(2026).
\doiurl{10.5281/zenodo.20433499} .
\url{https://doi.org/10.5281/zenodo.20433499}
\end{botherref}
\endbibitem

\bibitem[\protect\citeauthoryear{Cybenko}{1989}]{cybenko1989approximation}
\begin{barticle}
\bauthor{\bsnm{Cybenko}, \binits{G.}}:
\batitle{Approximation by superpositions of a sigmoidal function}.
\bjtitle{Mathematics of control, signals and systems}
\bvolume{2}(\bissue{4}),
\bfpage{303}--\blpage{314}
(\byear{1989})
\end{barticle}
\endbibitem

\bibitem[\protect\citeauthoryear{Alloghani
  et~al.}{2020}]{alloghani2020systematic}
\begin{botherref}
\oauthor{\bsnm{Alloghani}, \binits{M.}},
\oauthor{\bsnm{Al-Jumeily}, \binits{D.}},
\oauthor{\bsnm{Mustafina}, \binits{J.}},
\oauthor{\bsnm{Hussain}, \binits{A.}},
\oauthor{\bsnm{Aljaaf}, \binits{A.J.}}:
A systematic review on supervised and unsupervised machine learning algorithms
  for data science.
Supervised and unsupervised learning for data science,
3--21
(2020)
\end{botherref}
\endbibitem

\bibitem[\protect\citeauthoryear{Alsina et~al.}{2018}]{alsina2018use}
\begin{barticle}
\bauthor{\bsnm{Alsina}, \binits{E.F.}},
\bauthor{\bsnm{Chica}, \binits{M.}},
\bauthor{\bsnm{Trawi{\'n}ski}, \binits{K.}},
\bauthor{\bsnm{Regattieri}, \binits{A.}}:
\batitle{On the use of machine learning methods to predict component
  reliability from data-driven industrial case studies}.
\bjtitle{The International Journal of Advanced Manufacturing Technology}
\bvolume{94}(\bissue{5}),
\bfpage{2419}--\blpage{2433}
(\byear{2018})
\end{barticle}
\endbibitem

\bibitem[\protect\citeauthoryear{Kapp et~al.}{2023}]{kapp2023predictingEnergy}
\begin{barticle}
\bauthor{\bsnm{Kapp}, \binits{S.}},
\bauthor{\bsnm{Choi}, \binits{J.-K.}},
\bauthor{\bsnm{Hong}, \binits{T.}}:
\batitle{Predicting industrial building energy consumption with statistical and
  machine-learning models informed by physical system parameters}.
\bjtitle{Renewable and Sustainable Energy Reviews}
\bvolume{172},
\bfpage{113045}
(\byear{2023})
\end{barticle}
\endbibitem

\bibitem[\protect\citeauthoryear{Sharma et~al.}{2017}]{sharma2017activation}
\begin{barticle}
\bauthor{\bsnm{Sharma}, \binits{S.}},
\bauthor{\bsnm{Sharma}, \binits{S.}},
\bauthor{\bsnm{Athaiya}, \binits{A.}}:
\batitle{Activation functions in neural networks}.
\bjtitle{Towards Data Sci}
\bvolume{6}(\bissue{12}),
\bfpage{310}--\blpage{316}
(\byear{2017})
\end{barticle}
\endbibitem

\bibitem[\protect\citeauthoryear{Vinutha
  et~al.}{2018}]{vinutha2018detectionIQR}
\begin{bchapter}
\bauthor{\bsnm{Vinutha}, \binits{H.}},
\bauthor{\bsnm{Poornima}, \binits{B.}},
\bauthor{\bsnm{Sagar}, \binits{B.}}:
\bctitle{Detection of outliers using interquartile range technique from
  intrusion dataset}.
In: \bbtitle{Information and Decision Sciences: Proceedings of the 6th
  International Conference on Ficta},
pp. \bfpage{511}--\blpage{518}
(\byear{2018}).
\bcomment{Springer}
\end{bchapter}
\endbibitem

\bibitem[\protect\citeauthoryear{Jais et~al.}{2019}]{jais_adam_2019}
\begin{barticle}
\bauthor{\bsnm{Jais}, \binits{I.K.M.}},
\bauthor{\bsnm{Ismail}, \binits{A.R.}},
\bauthor{\bsnm{Nisa}, \binits{S.Q.}}:
\batitle{Adam {Optimization} {Algorithm} for {Wide} and {Deep} {Neural}
  {Network}}.
\bjtitle{Knowledge Engineering and Data Science}
\bvolume{2}(\bissue{1}),
\bfpage{41}
(\byear{2019})
\doiurl{10.17977/um018v2i12019p41-46}
\end{barticle}
\endbibitem

\bibitem[\protect\citeauthoryear{Abadi
  et~al.}{2015}]{tensorflow2015-whitepaper}
\begin{botherref}
\oauthor{\bsnm{Abadi}, \binits{M.}},
\oauthor{\bsnm{Agarwal}, \binits{A.}},
\oauthor{\bsnm{Barham}, \binits{P.}},
\oauthor{\bsnm{Brevdo}, \binits{E.}},
\oauthor{\bsnm{Chen}, \binits{Z.}},
\oauthor{\bsnm{Citro}, \binits{C.}},
\oauthor{\bsnm{Corrado}, \binits{G.S.}},
\oauthor{\bsnm{Davis}, \binits{A.}},
\oauthor{\bsnm{Dean}, \binits{J.}},
\oauthor{\bsnm{Devin}, \binits{M.}},
\oauthor{\bsnm{Ghemawat}, \binits{S.}},
\oauthor{\bsnm{Goodfellow}, \binits{I.}},
\oauthor{\bsnm{Harp}, \binits{A.}},
\oauthor{\bsnm{Irving}, \binits{G.}},
\oauthor{\bsnm{Isard}, \binits{M.}},
\oauthor{\bsnm{Jia}, \binits{Y.}},
\oauthor{\bsnm{Jozefowicz}, \binits{R.}},
\oauthor{\bsnm{Kaiser}, \binits{L.}},
\oauthor{\bsnm{Kudlur}, \binits{M.}},
\oauthor{\bsnm{Levenberg}, \binits{J.}},
\oauthor{\bsnm{Man\'{e}}, \binits{D.}},
\oauthor{\bsnm{Monga}, \binits{R.}},
\oauthor{\bsnm{Moore}, \binits{S.}},
\oauthor{\bsnm{Murray}, \binits{D.}},
\oauthor{\bsnm{Olah}, \binits{C.}},
\oauthor{\bsnm{Schuster}, \binits{M.}},
\oauthor{\bsnm{Shlens}, \binits{J.}},
\oauthor{\bsnm{Steiner}, \binits{B.}},
\oauthor{\bsnm{Sutskever}, \binits{I.}},
\oauthor{\bsnm{Talwar}, \binits{K.}},
\oauthor{\bsnm{Tucker}, \binits{P.}},
\oauthor{\bsnm{Vanhoucke}, \binits{V.}},
\oauthor{\bsnm{Vasudevan}, \binits{V.}},
\oauthor{\bsnm{Vi\'{e}gas}, \binits{F.}},
\oauthor{\bsnm{Vinyals}, \binits{O.}},
\oauthor{\bsnm{Warden}, \binits{P.}},
\oauthor{\bsnm{Wattenberg}, \binits{M.}},
\oauthor{\bsnm{Wicke}, \binits{M.}},
\oauthor{\bsnm{Yu}, \binits{Y.}},
\oauthor{\bsnm{Zheng}, \binits{X.}}:
{TensorFlow}: Large-Scale Machine Learning on Heterogeneous Systems.
Software available from tensorflow.org
(2015).
\url{https://www.tensorflow.org/}
\end{botherref}
\endbibitem

\bibitem[\protect\citeauthoryear{Chollet et~al.}{2015}]{chollet2015keras}
\begin{botherref}
\oauthor{\bsnm{Chollet}, \binits{F.}}, et al.:
Keras.
\url{https://keras.io}
(2015)
\end{botherref}
\endbibitem

\bibitem[\protect\citeauthoryear{Chicco et~al.}{2021}]{chicco2021coefficientR2}
\begin{barticle}
\bauthor{\bsnm{Chicco}, \binits{D.}},
\bauthor{\bsnm{Warrens}, \binits{M.J.}},
\bauthor{\bsnm{Jurman}, \binits{G.}}:
\batitle{The coefficient of determination r-squared is more informative than
  smape, mae, mape, mse and rmse in regression analysis evaluation}.
\bjtitle{Peerj computer science}
\bvolume{7},
\bfpage{623}
(\byear{2021})
\end{barticle}
\endbibitem

\bibitem[\protect\citeauthoryear{Gropp et~al.}{1994}]{gropp1994}
\begin{bbook}
\bauthor{\bsnm{Gropp}, \binits{W.}},
\bauthor{\bsnm{Lusk}, \binits{E.}},
\bauthor{\bsnm{Skjellum}, \binits{A.}}:
\bbtitle{Using MPI: Portable Parallel Programming with the Message-passing
  Interface}.
\bpublisher{MIT Press},
\blocation{Cambridge, MA, USA}
(\byear{1994})
\end{bbook}
\endbibitem

\bibitem[\protect\citeauthoryear{Barker}{2015}]{barker2015MPI}
\begin{bchapter}
\bauthor{\bsnm{Barker}, \binits{B.}}:
\bctitle{Message passing interface (mpi)}.
In: \bbtitle{Workshop: High Performance Computing on Stampede},
vol. \bseriesno{262}
(\byear{2015}).
\bcomment{Cornell University Publisher Houston, TX, USA}
\end{bchapter}
\endbibitem

\bibitem[\protect\citeauthoryear{Masoudnia and
  Ebrahimpour}{2014}]{masoudnia2014mixture}
\begin{barticle}
\bauthor{\bsnm{Masoudnia}, \binits{S.}},
\bauthor{\bsnm{Ebrahimpour}, \binits{R.}}:
\batitle{Mixture of experts: a literature survey}.
\bjtitle{Artificial Intelligence Review}
\bvolume{42}(\bissue{2}),
\bfpage{275}--\blpage{293}
(\byear{2014})
\end{barticle}
\endbibitem

\bibitem[\protect\citeauthoryear{Oulghelou et~al.}{2025}]{oulghelou2025machine}
\begin{barticle}
\bauthor{\bsnm{Oulghelou}, \binits{M.}},
\bauthor{\bsnm{Cherroud}, \binits{S.}},
\bauthor{\bsnm{Merle}, \binits{X.}},
\bauthor{\bsnm{Cinnella}, \binits{P.}}:
\batitle{Machine-learning-assisted blending of data-driven turbulence models}.
\bjtitle{Flow, Turbulence and Combustion}
\bvolume{115}(\bissue{3}),
\bfpage{1095}--\blpage{1132}
(\byear{2025})
\end{barticle}
\endbibitem

\bibitem[\protect\citeauthoryear{Rubinstein and
  Marcus}{1985}]{rubinstein_efficiency_1985}
\begin{barticle}
\bauthor{\bsnm{Rubinstein}, \binits{R.Y.}},
\bauthor{\bsnm{Marcus}, \binits{R.}}:
\batitle{Efficiency of {Multivariate} {Control} {Variates} in {Monte} {Carlo}
  {Simulation}}.
\bjtitle{Operations Research}
\bvolume{33}(\bissue{3}),
\bfpage{661}--\blpage{677}
(\byear{1985})
\doiurl{10.1287/opre.33.3.661}
\end{barticle}
\endbibitem

\end{thebibliography}
